\documentclass[]{aastex631}%linenumbers
\usepackage{ gensymb }

\shorttitle{Modified microvariability model}
\shortauthors{Xu et al.}
\graphicspath{{./}{figures/}}
\begin{document}

\title{A small scale structure model of jet based on the observation of microvariability}

\author[0000-0003-2477-3430]{Jingran Xu}
\affiliation{Shandong Provincial Key Laboratory of Optical Astronomy and Solar-Terrestrial Environment, \\ School of Space Science and Physics, Institute of Space Sciences, Shandong University, Weihai, 264209, China}

\author[0000-0003-3217-7794]{Shaoming Hu}
\affiliation{Shandong Provincial Key Laboratory of Optical Astronomy and Solar-Terrestrial Environment, \\ School of Space Science and Physics, Institute of Space Sciences, Shandong University, Weihai, 264209, China}

\author[0000-0001-5603-7521]{Xu Chen}
\affiliation{Shandong Provincial Key Laboratory of Optical Astronomy and Solar-Terrestrial Environment, \\ School of Space Science and Physics, Institute of Space Sciences, Shandong University, Weihai, 264209, China}

\author[0000-0003-2679-0445]{Yunguo Jiang}
\affiliation{Shandong Provincial Key Laboratory of Optical Astronomy and Solar-Terrestrial Environment, \\ School of Space Science and Physics, Institute of Space Sciences, Shandong University, Weihai, 264209, China}

\author[0000-0002-8709-4665]{Sofya Alexeeva}
\affiliation{CAS Key Laboratory of Optical Astronomy, National Astronomical Observatories, \\ Chinese Academy of Sciences, Beijing, 100102, China}

\correspondingauthor{Shaoming Hu}
\email{husm@sdu.edu.cn}

\begin{abstract}

We developed a multi-region radiation model for the evolution of flux and spectral index with time. In this model, each perturbation component in the jet produces an independent flare. The model can be used to study the decomposition of microvariability, the structural scale of the perturbed components, and the physical parameters of the acceleration processes. Based on the shock acceleration model in relativistic jet, the influence of acceleration parameters on multiband flare parameters is calculated. We present the results of multiband optical microvariability of the blazar BL Lacertae observed performed during 89 nights in the period from 2009 to 2021, and use them as a sample for model fitting. The results show that both the amplitude and duration of flares decomposed from the microvariability light curves confirm a lognormal distribution. The time delays between the optical bands follow the normal distribution and amount to several minutes, that corroborates with both predictions from the theoretical model and the calculation of the discrete correlation function (DCF). Using the spectral index evolution and the simultaneous fitting of the multiband variability curves, we obtain the acceleration and radiation parameters to constrain and distinguish the origins of different flares. Based on the flare decomposition, we can well reproduce the time-domain evolution trends of the optical variation and energy spectrum, and explain the various redder-when-brighter (RWB) and/or bluer-when-brighter (BWB) behavior.

\end{abstract}


\section{Introduction} \label{sec:intro}

Blazars belong to a special subclass of active galactic nuclei (AGNs), where the trajectory of plasma in relativistic jet is closely aligned to the line of sight. Their ultraluminous emission, high amplitudes of variability on short timescales at various wavelengths, the high degree of polarization and other observational features will facilitate the study of the physical properties of the blazar jet. Variability, as one of the most obvious features, can help us to study the acceleration processes and radiation mechanisms in the jet, and provide a clearer understanding of the location, size, structure and dynamic evolution of the emission region.

Blazars are divided into flat spectrum radio quasars and BL Lacertae objects (BL Lacs). As the prototype of BL lacs, BL Lacertae \citep[hereafter BL Lac; 1ES 2200+42; redshift z = 0.0686;][]{Mil78} is usually classified as low frequency peaked Blazar \citep[LBL;][]{Nil18}, but it is sometimes listed as an intermediate frequency peaked Blazar \citep[IBL;][]{Ack11}. Long term multiwavelength monitoring campaigns have been carried out on BL Lac, like the past campaigns of the Whole Earth Blazar Telescope \citep{Vil02} dedicated to this source, providing abundant data sets \citep[e.g.][]{Blo97, Mad99, Sam99, Mar08, Vil09, Bot13, Rai13, Aga15, Gau15, Abe18, Bha18a}.

A number of studies have been performed to analyze the properties of BL Lac \citep[e.g.][]{Hag02, Sak13}. The multiwavelength observation data are analyzed by single or multi-region radiation model, and the physical properties of the emission region were obtained. The increasing of multiband observations and improvement of time resolution allow us to refine the analysis of jet emission region better and better. Some studies with short term variability (STV; its timescale ranges from days to months) analysis show that the multi region model can better reproduce the observations compared to the case, which the single region model is applied. The X-ray variability of BL Lac, except for a few major flaring events, has a lognormal distribution \citep{Gie09}, meaning that the emission is the product of the superposition of a large number of independent random events.

To explore the radiation mechanism, geometric structure and complex acceleration process of blazar jet, observations of variability on both short and long timescales are needed. According to timeseries studies of long term variability (LTV; its timescale ranges from months to years, or even decades) for a lot of blazars, there is a correlation between light curves in different bands and the time delays are ranging from zero to several days \citep[e.g.][]{Cha12, Jor13, Rai13, Dam19}. The delays obtained by using LTV to calculate correlations are usually comparable to their uncertainties. This limitation can be overcome by STV observation. The high cadence observations of STV will ensure to improve the correlation analysis accuracy and explores details of the characteristic time scales and variations at different wavelengths \citep{Utt02, Wea20}.

According to the light travel time of microvariability and the causality argument, it can be concluded that the emission region should be highly compact space volumes. This dense small scale emission region can not be spatially resolved by any existing instruments, so multiwavelength microvariability research may be one of the most powerful tools, which limits the properties of general physical processes, such as particle acceleration and energy dissipation mechanism, magnetic field geometry, jet composition and so on. In order to explain STV behavior, several models have been proposed. Most of them connect the source of intraday variability (IDV; or known as microvariability) with some physical processes in the accretion disk and jet. These models include the emission region rotating around the central source, variable shielding, various magneto-hydrodynamic instabilities, the propagation of shock along turbulence in the jet, and the projection effect of relativistic jet relative to the line of sight \citep[e.g.][]{Mar85, Cam92, Bha13}. However, the specific processes and the origin of micro variability in details are still unclear.

We present the results of the microvariability light curves analysis from continuous monitoring of BL Lac in Weihai Observatory of Shandong University during 89 night covering the 2009-2021 period. Taking the model of \cite{Kir98} and \cite{Bha18a} as reference, we improved a small scale model to explain the origin of microvariability. We assume that the microvariability originates from each independent flare event, which may occur in different emission regions, so the flare parameters may be different. The possible physical parameters of each flare are set under the condition if the observed flux and spectral time domain evolution of variability are reproduced by modelling. It allows us to reach self consistency in explanation of various observed behavior. 

This paper is organised as follows. In section \ref{sec:obs}, the observations and data reduction methods of BL Lac are introduced. The improved model and parameters on flare and related theoretical analysis are described in section \ref{sec:model}. In section \ref{sec:result}, based on theoretical analysis, we fit and reproduce the physical model with the results obtained from the observation data, and carry out statistical analysis and related discussion. Section \ref{sec:sum} is dedicated to the summary and discussion.

\section{Observations and Data Reductions} \label{sec:obs}

Photometry observations were carried out with the 1.0 m telescope at Weihai Observatory of Shandong University \citep{Hu14} from 2009 to 2021 (MJD: 55070-59352). The telescope is equipped with  a back-illuminated PIXIS 2048B CCD camera and standard $B$, $V$ (Johnson) and $R$, $I$ (Cousins) filters. The telescope is a classical Cassegrain design with a focal ratio of $f/8$. Quasi-simultaneous optical multiband observations provide colour behavior and time lags between different bands. Standard differential photometry method was used to reduce all frames. Comparison stars were taken from \cite{Fio96} for the $V$ , $R$ and $I$-band, while for the $B$-band we adopted the values by \cite{Ber69}. Stars B, C, K and H of the finding chart were selected as comparison stars.

\begin{figure*}
	\begin{minipage}{\textwidth}
		\centering
		\includegraphics[trim=2.5cm 1.6cm 0cm 0cm,width=0.48\textwidth,clip]{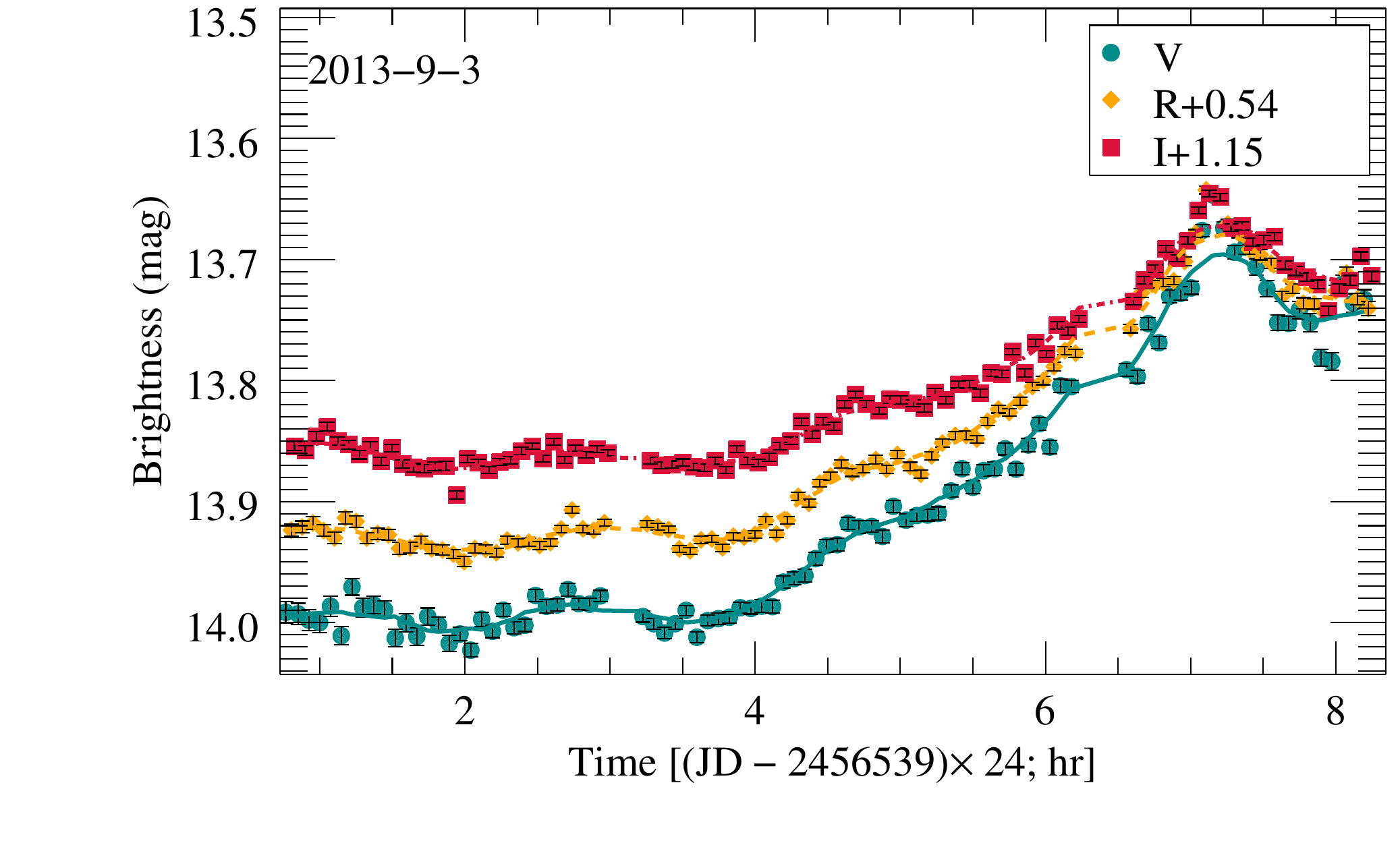}
		\includegraphics[trim=2.5cm 1.6cm 0cm 0cm,width=0.48\textwidth,clip]{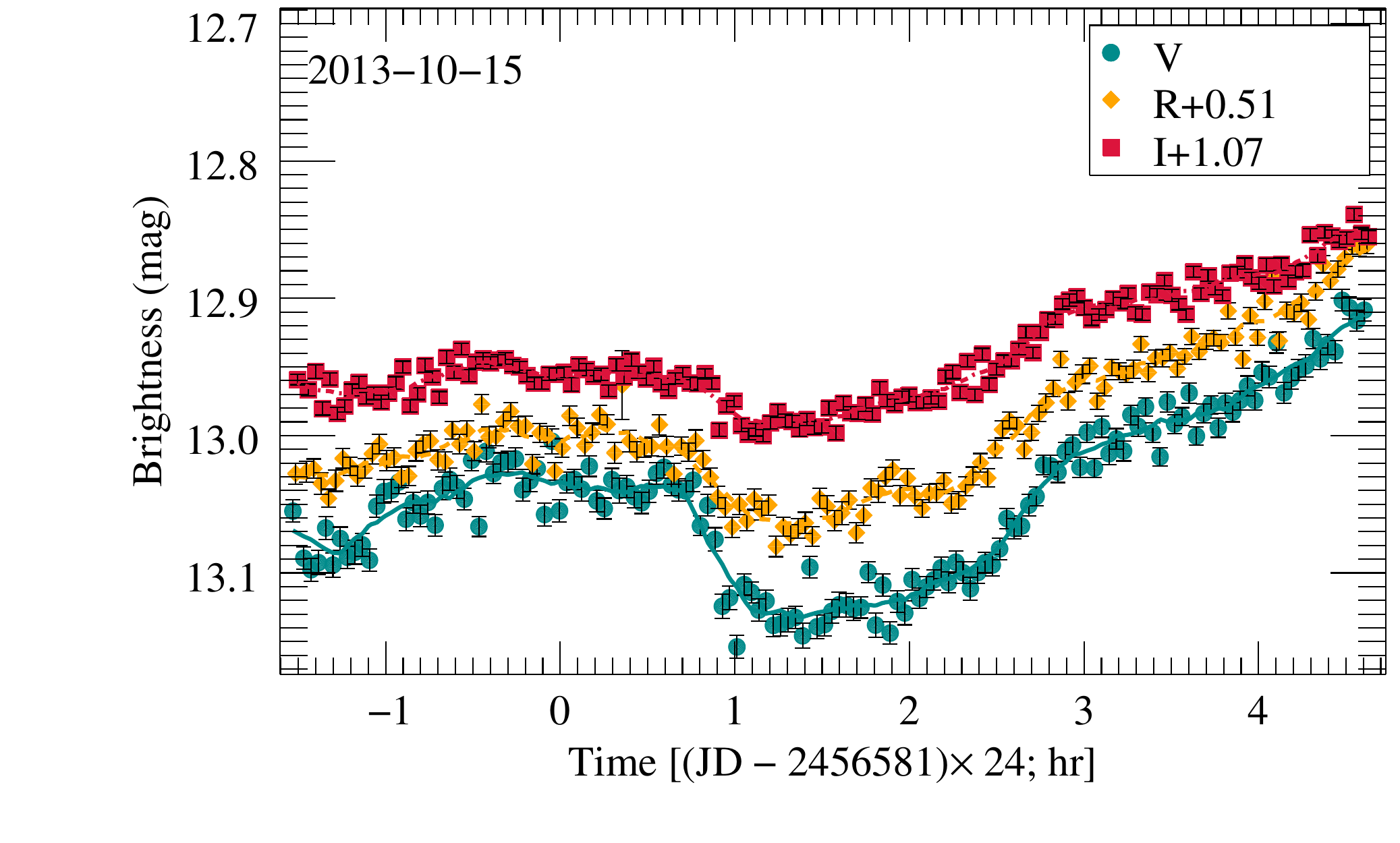}
		\includegraphics[trim=2.5cm 1.6cm 0cm 0cm,width=0.48\textwidth,clip]{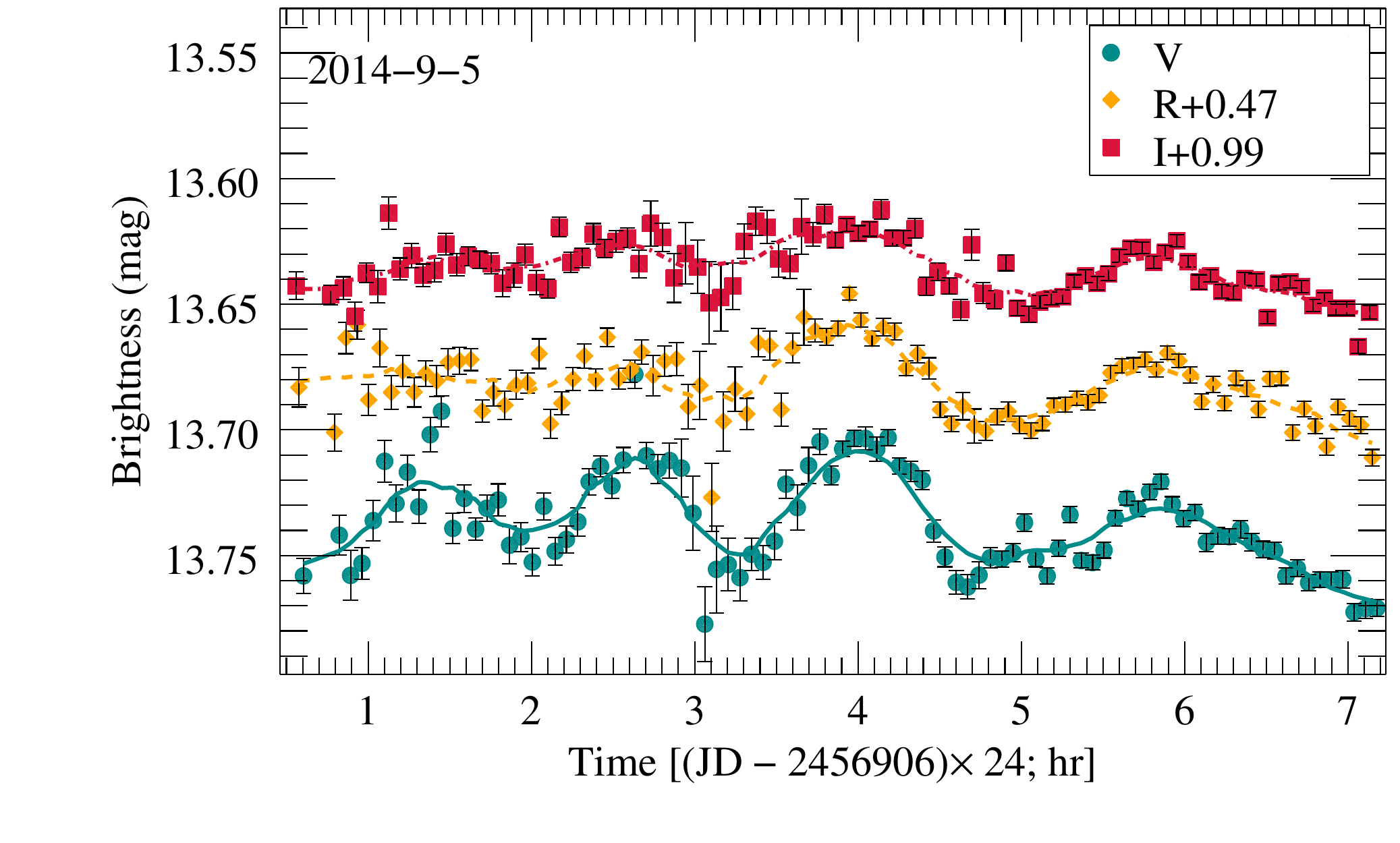}
		\includegraphics[trim=2.5cm 1.6cm 0cm 0cm,width=0.48\textwidth,clip]{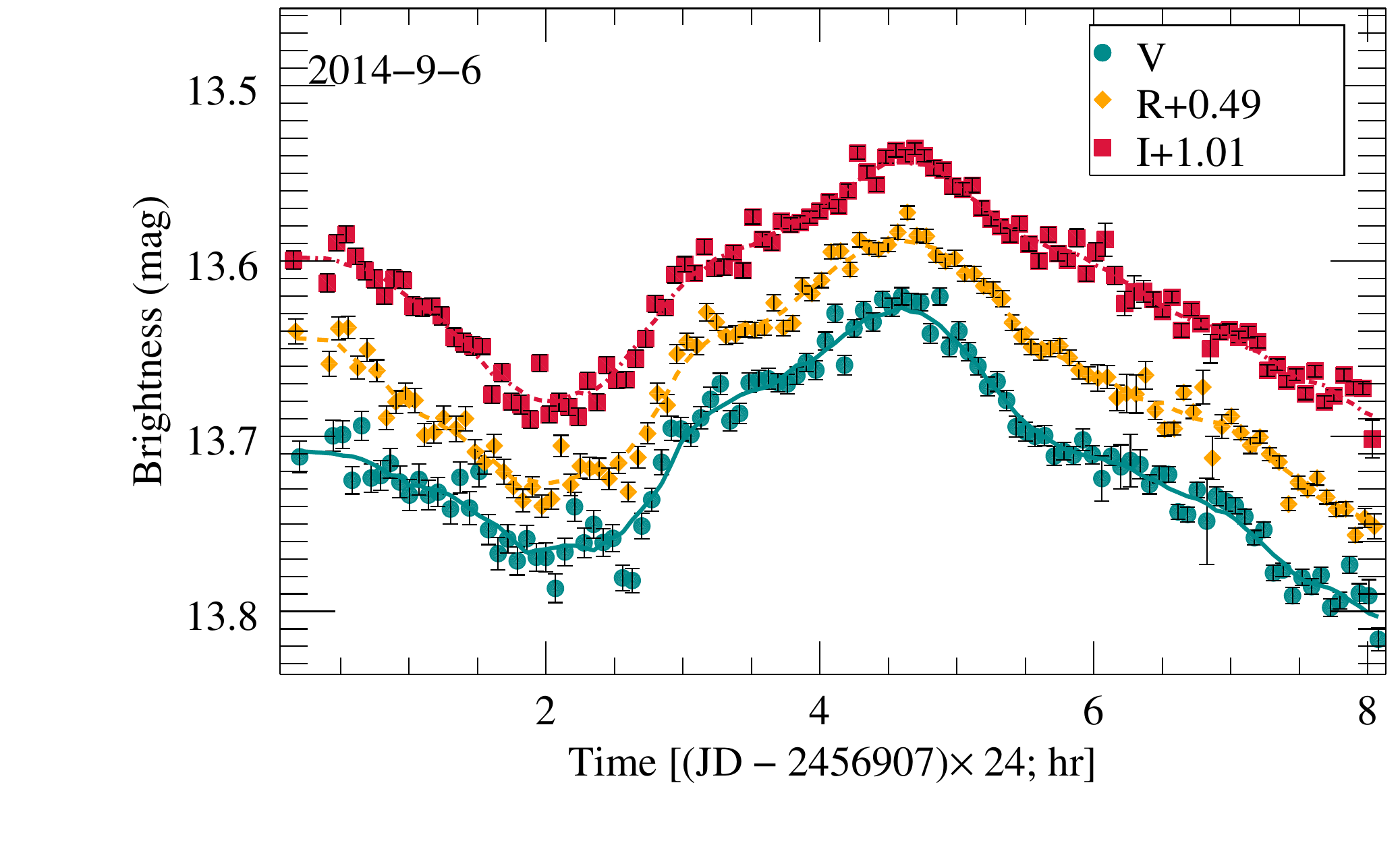}
	\end{minipage}\vspace{0.001cm}
	\caption{Four examples of IDV light curves of BL Lac from Weihai observatory of Shandong University. The circles, rhombuses and squares represent the IDV light curves in $V$, $R$ and $I$-band for which galactic extinction corrections were performed, respectively. For easy comparison of the variations between different bands, the $R$ and $I$-band data are shifted with respect to the $V$-band data, respectively. The solid lines of cyan, yellow and red represent the light curves of $V$, $R$ and $I$-band after smoothing and interpolation resampling, respectively.}
	\label{fig:lightcurve}
\end{figure*}

All object frames were automatically processed by the aperture photometry program compiled by \cite{Che14}. Standard image processing (bias subtraction and flat fielding using twilight-sky exposures) was applied to all frames. The photometry data are divided into different IDV light curves according to the date, and select light curves with high observation accuracy and dense time sampling. Ensuring at least 50 data points per day and per band, we screened 89 nights of data from the observed LTV, including 2 nights in $B$-band, 86 nights in $I$-band, and 87 nights in both $R$ and $V$-band. Four IDV light curves are shown as examples in Figure \ref{fig:lightcurve}. The complete IDV figure set (89 images) is available in the online journal. The observations with photometry error less than 0.05 were used only, to ensure the accuracy of the model fits. The source brightness in magnitudes was converted into the flux in Jansky following \cite{Bes79}. Galactic extinction along the line of sight to BL Lac was calculated according to \cite{Car89}, and we used the extinction (in magnitudes) in the $B$, $V$, $R$, and $I$ filters: $A_{B} = 1.16$, $A_{V} = 0.87$, $A_{R} = 0.74$, and $A_{I} = 0.53$ from \cite{Sch11}. 

The observed data were smoothed under assumption that there are no instantaneous flares ($\textless 1$ min). The smoothed IDV curves are shown by the lines in Figure \ref{fig:lightcurve}. The employed smoothing algorithm eliminates the local features of time sampling interval scale, and does not affect the overall  variability curves.  In order to be sure that there is no universal periodicity in the light curves, we carried out the discrete fourier transform analysis (DFT). In addition, the DFT method was used to analyze the frequency domain characteristics of several IDV curves. 

Due to the limitations of observation, the observation time of data in different bands is quasi-simultaneous rather than completely simultaneous. We used cubic spline interpolation to reproduce the IDV curves and calculate the evolution of the spectral indices with time. Interpolation processing is also applied to make continuous interval of data. The purpose is to reduce the error caused by the boundary value of discontinuous points in the fitting process. The fitting parameters obtained by this interpolation are not included in the statistical results, so this interpolation will not affect the final model fitting results.

\section{Model} \label{sec:model}

Many investigations have studied the microscopic variability of blazars \citep[e.g.][]{Qui91, Bha18b, Web21}. These works indicate that the microvariability can be decomposed into pulses or snapshots. Here, we develop a flare model that is consistent with the observations.

\subsection{Physical Model} \label{subsec:physical}

It is assumed that the emission from blazar jet consists of the stationary emission and flaring events \citep{Kir98, Pot18, Bha21}. The phenomenon of optical variation can be attributed to various factors, such as alterations in the Doppler factor $\delta$ resulting from changes in visual angle, and variations in the radiation flux stemming from the evolution of particle distribution, among other contributing factors. The model considers that flares emanate from the radiation cooling of particles within the jet, subsequent to their acceleration.

\begin{figure*}
	\centering
	\includegraphics [trim=0.0cm 0.0cm 0.0cm 0.0cm,width=0.99\textwidth,clip]{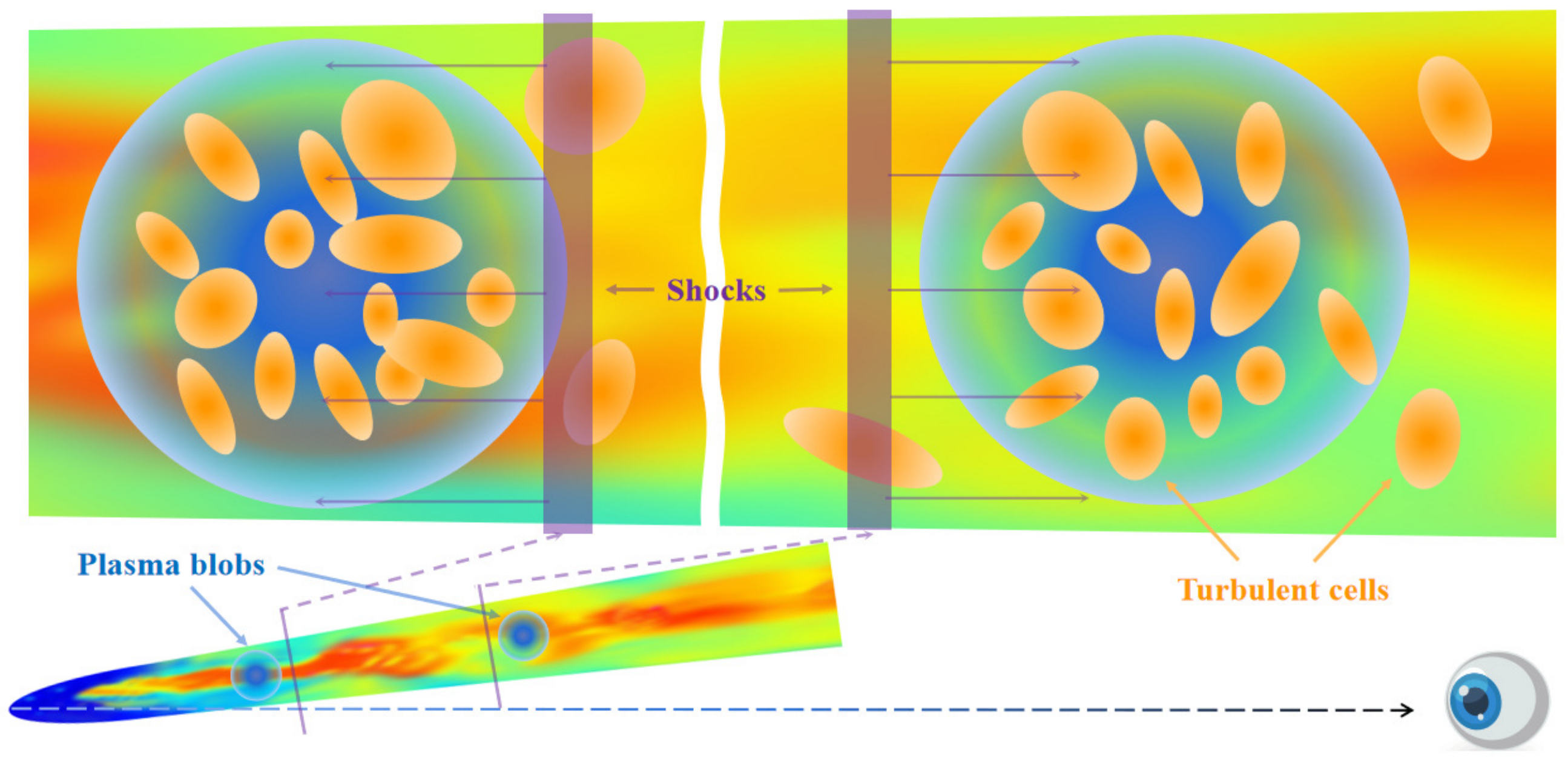}
	\caption{Sketch of the geometry used to describe the origin of flares that may originate from different emission regions. The model decomposes the emission into stationary and flaring emission. The static emission is considered to be the overall emission of the blob or jet extension component. LTV arises from an overall variability (e.g., visual angle) in the blob or a variation in the jet’s total power due to some factor. Flare events are considered to originate from local turbulent cells, thus forming a ``flare-in-flare" pattern.}
	\label{fig:schematic}
\end{figure*}

The sketch of jet geometry and emission zones is presented in Figure \ref{fig:schematic}. Based on the light travel time argument, the observed light curves at different timescales may originate from excitation sources at different scales. We assume that the STV/LTV originate within the kinematic blobs or large scale extensions in the jet, where small scale turbulence components may be present. Within jet-extended structures or large spheres, localized regions of perturbation known as turbulent cells may be present, characterized by parameters like particle number density or magnetic field that diverge from those of the surrounding environment. The emission emitted by these cells excited by the shock would correspond to the numerous flare events in the IDV. Overall, local inhomogeneous structures in these blobs will form ``flare-in-flare" patterns, and the variability light curves can be simulated as the result of multiple flares superimposed on the LTV. The jets-in-a-jet model proposed by \cite{Gia09} based on high-energy fast variability is an alternative interpretation of the ``flare-in-flare" patterns. However, the applicability and comparison of these models will require further work for elaboration.

\cite{Mar92} studied a turbulence model of the plasma in radio jets and interpreted radio flickering as a result of a shock encountering this turbulence. \cite{Bha13} based on the previous works and explained the microvariability phenomenon with a model of shock passing through turbulent cells, which was first proposed by \cite{Web10}. The formation of the turbulent cells is a stochastic process, thus the microvariability light curve does not show repeatable periodicity. Each independent turbulent cell may have a different density, size, and magnetic field direction. The model assumes that the microvariations are the result of a shock propagating down a turbulent jet and energizing individual turbulent cells that cool by emitting synchrotron radiation in the form of pulses. 

We improved the model by adding the assumption that there may be no causal connection between flares, that is, adjacent flares in time may also come from distant emission regions. It means that there may be more than one excitation source in the jet, and the difference between their excitation moments is similar to the light-travel-time for the spatial distance between them. So, each flaring event at different times is independent of each other and originates from the region of the blob corresponding to the dominant radiation. By breaking down the IDV profile into separate flares, we can gain insight into the turbulence distribution and physical parameters within the jet. For this purpose, it is imperative to obtain a pulse model derived from the turbulent cells.

\subsection{Pulse Profile} \label{subsec:pulsemodel}

For the pulse profile in the microvariability light curves, we used the equation of \cite{Kir98}, hereafter KRM. The KRM equation shows that when a plane shock encounters a cylindrical density enhancement region, particles with various magnetic field directions and particle density are accelerated in front of the shock. An increase in magnetic field strength or particle number density will give rise to a corresponding increment in particle injection rate. Once the shock has fully traversed the turbulent cell, the particles contained within it cease to undergo energy gain. Whenever the cooling efficacy of particles surpasses their acceleration capability, the radiative flux will gradually diminish to the degree of the underlying emission region. The observed microvariability curves are the result of the turbulent cells interacting with propagating shock waves in turn. 

Here, we develop the time-domain multi-region radiation model in order to explain more observational characteristics. Based on the geometry proposed above, the acceleration process and radiation mechanism of particles in a single emission region, where the magnetic field is uniform, are analytically studied. The model is computed by assuming the observer lies in the direction of the normal to the shock front. Predecessors have made substantial research on the analysis of KRM equation. In this section, we perform a semi-analytical study on the effect of various parameters on the pulse profile, based on the analytical work of the KRM equation and following the results of previous research \citep{Xu19}. We use the derived baseline model as a comparison to establish a theoretical foundation for examining the effects of different parameters.

Based on the model assumptions mentioned above, kinetic equations are solved for the time, space and energy dependence of the particle distribution functions in the acceleration region and the emission region respectively. In the acceleration region, particles are accelerated at a shock front and cool by synchrotron radiation in the emission region behind it. We assumed that the electrons accelerated by the shock obey the particle distribution function given by \cite{Bal92}:
\begin{equation}
\frac{\partial N(\gamma, t)}{\partial t} + \frac{\partial}{\partial \gamma}[(\frac{\gamma}{t_{acc}} - \beta_{s}\gamma^{2})N] + \frac{N(\gamma, t)}{t_{esc}} = Q\delta (\gamma - \gamma_{0})
\label{equ:KRM}
\end{equation}
with
\begin{equation}
\beta_{s} = \frac{4}{3} \frac{\sigma_{T}}{m_{e}c} (\frac{B^{2}}{2\mu_{0}})
\label{equ:betas}
\end{equation}
where $N$ is the number density of the electrons in the energy space represented by Lorentz factor $\gamma$. In the equation, the parameters $t_{acc}^{-1}$, $t_{esc}^{-1}$, $B$, $\sigma_{T}$, $m_{e}$, $\mu_{0}$ and $c$ are acceleration rate, escape rate, magnetic field strength, Thomson's scattering cross-section, mass of an electron, permeability of the free space and the speed of light respectively. Particles are assumed to be injected into the acceleration process with Lorentz factor $\gamma_{0}$ at a rate of $Q$ particles per second. After acceleration, the particles escape to the downstream plasma, where they are cooled by synchrotron radiation. The time-domain evolutionary pattern of the particle population distribution can be obtained by solving the semi-analytic numerical equations.

We consider the superposition of the total emission obtained by the shock front passing through these large-scale plasma blobs as the global emission. For the flaring event under the STV, LTV described above can be regarded as static emission, that is, the background laminar flow component. This paper focuses on the flare characteristics under the STV, while the detailed discussion of the unified model under the multi-timescale, such as ``flare-in-flare", will be elaborated in future work.

Possible causes of microvariations or flares are diverse, including the existence of a perturbative component in a localized region of the magnetic field or particle number density. The observed emission is considered to be the superposition of static emission and local flaring events, and an increase of the injection rate by a factor $1 + \eta_{f}$ for a time $t_{f}$ is assumed. Here we consider the Eq.(24) of KRM as a superposition of these components. After the semi-analytical numerical calculation of the above model, the flare profiles at different frequencies can be obtained. 

\begin{figure*}
	\begin{minipage}{\textwidth}
		\centering
		\includegraphics[trim=0.3cm 0.3cm 0.8cm 0.5cm,width=0.48\textwidth,clip]{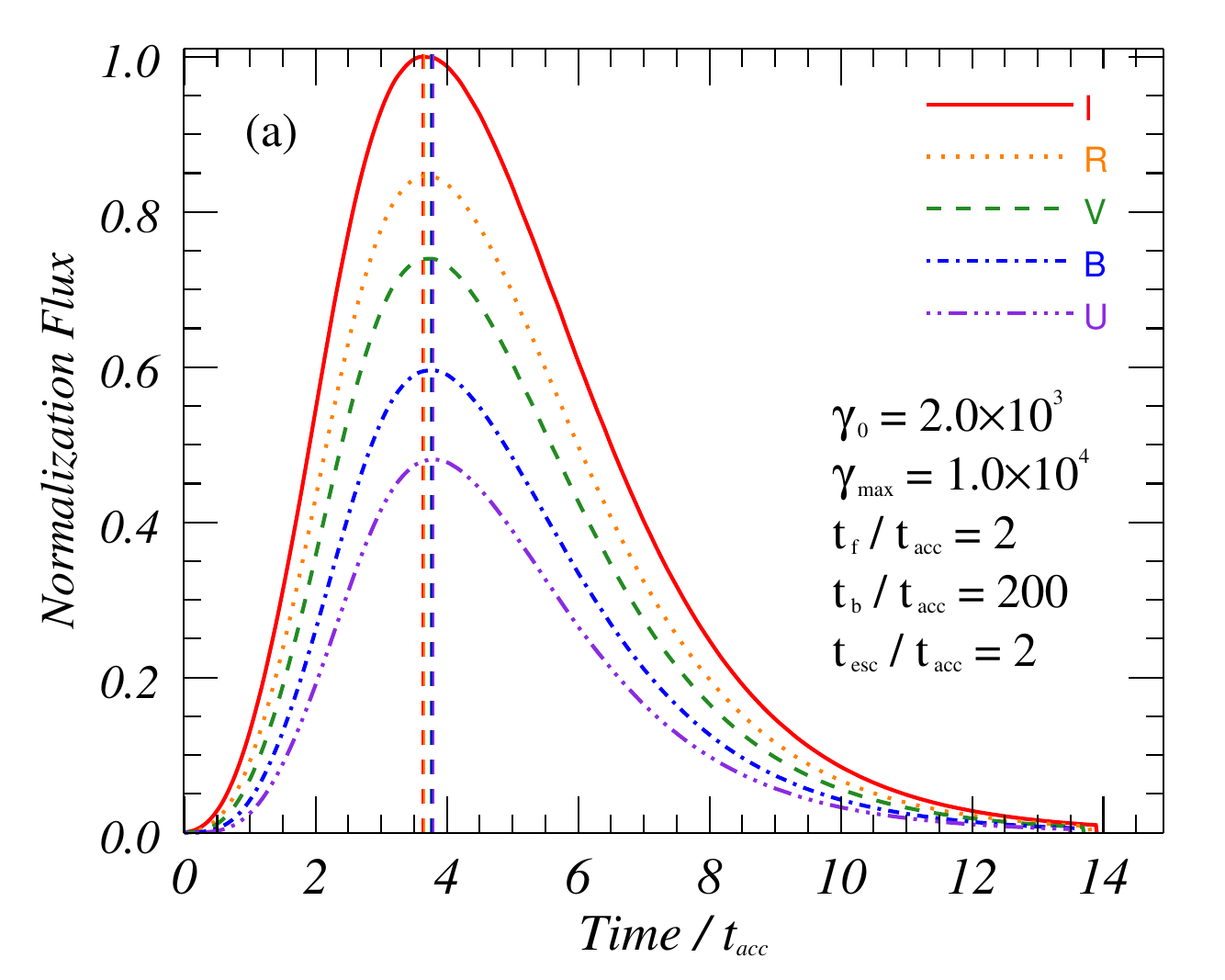}
		\includegraphics[trim=0.3cm 0.3cm 0.8cm 0.5cm,width=0.48\textwidth,clip]{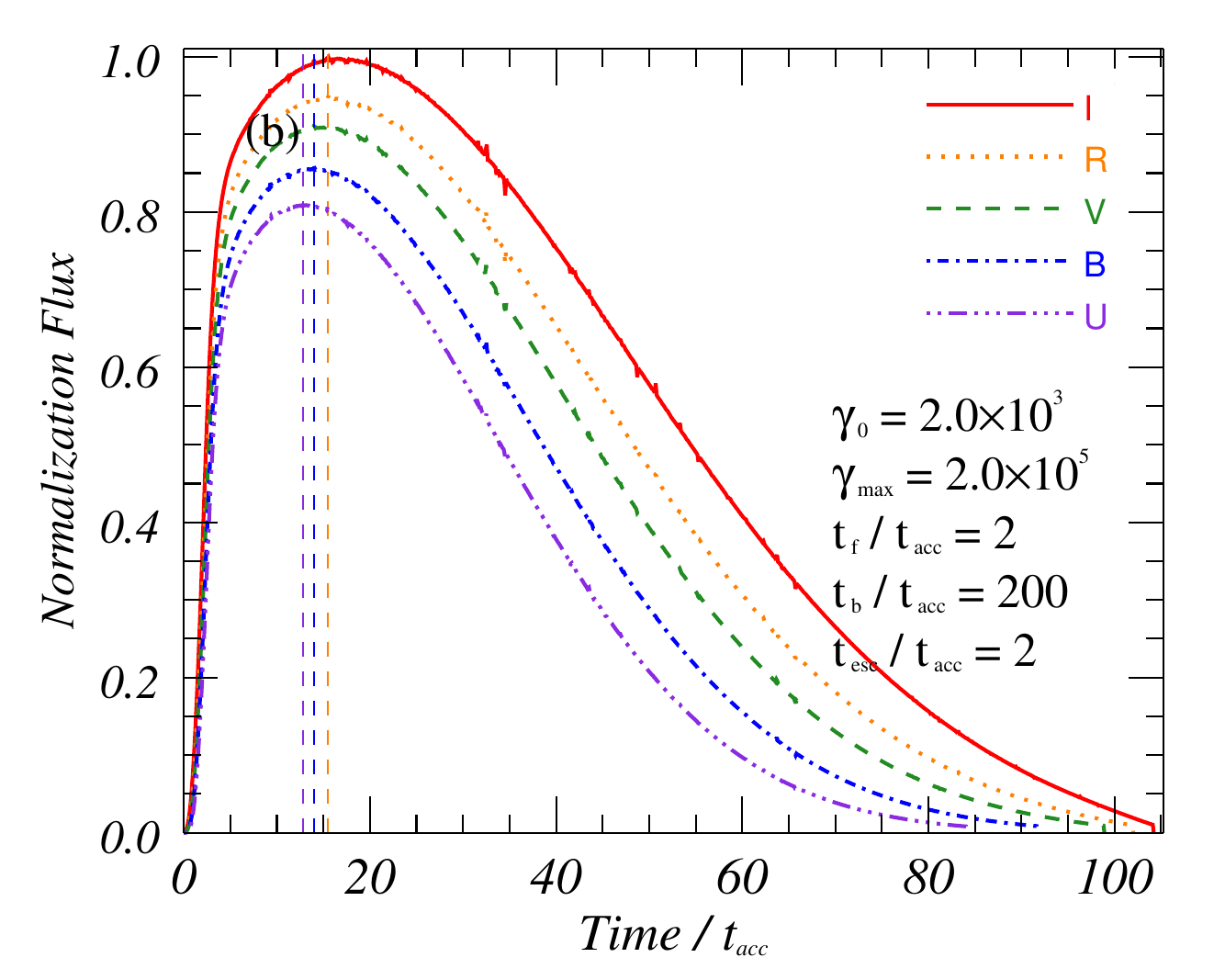}
		\includegraphics[trim=0.6cm 0.4cm 1.0cm 0.8cm,width=0.48\textwidth,clip]{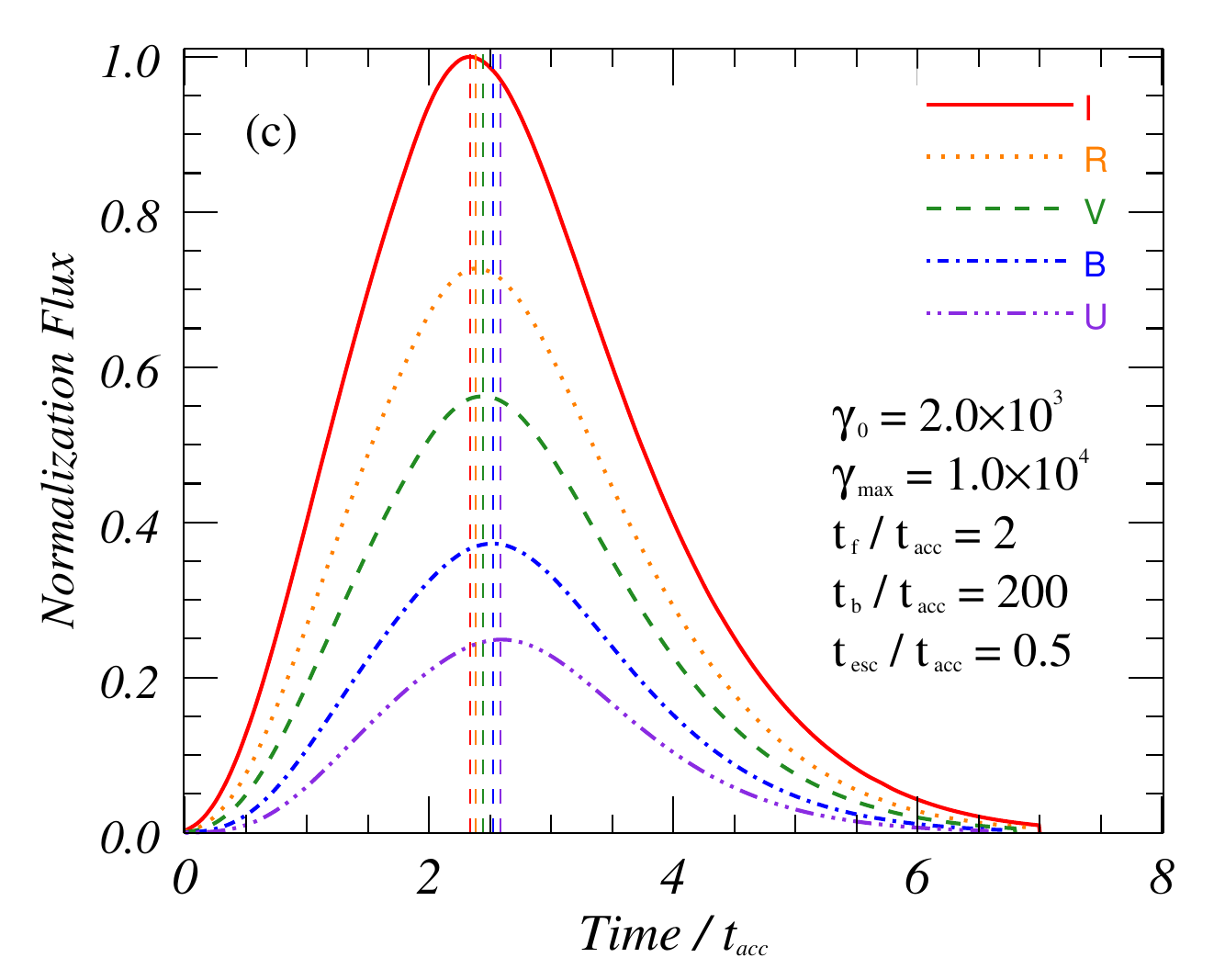}
		\includegraphics[trim=0.6cm 0.4cm 1.0cm 0.8cm,width=0.48\textwidth,clip]{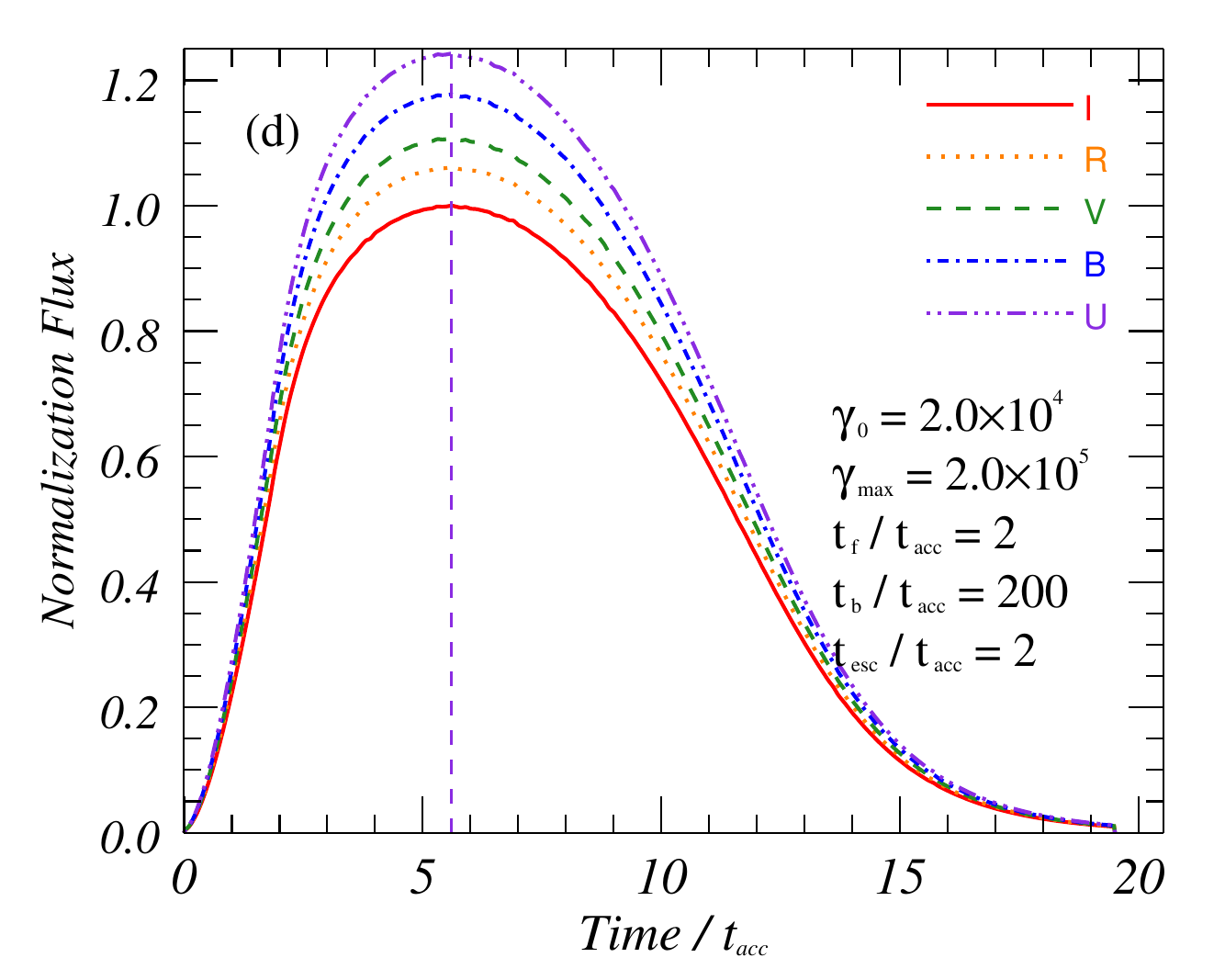}
		\includegraphics[trim=0.6cm 0.4cm 1.0cm 0.8cm,width=0.48\textwidth,clip]{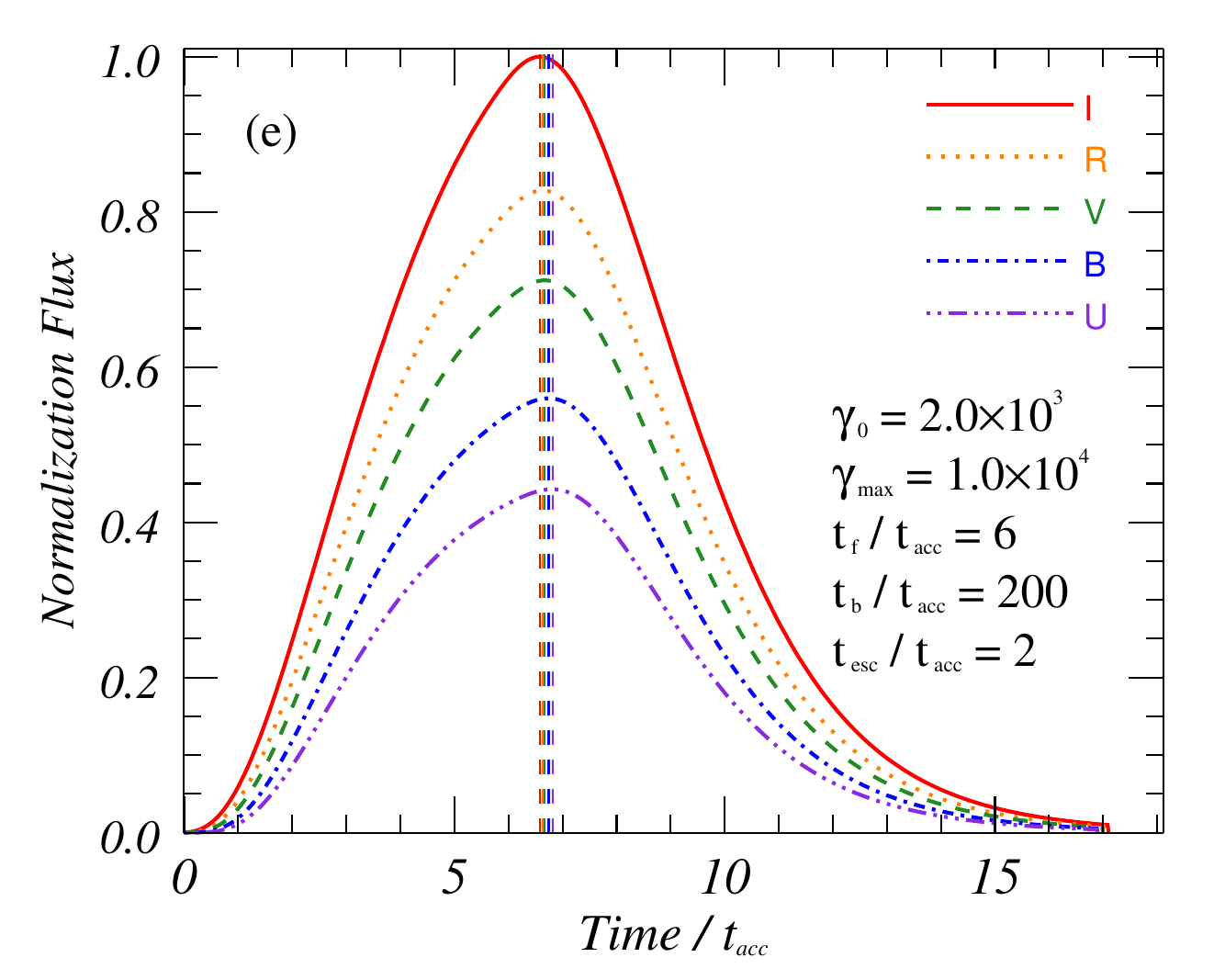}
		\includegraphics[trim=0.6cm 0.4cm 1.0cm 0.8cm,width=0.48\textwidth,clip]{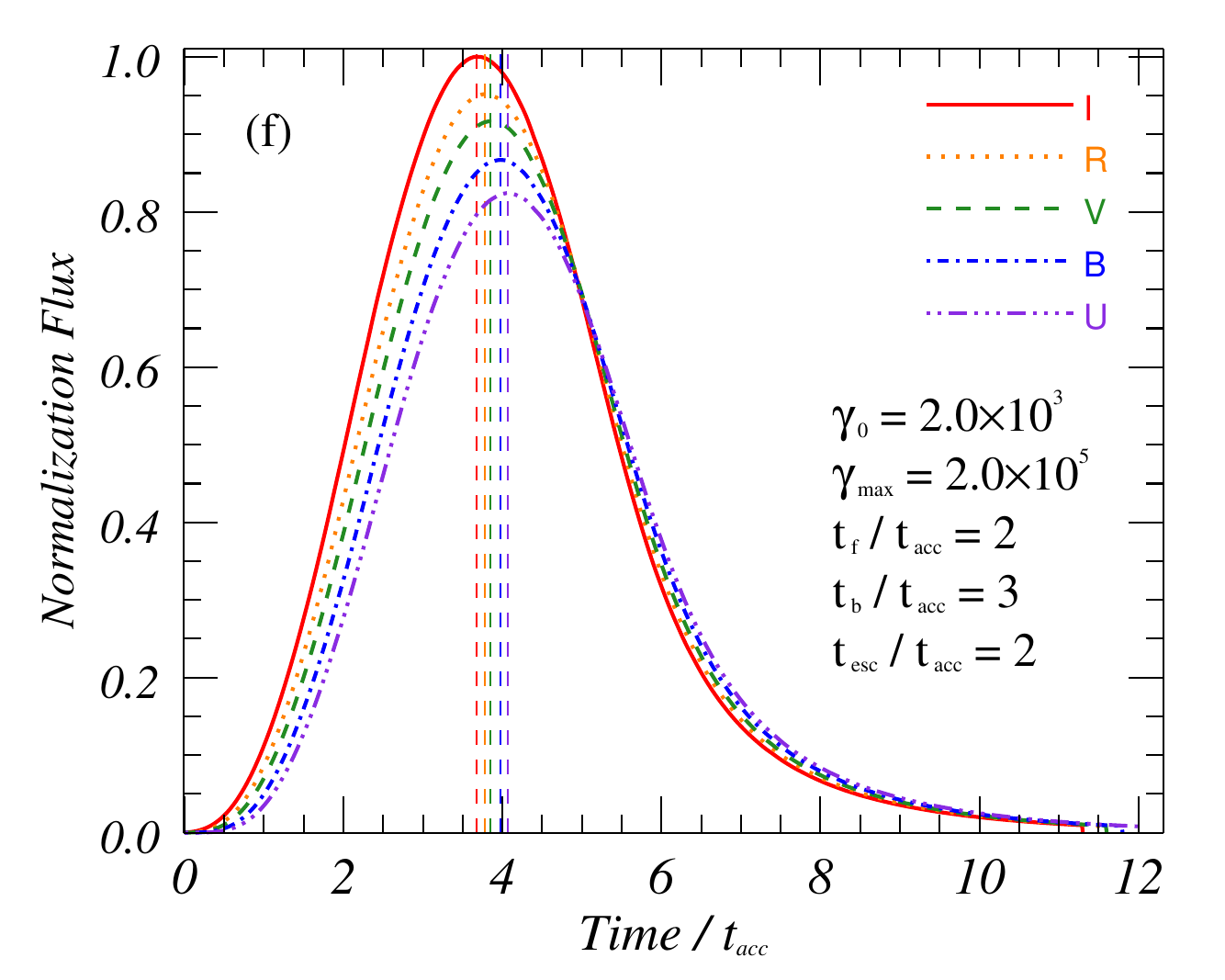}
	\end{minipage}\vspace{0.001cm}
	\caption{(a) The baseline model used to describe the simulated pulse profile, which is normalized with the $I$-band flux. The colored vertical dashed lines mark the peak times. (b)-(f) Comparison examples of multi-band flare profiles under different parameters. Simulation of flare profiles demonstrates notably distinct profiles for various parameter sets, which result in dissimilar amplitude ratios and band-specific time delays. While $\gamma_{max} = 10^{4}$ is assumed in the baseline model, and variations of $\gamma_{max}$ up to $2 \times 10^{5}$ are explored.}
	\label{fig:nompulse}
\end{figure*}

Figure \ref{fig:nompulse}(a) shows the baseline model used in this article, the parameters used in the model are listed in Table \ref{tab:nompara}. The correlation coefficients between $I$-band and other bands are about $99.71^{+0.25}_{-0.37}\%$, which shows that the multi-band flare profiles obtained by this model under the same parameters have high similarity. These patterns show that there are obvious differences in the amplitude, duration and peak time of flare in different bands. The frequencies used in this study to calculate the flare profiles are the typical frequencies that are commonly used by telescopes for observations in optical bands. To obtain the theoretical flare profile for each observed frequency, we perform a Doppler transformation and convert the observed frequencies into the frequencies within the plasma rest frame.

\begin{table}[]
	\centering
	\caption{Parameters used in the Baseline Model}
	\label{tab:nompara}
	\begin{tabular}{lcc}
		\hline\hline
		Parameter                      & Symbol & Value     \\
		\tableline
		Redshift                       & $z$      & 0.069$^{\rm a}$     \\
		Minimum electron energy        & $\gamma_{0}$     & $2 \times 10^{3}$        \\
		Maximum electron energy        & $\gamma_{max}$   & $1 \times 10^{4}$        \\
		Magnetic field intensity       & $B$      & 0.4 Gauss$^{\rm b}$ \\
		viewing angle                  & $\theta$  & $3.3\degree$$^{\rm c}$      \\
		Shock speed                    &  $\beta_{s} $      & $0.1c$      \\
		Bulk Lorentz Factor of the jet &   $\Gamma$     & 14$^{\rm c}$      \\
		\hline    
	\end{tabular}
    \tablecomments{$^{\rm a}$Redshifts given in \cite{Mil78}. \\
    $^{\rm b}$The magnetic field intensity from \cite{Pot12}.\\
$^{\rm c}$The viewing angle and the bulk Lorentz Factor of the jet from \cite{Cel08}.}
\end{table}

From the theoretical analysis, it can be known that under the condition of the same acceleration region and emission region, the flare amplitude, duration and peak time at different frequencies are not exactly the same. Differences in flare amplitudes at different frequencies will lead to time-domain evolution patterns of spectral indices, which will be analyzed in Section \ref{subsec:ParameterA}. As shown in the right panel of Figure \ref{fig:nompulse}, different peak time of flare at different frequencies will lead to time delay of light curves at different frequencies, which is caused by different acceleration efficiency of particles with different energies. Although there is little difference in flare duration scales at different frequencies of optical bands, it can still be concluded theoretically that flare duration at lower frequencies are longer. This characteristic indicates that for low-frequency flares, their duration scales usually correspond to the range of large-scale emission, so they are limited by the physical properties of the large-scale emission, and are less affected by the local small-scale structure. In order to further analyze the influence of these acceleration and radiation parameters on flare, the influences of different parameters on multifrequency flare characteristics will be analyzed in details below.

\subsection{Parameter Analysis} \label{subsec:ParameterA}

Based on theoretical analysis and numerical simulation, the effects of kinematic parameters (e.g., $\gamma_{0}$, $\gamma_{max}$, and $t_{esc}/t_{acc}$) on flux level, profile and spectral hardness evolution of flare at multifrequencies were investigated. The influences of different parameters on flare observation characteristics at different frequencies, such as flux level, center time and duration, are analyzed below. On this basis, flare symmetry, peak amplitude ratio and time delay between different frequencies, time domain evolution of spectral index ($\alpha$; we assume that the spectral energy distribution (SED) follows $\nu F_{\nu} \propto \nu^{-\alpha}$) and other characteristics are further compared.

Based on theoretical analysis, the evolution of particle number density distribution, and ultimately, the profile of flare, are impacted by several factors including the acceleration rate ($t_{acc}^{-1}$), escape rate ($t_{esc}^{-1}$), initial ($\gamma_{0}$) and final energy ($\gamma_{max}$) of particles, flare timescale ($t_{f}$), and the time it takes for a shock to pass through the plasma blob ($t_{b}$). The results of the numerical simulations align with the aforementioned theoretical expectations. Examples of flare curves for different parameter sets are illustrated in Figure \ref{fig:nompulse}. The distribution of particles with different energies is not uniform during the acceleration process, so the flare profile responding to different frequencies is not exactly the same, which will lead to systematic differences in flare at different frequencies. 

The simulation results reveal high correlation between flare profiles at different frequencies, which indicates that frequency is not a main parameter affecting the shape of the flare profile. However, there are differences in flare amplitude, peak time and duration at different frequencies, which we believe will lead to differences in observed characteristics at different frequencies. Different combinations of parameters will lead to the asymmetry of the ascending and descending trend of flare profile, the lag or advance of flare center time, and the sharpening or flattening of flare shape. The flare characteristics caused by these different sets of parameters can provide a theoretical foundation for the observed variability curves.

When the escape rate is much faster than the acceleration rate, i.e., $t_{esc}/t_{acc} \ll 1$, the radiation intensity will be too low to observe the substantial variability characteristics. In contrast, the particles do not have enough time to cool before leaving the source. When $t_{esc}$ and $t_{acc}$ are comparable, the simulation shows that the flare profile’s rising and declining times have a similar range of values, so the values used here are similar. Numerical simulations demonstrate that as the ratio of $t_{esc}/t_{acc}$ increases, the acceleration becomes more efficient compared to the escape, resulting in a shorter time to reach the peak. This indirectly affects the frequency at which the flare peaks at, and therefore the time delay between each frequency peak.

Simulations conducted under the baseline model for 1-hour flares in the observer frame demonstrate that the peak times for $I$-band flares occur ahead of the $R$, $V$, $B$, and $U$-band by $0.65$ min, $1.05$ min, $1.74$ min, and $1.81$ min, respectively. The time delay between bands is not an intrinsic constant value, as it is influenced by various other parameters. Figure \ref{fig:nompulse} shows that low-frequency flares can lead high-frequency flares for some parameters, while for other parameters the opposite may occur. Therefore, we further analyze the influence of different parameters on the time delay between different bands through simulation. Here, some quantitative simulation results are presented based on the comparison under the baseline model.

\begin{figure*}
	\begin{minipage}{\textwidth}
		\centering
		\includegraphics[trim=2.0cm 0.3cm 0.3cm 0.4cm,width=0.49\textwidth,clip]{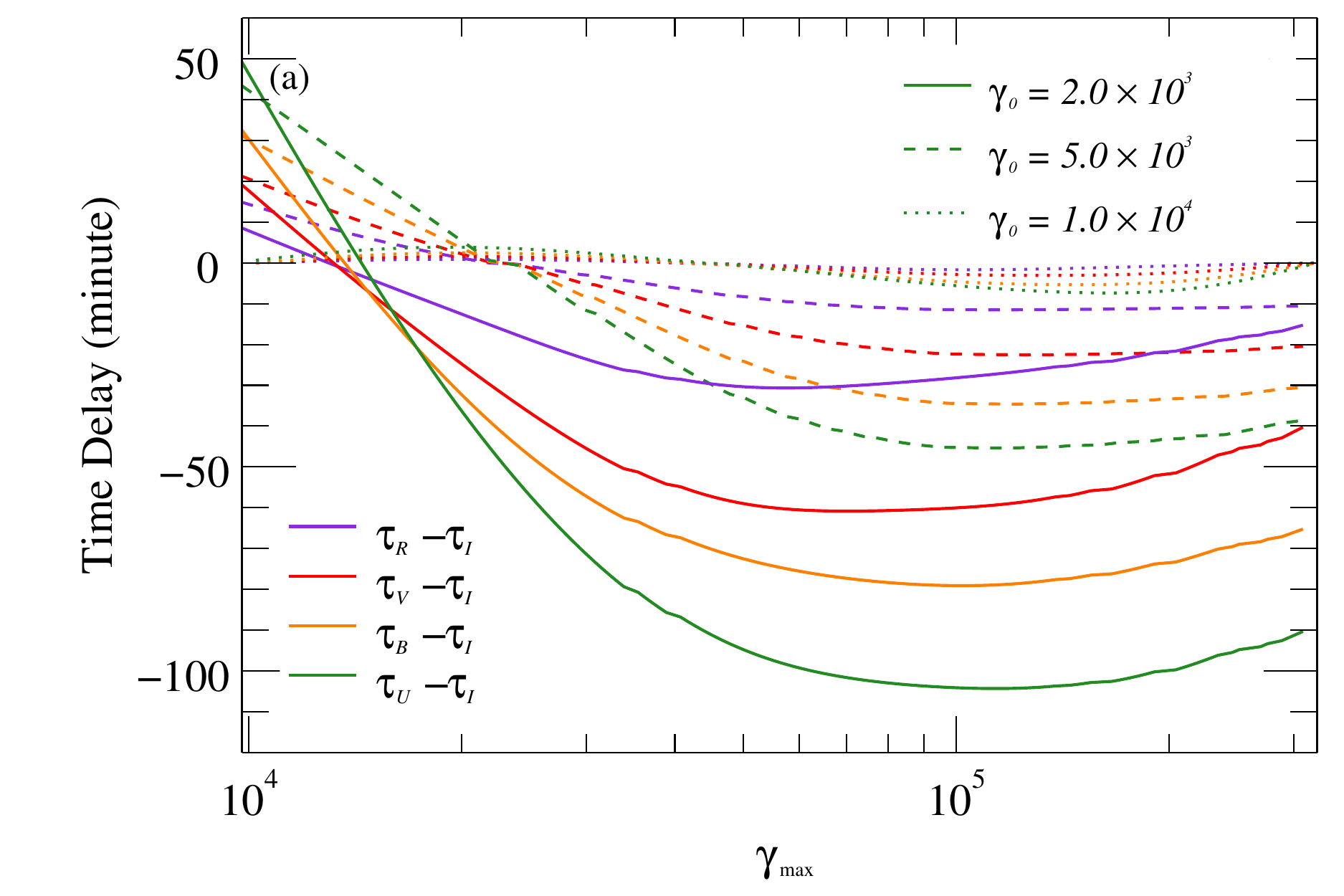}
		\includegraphics[trim=1.3cm 0.3cm 0.3cm 0.4cm,width=0.49\textwidth,clip]{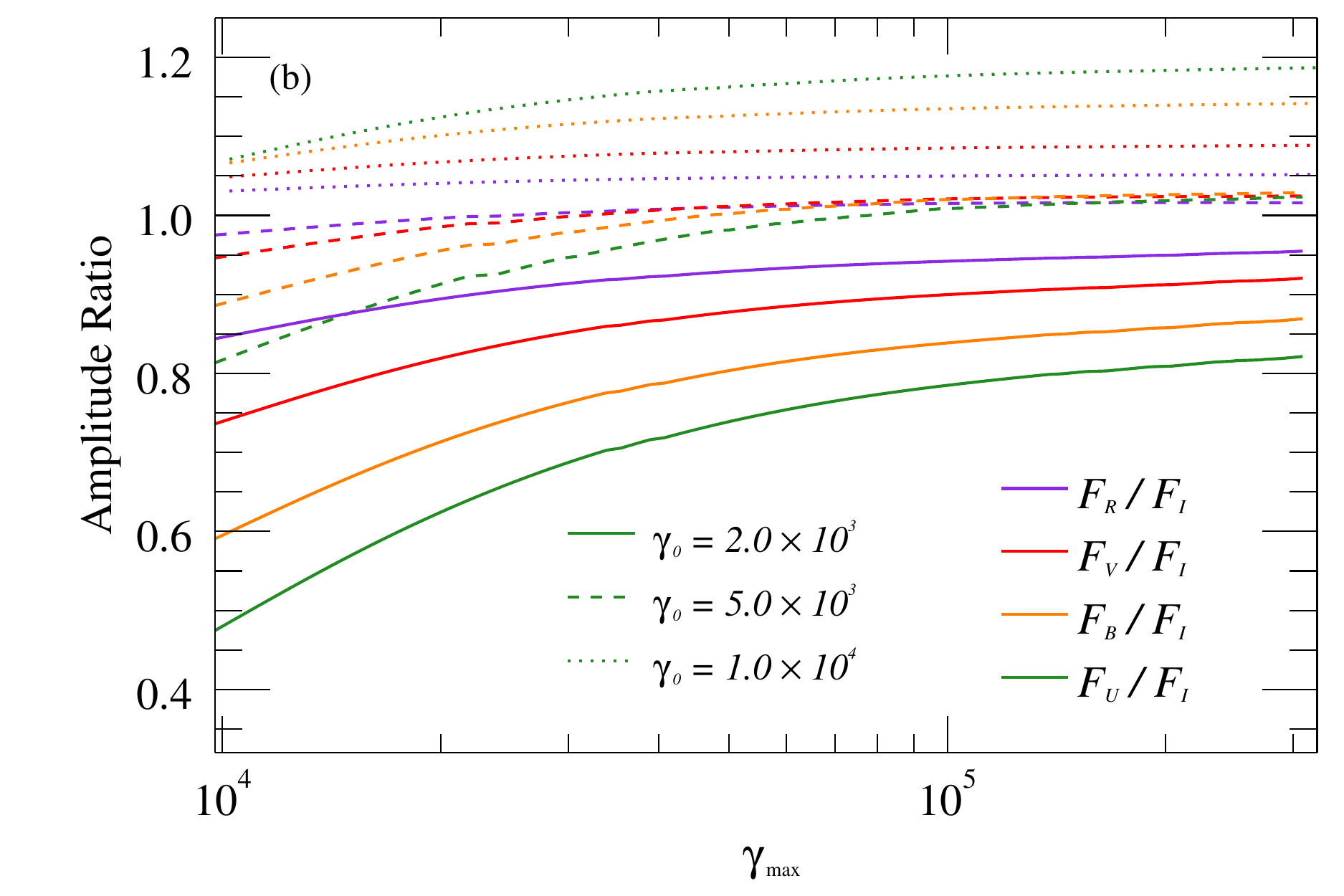}
		\includegraphics[trim=2.0cm 0.3cm 0.25cm 0.4cm,width=0.49\textwidth,clip]{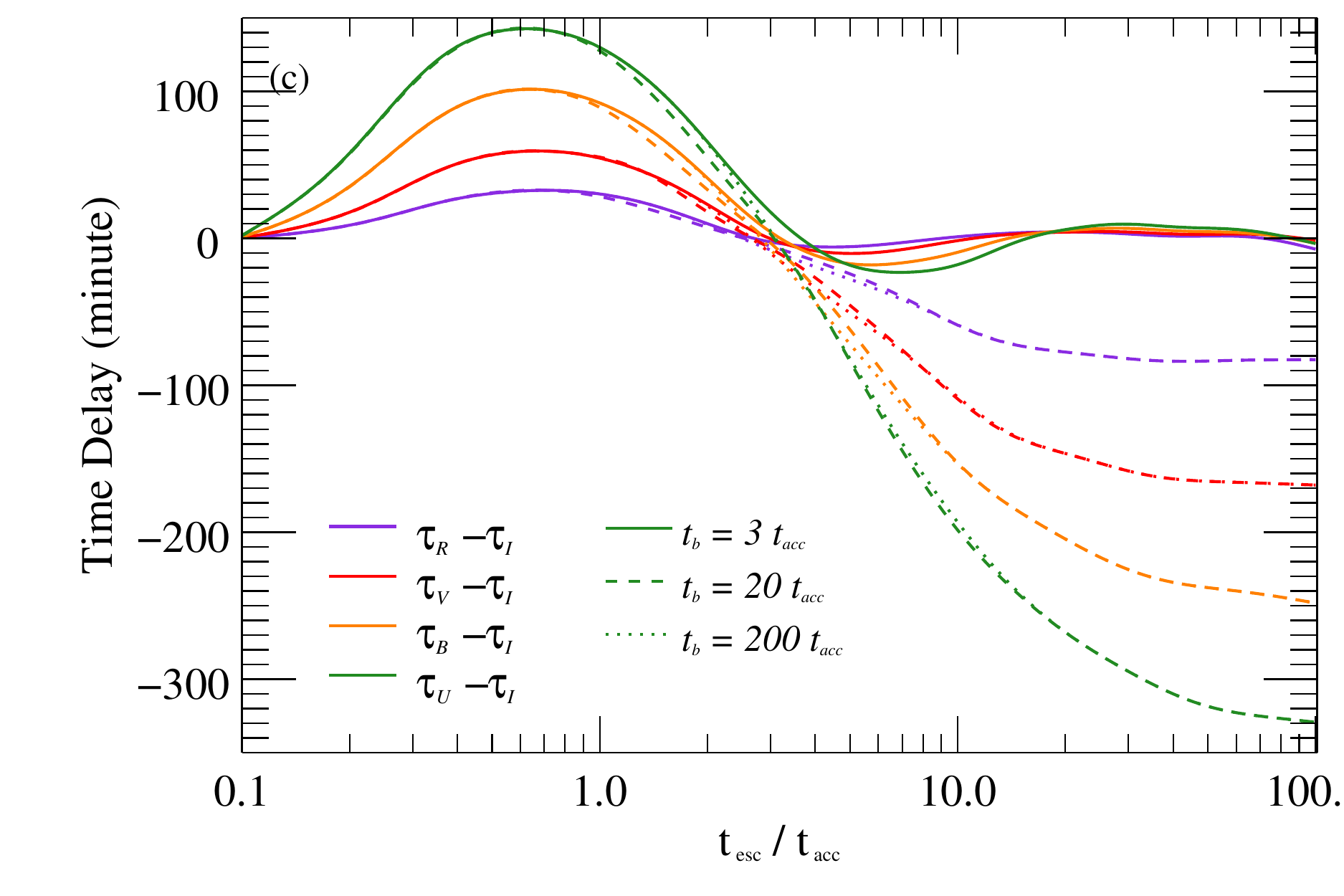}
		\includegraphics[trim=1.3cm 0.3cm 0.25cm 0.4cm,width=0.49\textwidth,clip]{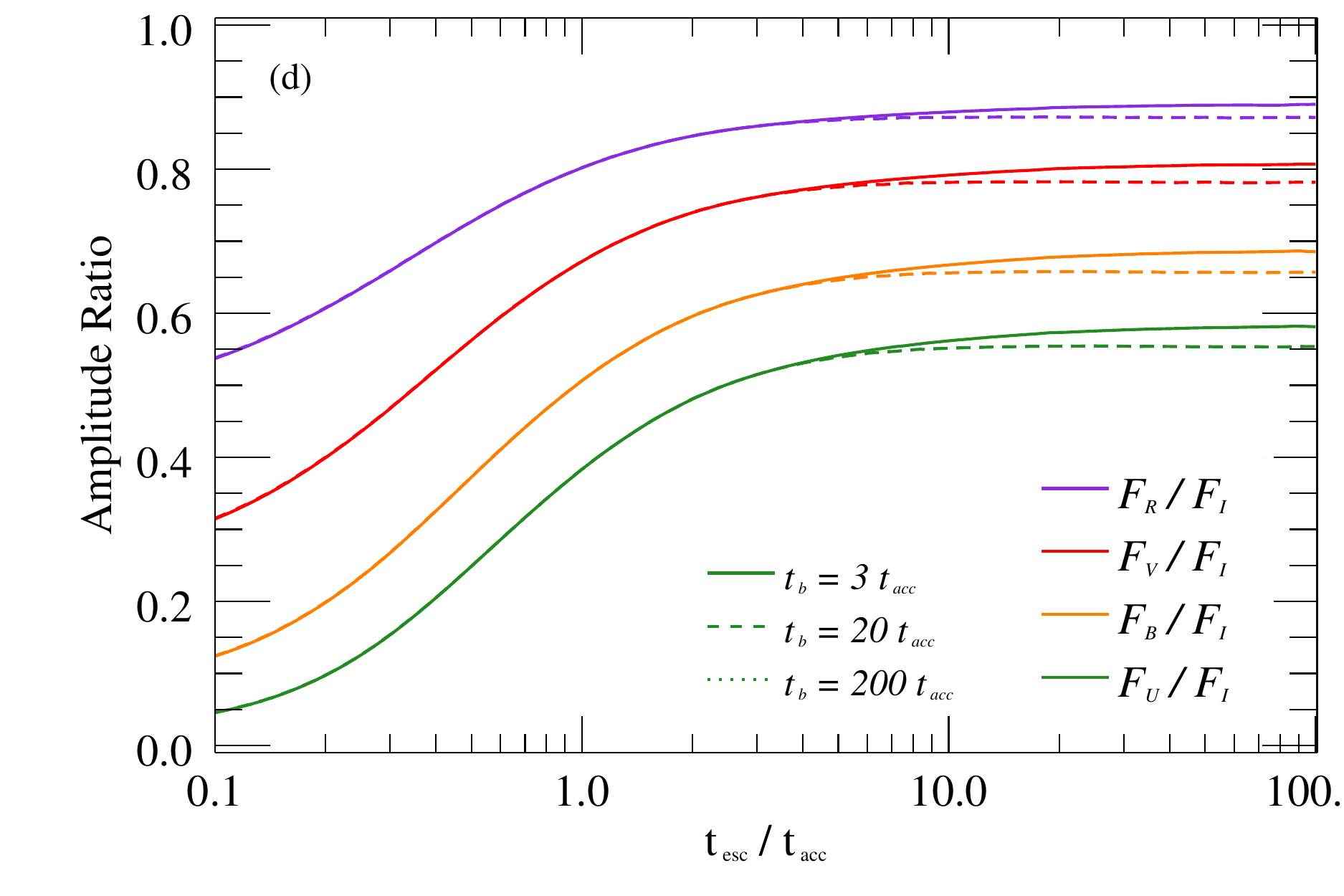}
		\includegraphics[trim=2.0cm 0.3cm 0.1cm 0.4cm,width=0.49\textwidth,clip]{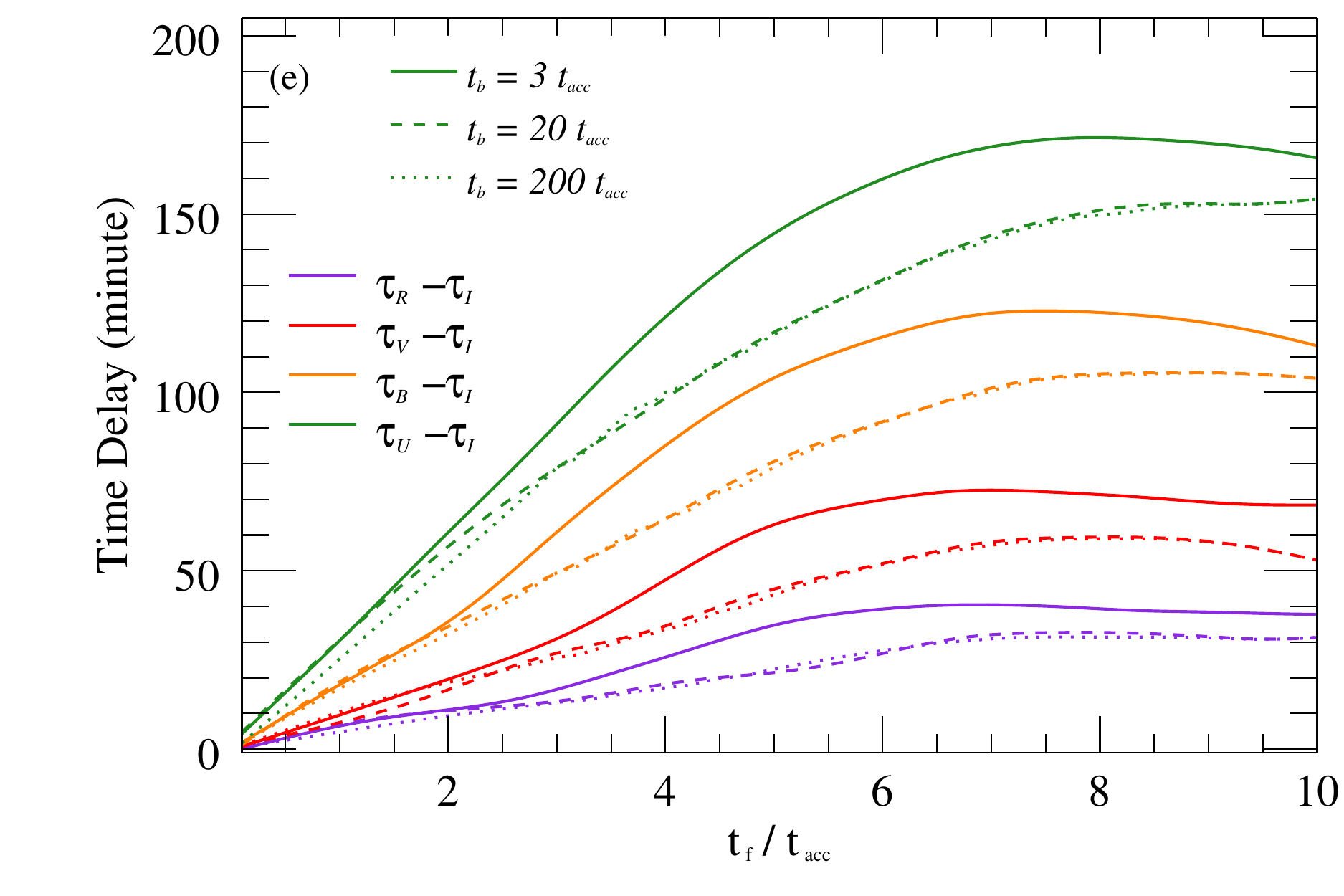}
		\includegraphics[trim=1.3cm 0.3cm 0.1cm 0.4cm,width=0.49\textwidth,clip]{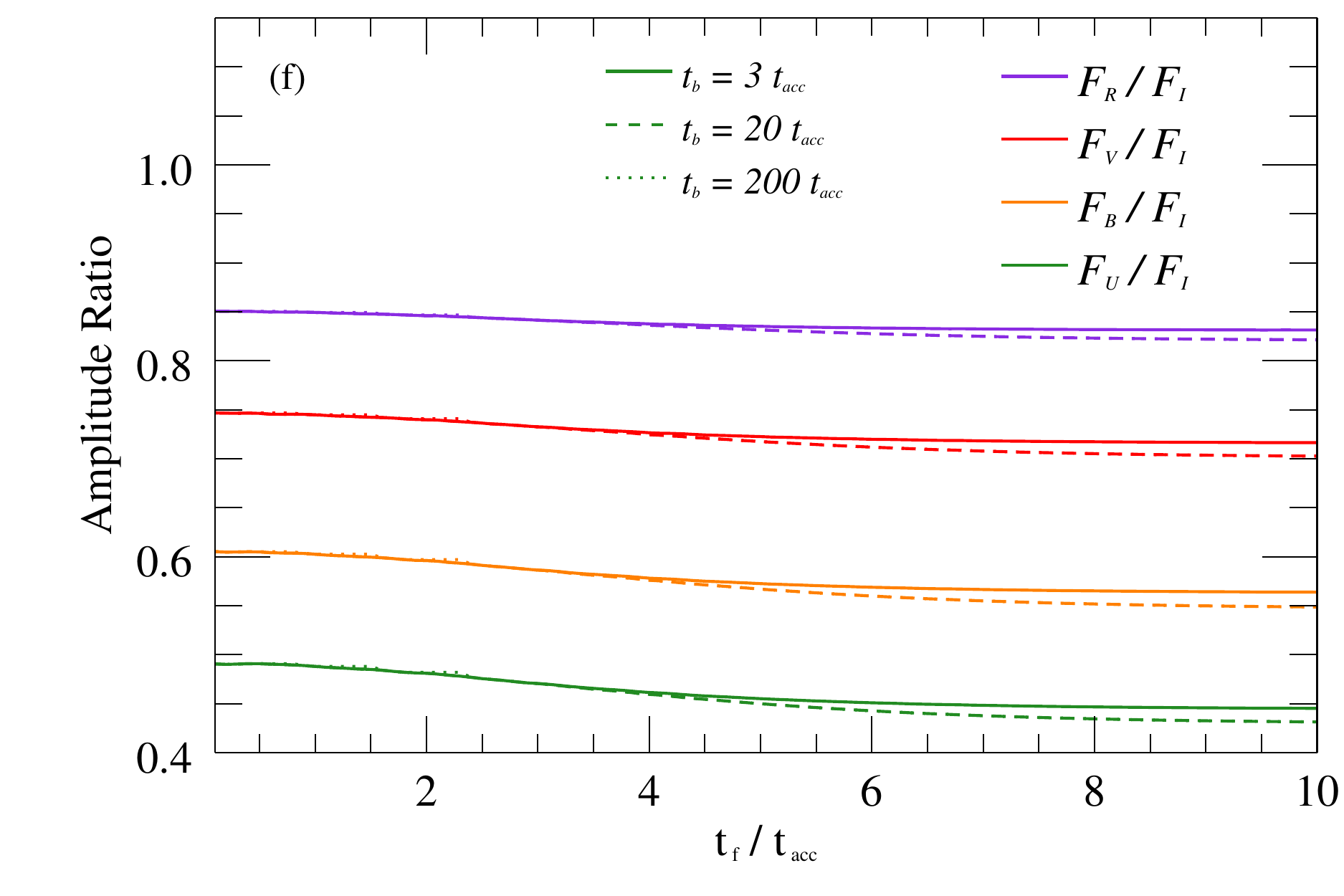}
	\end{minipage}\vspace{0.001cm}
	\caption{Examples of quantified comparisons of individual parameters on time delays (left panel) and amplitude ratios (right panel) between different bands. In these examples, all parameters align with the baseline model (Figure \ref{fig:nompulse}a), except for those specifically highlighted in the legend. The colored curves illustrate the distribution of time delays or amplitude ratios across different bands, while the various line shapes indicate the effects of specific parameters. In panels (c)-(f), the curves for $t_{b}=20t_{acc}$ and $t_{b}=200t_{acc}$ are almost identical.}
	\label{fig:timelag}
\end{figure*}

Figure \ref{fig:timelag} presents sample simulation results of the effects of parameters such as $\gamma_{0}$, $\gamma_{max}$, $t_{b}$, $t_{f}$ and $t_{esc}/t_{acc}$ on the time delays or amplitude ratios of flares in different bands. In these examples, all parameters align with the baseline model (Figure \ref{fig:nompulse}a) except for those utilized for quantile testing. The numerical simulation results show that the time delay of flares in different optical bands depends on the various sets of parameters. This implies that flares originating from diverse physical environments may manifest as either low-frequency or high-frequency lag modes, contingent upon the time-containing evolution of the particle distribution. If a variability curve contains multiple flares with similar origins, their superposition will produce a curve with high similarity between bands. In contrast, if the flares have distinct origins, the inter-band time delays, amplitude ratios, and other characteristics possess non-systematic differences, leading to weak correlation between the bands. Hence, this dissimilarity serves to differentiate the source of the flare.

The simulation results indicate that the amplitude ratio of the flare between the specific bands is dependent on the parameter sets. Illustrated in the right panel of Figure \ref{fig:timelag}, the diverse parameters have varying impacts on the amplitude ratio, with the most pronounced influence seen from the $t_{esc}/t_{acc}$, $\gamma_{0}$ and $\gamma_{max}$. This suggests that the initial and maximum energies, as well as the acceleration and escape rates, are critical factors that impact the distribution and evolution of particles, leading to changes in the energy spectrum index over time. However, parameters such as $\delta$, $t_{f}$ and $t_{b}$ have little effect on the amplitude ratio between the optical bands. In addition, consider that the peak frequency of synchrotron radiation grows with increasing $\gamma_{max}$. Thus, if the synchrotron peak frequency is near the optical observation frequency, observed high-frequency flares may grow more than low-frequency flares, but it could also be the opposite. Essentially, the nonlinear evolution of the particle number distribution under different parameter sets results in a dissimilar growth rate of the flare fluxes across each band. As a consequence, there is a non-linear development of the spectral indices in the SED. The time-dependent evolution of the spectral indices will help to distinguish the origins of the variability, which will be analyzed and discussed in detail in the next subsection.

\subsection{Spectral Variability Pattern} \label{subsec:SVP}

\begin{figure*}
	\begin{minipage}{\textwidth}
		\centering
		\includegraphics[trim=1.8cm 0.6cm 0.5cm 0.4cm,width=0.49\textwidth,clip]{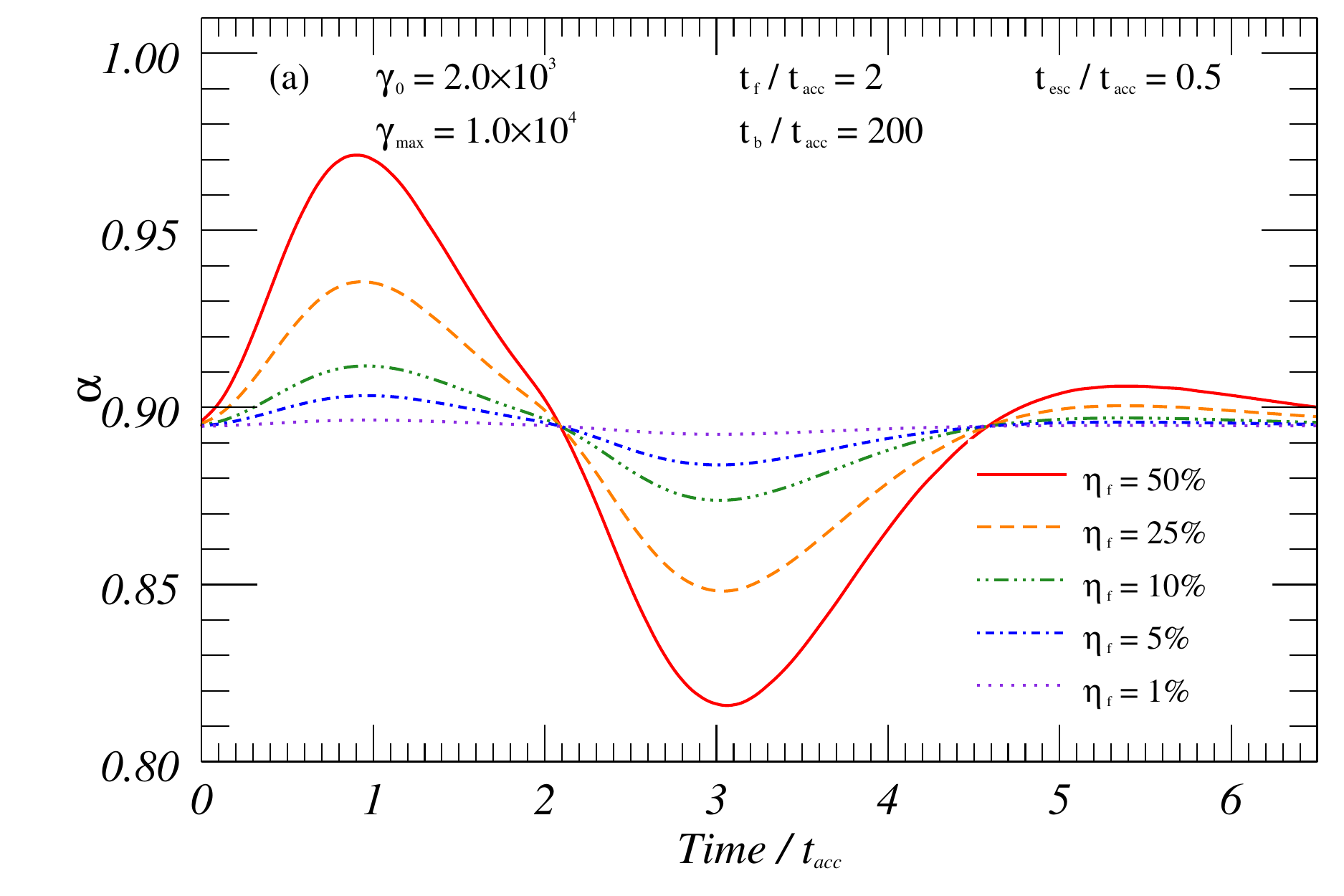}
		\includegraphics[trim=1.8cm 0.6cm 0.5cm 0.4cm,width=0.49\textwidth,clip]{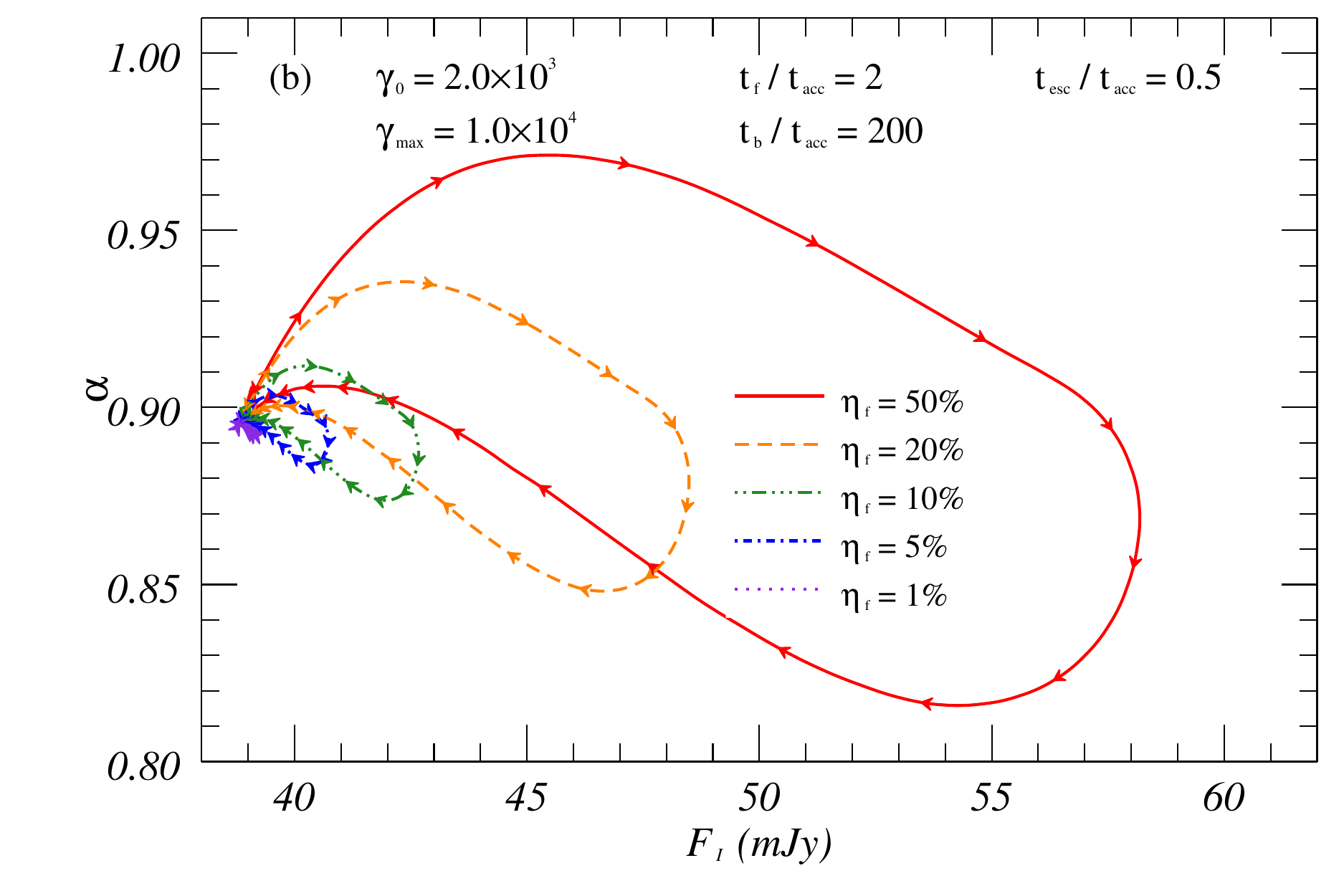}
		\includegraphics[trim=1.8cm 0.6cm 0.5cm 0.4cm,width=0.49\textwidth,clip]{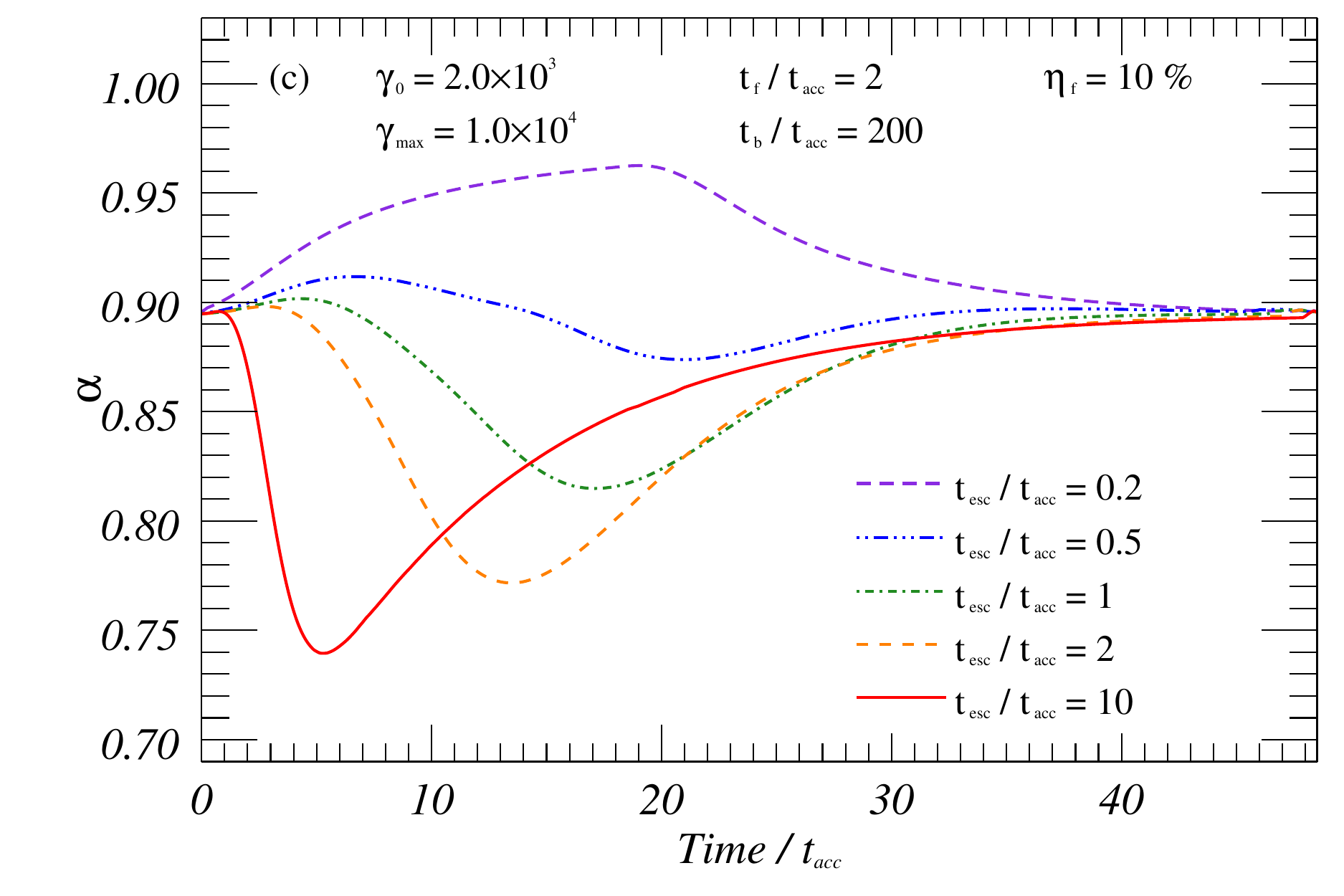}
		\includegraphics[trim=1.8cm 0.6cm 0.5cm 0.4cm,width=0.49\textwidth,clip]{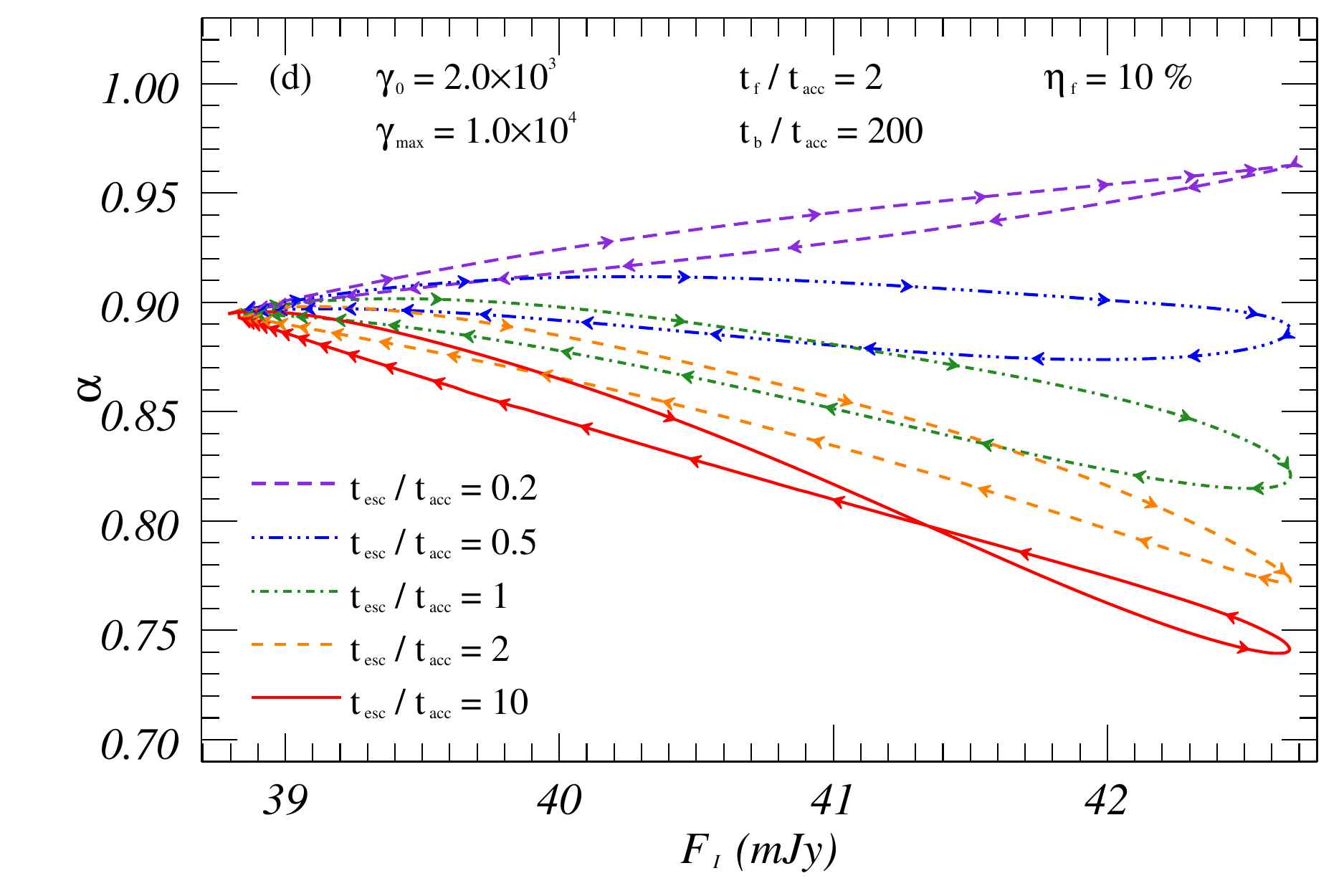}
		\includegraphics[trim=1.8cm 0.6cm 0.5cm 0.4cm,width=0.49\textwidth,clip]{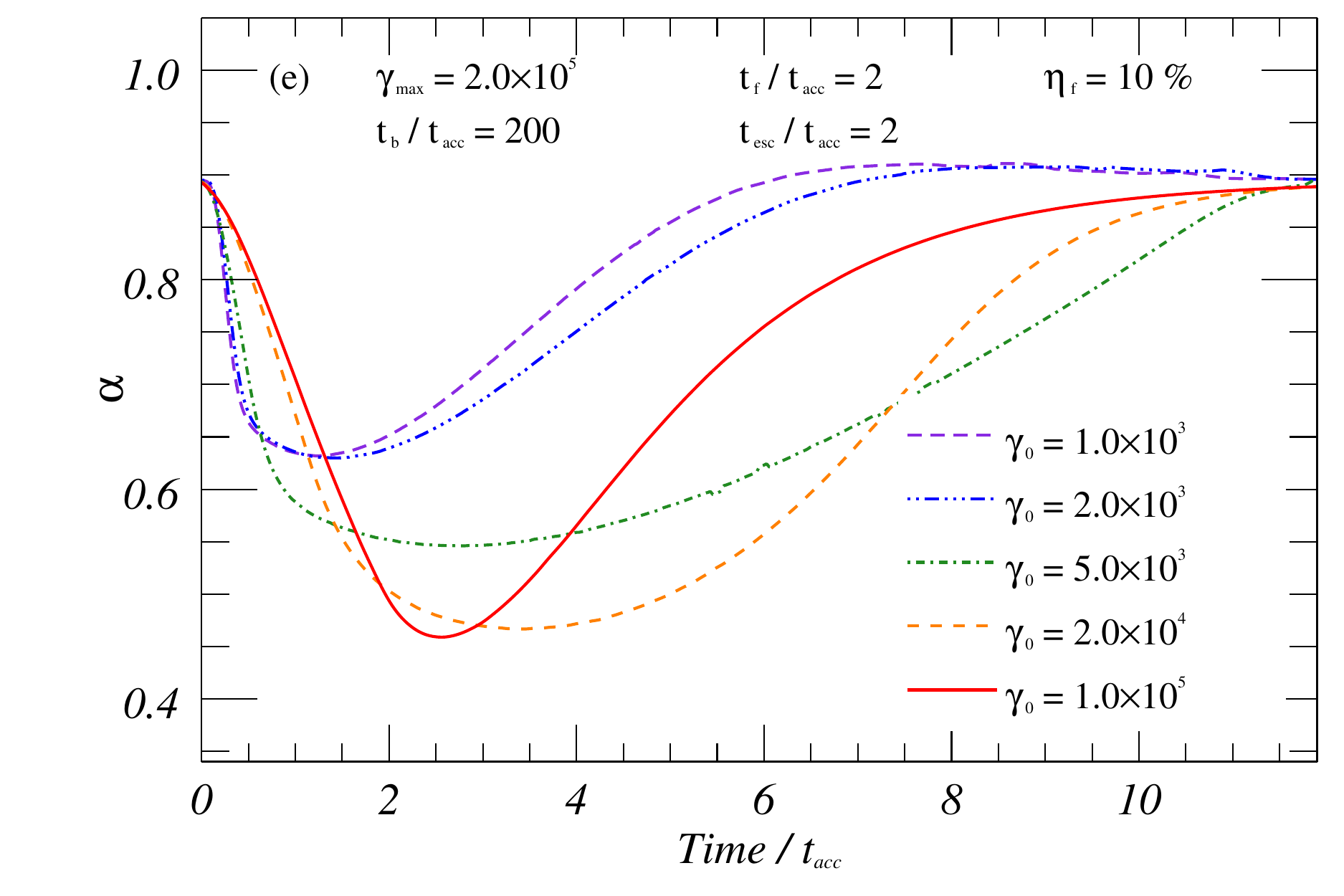}
		\includegraphics[trim=1.8cm 0.6cm 0.5cm 0.4cm,width=0.49\textwidth,clip]{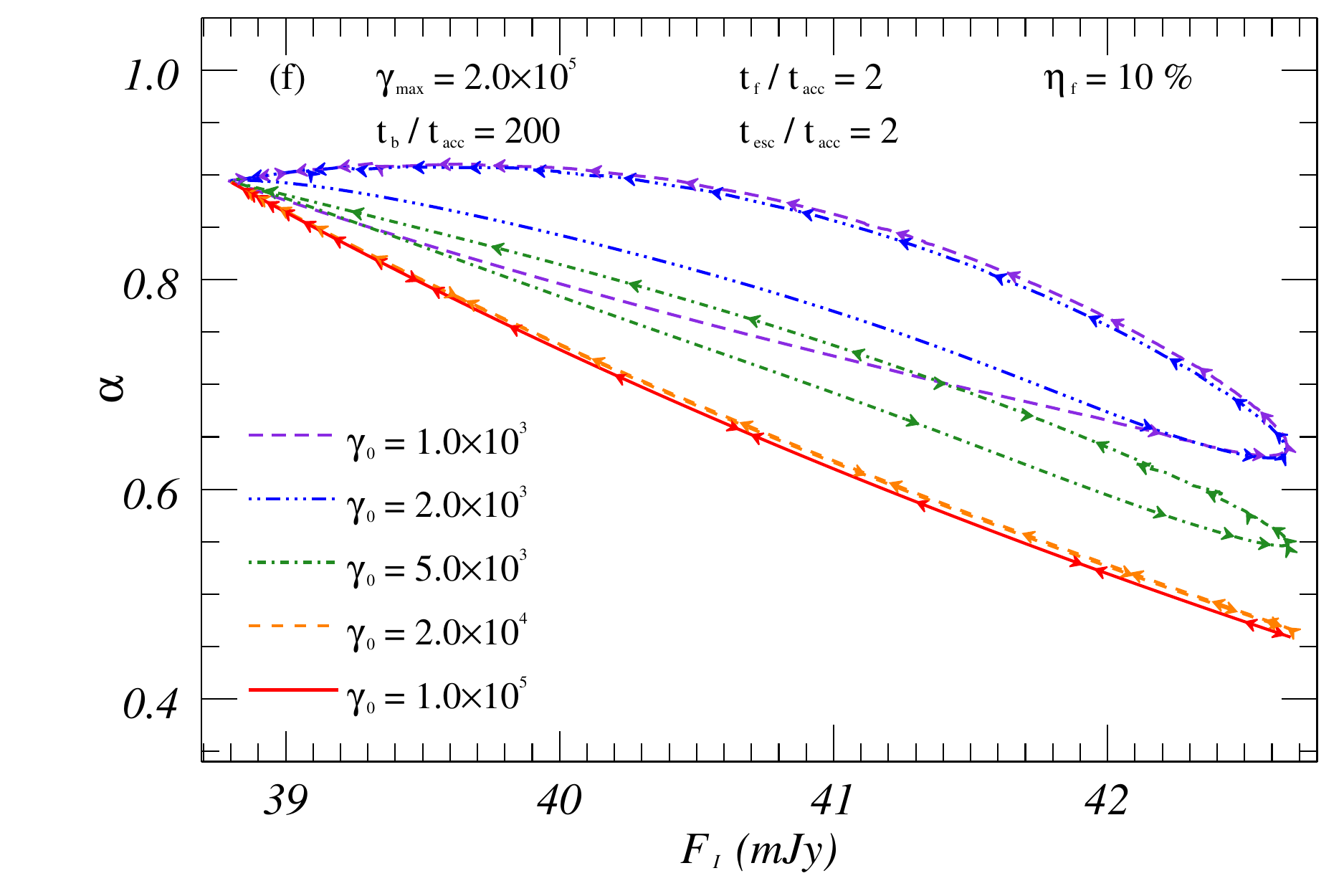}
	\end{minipage}\vspace{0.001cm}
	\caption{Examples of simulations of SED spectral indices ($\nu F_{\nu} \propto \nu^{-\alpha}$) for flare events with various parameter sets. Dependencies of $\alpha$ on time (left panels) and flux in the $I$-band (right panels) for different sets of physical parameters, such as $\eta_{f}$, $t_{esc}/t_{acc}$, and $\gamma_{0}$ are shown in the upper, middle, and lower panels, respectively. The simulation parameters in this diagram are based on the baseline model, except for the parameters indicated by the legends.}
	\label{fig:svp}
\end{figure*}

We have used the synchrotron radiation model \citep{Kir98} to simulate time-evolving optical spectral variation patterns in a specific physical environment. As described in Section \ref{subsec:physical}, we consider that the jet radiation component originates from the superposition of various components such as the global radiation and the local flare radiation. The global radiation is assumed to be constant at STV, and the flare model parameters here are based on the baseline model parameters (Figure \ref{fig:nompulse}a). We utilize the aforementioned simulation as the fundamental framework for analyzing how varying parameter sets impact the time-contained evolution of the SED spectral index and the intensity hardness pattern. The parameter set in the baseline model was configured to achieve $\alpha \sim 0.9$, based on calculations derived from observation data. Several hypotheses attempt to explain the steep SED in the optical regime, including its proximity to the exponential cut-off region, specific acceleration processes \citep{Sum12}, and the fine-tuned superposition of components, etc. Nevertheless, further studies are needed to establish the exact origin of the steep spectrum.

From the Eq.(24) of KRM, it is shown that the ratio of the flux of the flare to the flux of the static emission ($\eta_{f}$; $\eta_{f}$ is defined in relation to the I-band flux, noting that the ratio of the flux increment to the basal flux differs among various bands.) significantly affects the fluctuation level of the amplitude in the active state, which in turn affects the specific value of the spectral index of the spectral energy distribution. Figures \ref{fig:svp}(a) and \ref{fig:svp}(b) illustrate the time-contained evolution of the $\alpha$ and $\alpha$-intensity maps for different values of $\eta_{f}$, respectively. Figure \ref{fig:svp}(a) shows that $\alpha$ can decrease or increase during the flare event, and the flare’s intensity only affects the range of the $\alpha$ variation, not its overall trend over time. In Figure \ref{fig:svp}(b), the $\alpha$-intensity loop pattern suggests multiple transitions between RWB and BWB behavior during the flare event. Similarly, the comparable loop patterns of other $\eta_{f}$ parameters indicate that changes in only flare intensity do not affect the hardness-intensity trend. This points to radiation intensity changes not being the primary cause of the shift between the BWB and RWB phenomena. It is worth noting that this does not mean that parameters such as $\eta_{f}$ do not affect RWB and BWB phenomena, and they control the significance level of BWB and RWB variation during the flare period: if $\eta_{f} \ll 1$, i.e., the static emission is significantly larger than the flare flux, then $\alpha$ is almost constant at this time; if $\eta_{f} \sim 1$, then significant BWB and/or RWB phenomena can be observed.

Based on the parametric analysis in Section \ref{subsec:ParameterA}, the evolution of the particle number distribution and multi-band flare amplitude ratio are significantly impacted by the ratio of $t_{esc}/t_{acc}$, which makes it a critical factor in the time-containing evolution with $\alpha$. As illustrated in Figure \ref{fig:svp}(c), the softening trend in the evolution of $\alpha$ becomes more apparent as the ratio of $t_{esc}/t_{acc}$ decreases, and vice versa. Figure \ref{fig:svp}(d) shows that the $t_{esc}/t_{acc}$ ratio’s variation affects not only the $\alpha$ profile over time but also the $\alpha$-intensity trend. For a specific flare event, the evolution of $\alpha$ may demonstrate a monotonic trend of either BWB or RWB.

In essence, we consider that the evolution of the spectral index over time at STV is due to the non-uniformity and non-simultaneity of the evolution of the particle number between different frequencies. Assuming that the ratio of flare amplitudes between the variability curves at different frequencies remains constant, it can be inferred that the $\alpha$ does not change with time. Thus, the observed variability of the $\alpha$ implies that the ratio of the intensity between the flares at different frequencies is changing with time. Furthermore, assuming similar flare profiles at different frequencies, if the acceleration and cooling processes of particles between different frequencies occur simultaneously, i.e., there is zero time delay between the flares at different frequencies, the pattern of the $\alpha$ evolution with time in a single flaring event will only show a monotonic RWB/BWB trend, while the presence of time delay will make it possible for both BWB and RWB to co-exist in the same flaring event. In summary, the amplitude ratios and time delays between multi-band flares can be used as measurements to distinguish differences in flares and track particle distribution evolution in accelerated and radiated regions.

From the theoretical analysis in Section \ref{subsec:ParameterA}, it is shown that the main parameters affecting the time delay and amplitude ratio are $t_{esc}/t_{acc}$, $\gamma_{0}$, and $\gamma_{max}$. Taking Figure \ref{fig:svp}(e) and \ref{fig:svp}(f) as examples, we show the variability patterns of the $\alpha$ under different $\gamma_{0}$. The $\alpha$ changes significantly in these cases,  potentially due to the dominant presence of the hard component (flare) over the steady component. Besides, according to the Eq.(6) of KRM and $t_{acc} = (\beta_{s}\gamma_{max})^{-1}$, it is known that the variability of parameters such as $B$ as well as $\gamma_{max}$ will lead to the change of $t_{acc}$, which in turn will affect the change of $\alpha$. In addition to the above parameters, we also simulated the effects of other parameters in the Eq.(1) of KRM, and the results are consistent with the above theoretical analysis. To refine the theoretical model, we analyze it in the next section by fitting the observed data with statistical methods.

\section{Fitting and Results} \label{sec:result}

Based on the theoretical model described in Section \ref{sec:model} and the pulse solutions given by equations, the observed IDV light curves can be interpreted as the convolution sum of the flares generated by these independent radiant regions. The radiation parameters corresponding to a single pulse can be obtained by decomposition of the observed curves. However, the observed light curve of one-dimensional time series can only determine the arrival time of pulses, but not the position of individual cells in the three-dimensional jet. The simulation of three-dimensional fine structure in the jet will be carried out in future further studies. 

\subsection{Fitting Pulses} \label{subsec:fit}

\begin{figure*}
	\begin{minipage}{\textwidth}
		\centering
		\includegraphics[trim=3.5cm 1.5cm 0.2cm 0.0cm,width=0.49\textwidth,clip]{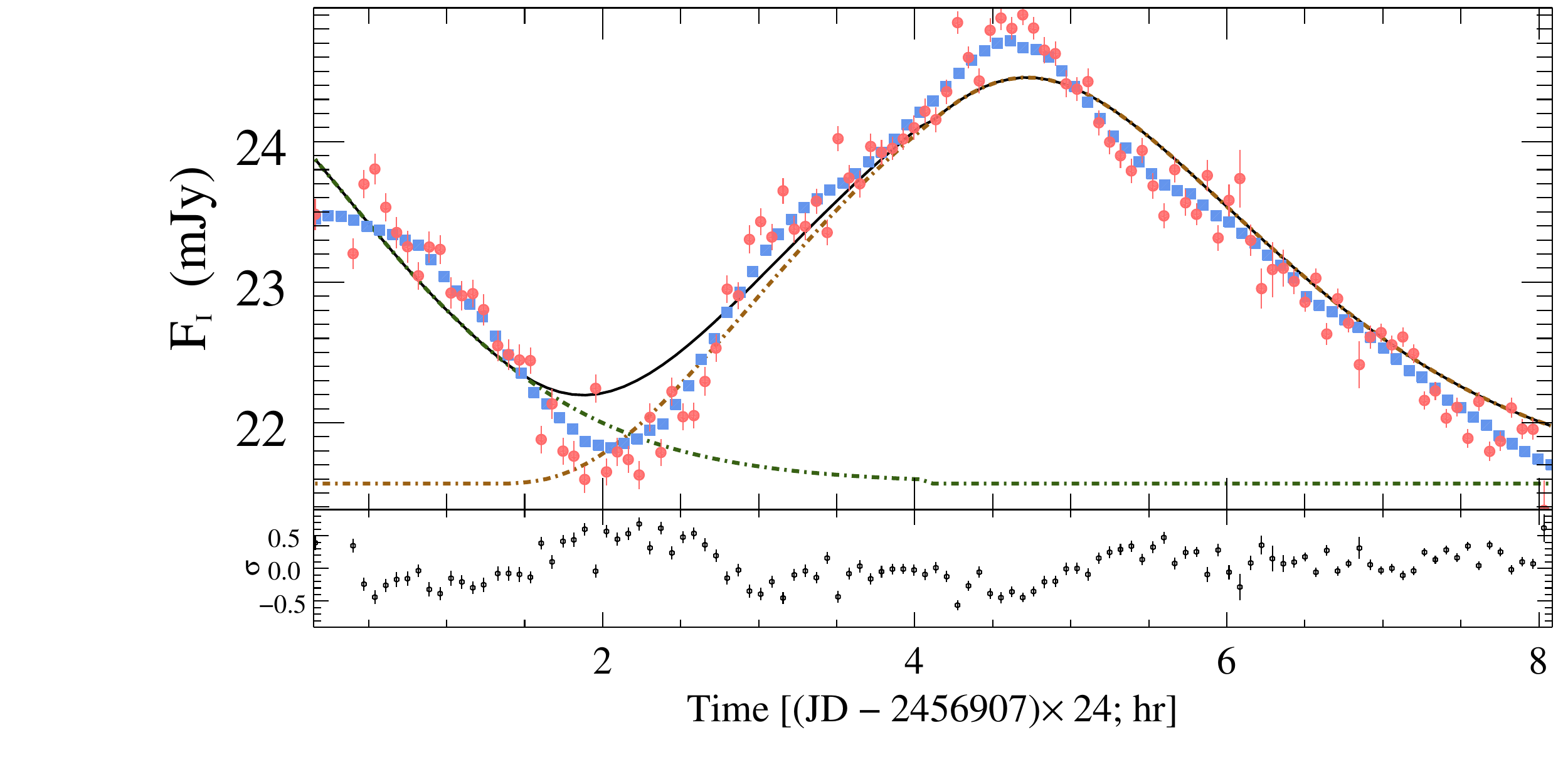}
		\includegraphics[trim=3.5cm 1.5cm 0.2cm 0.0cm,width=0.49\textwidth,clip]{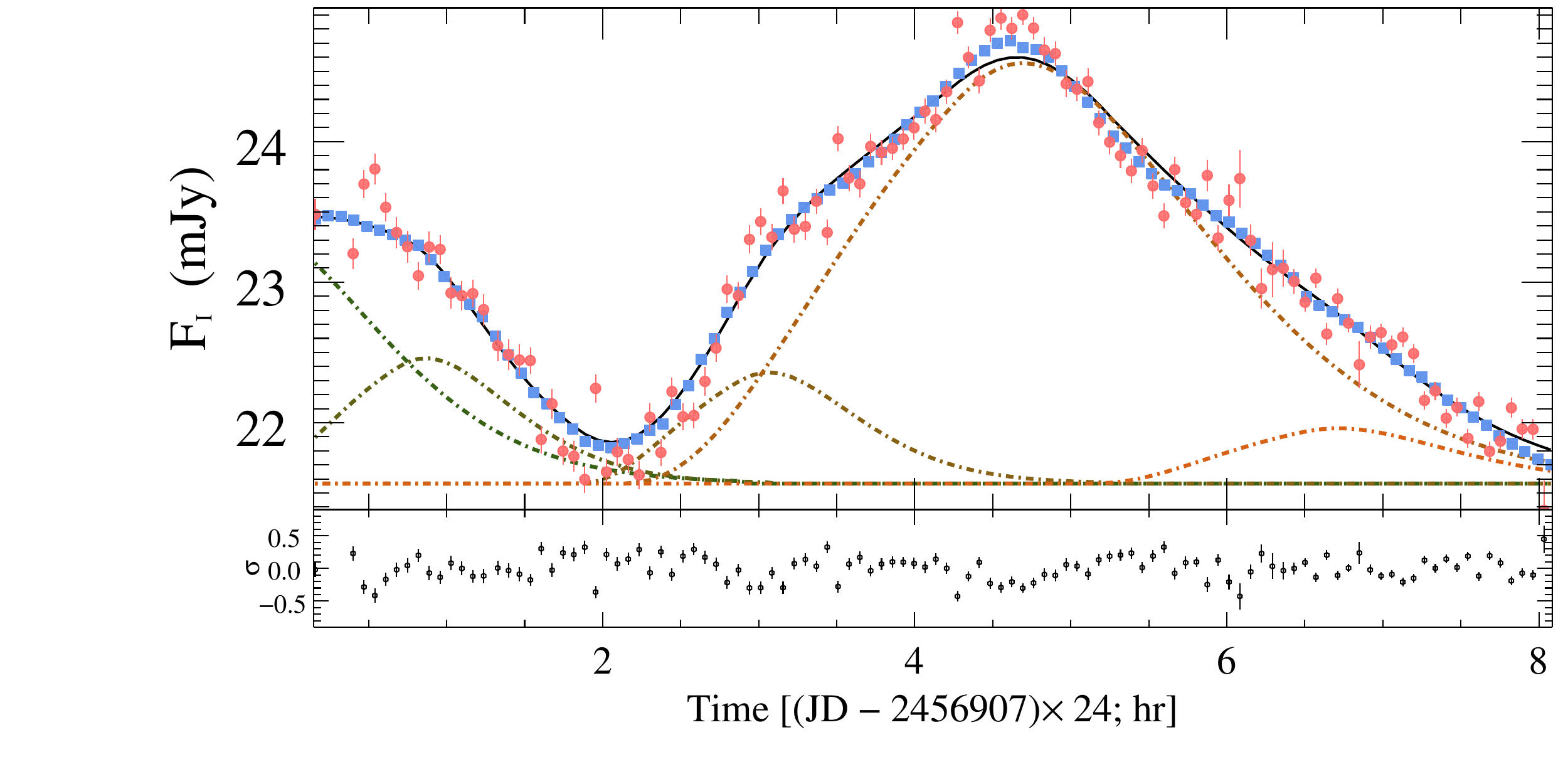}
		\includegraphics[trim=3.5cm 1.5cm 0.2cm 0.0cm,width=0.49\textwidth,clip]{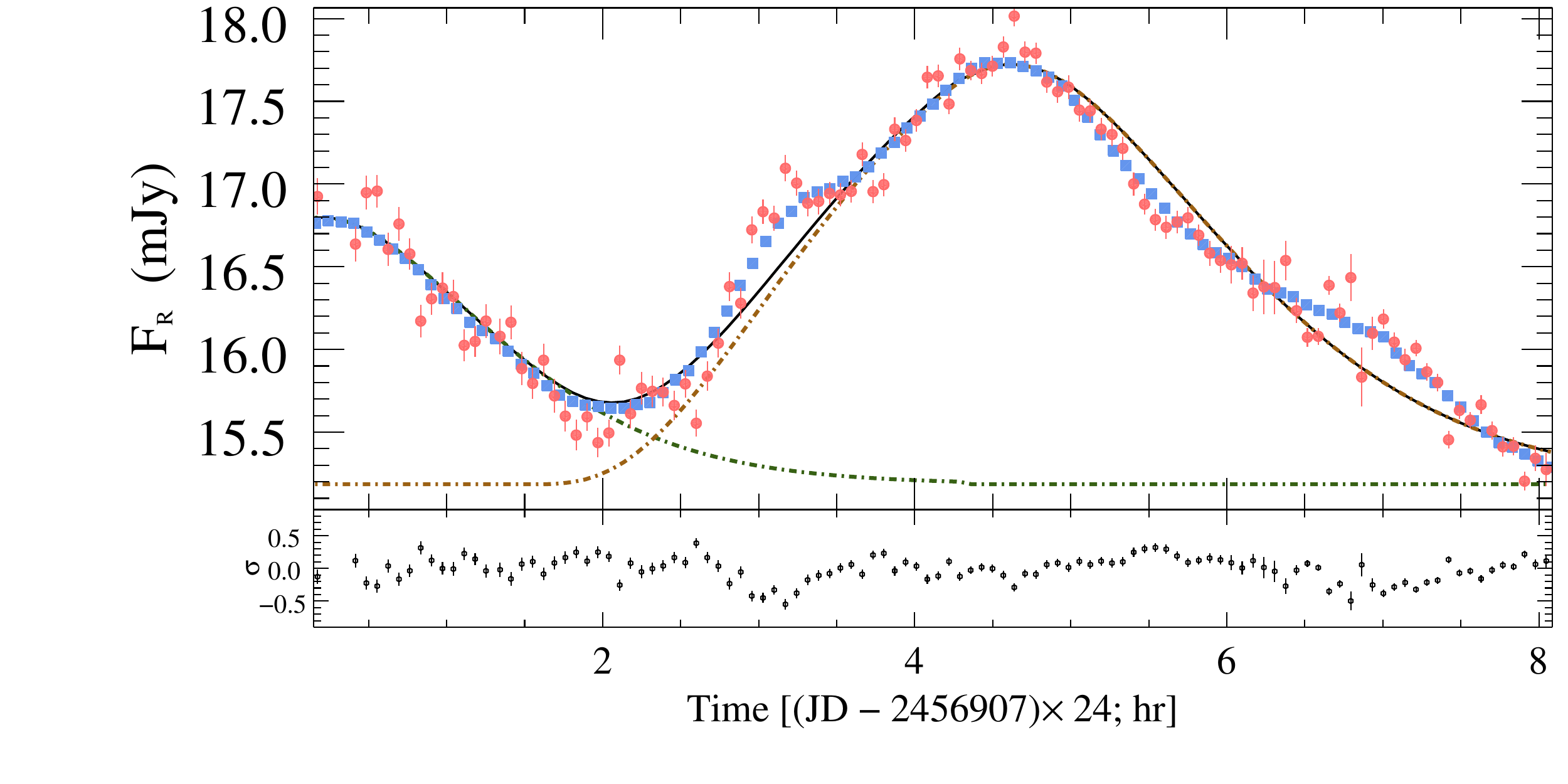}
		\includegraphics[trim=3.5cm 1.5cm 0.2cm 0.0cm,width=0.49\textwidth,clip]{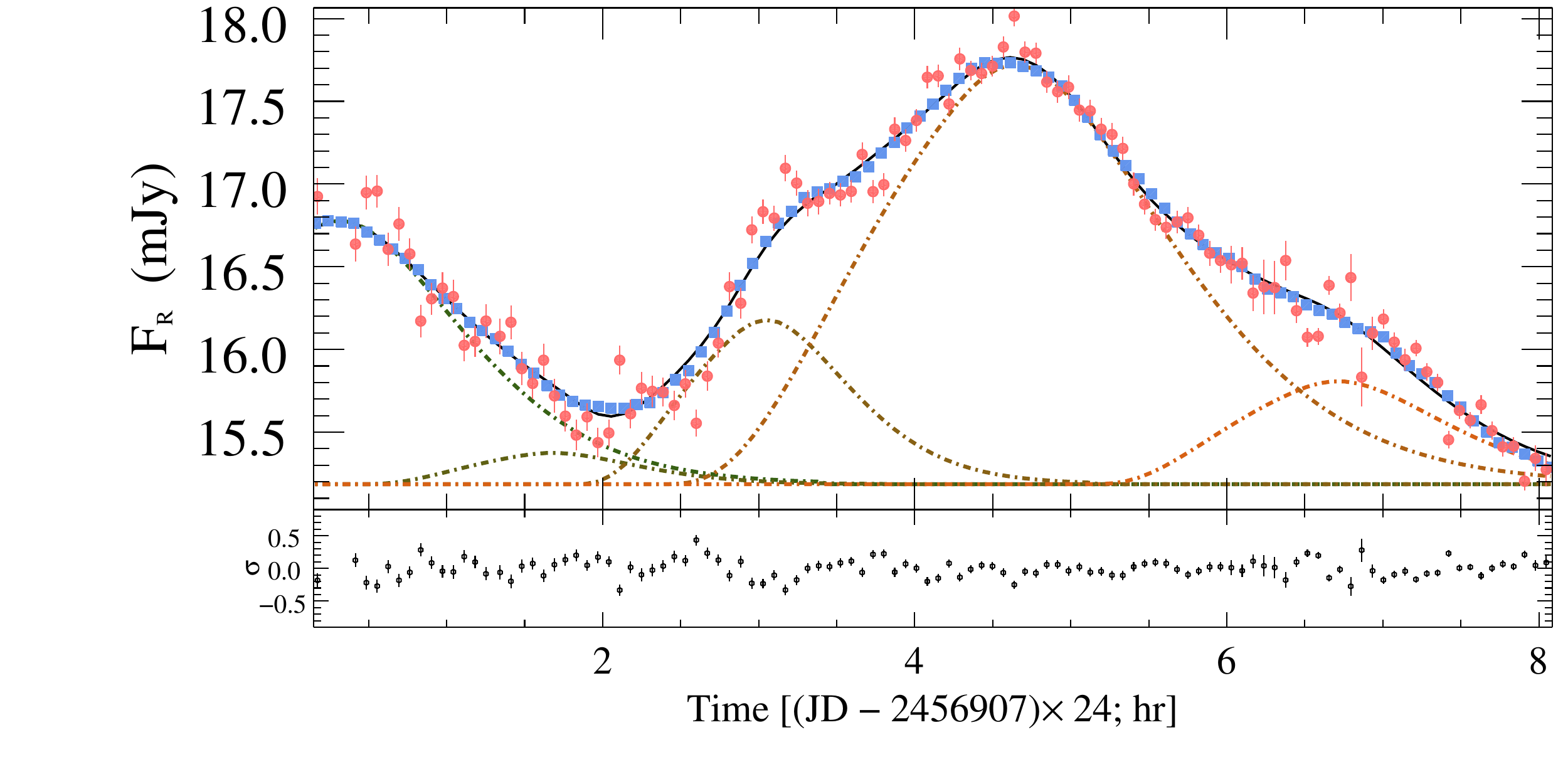}
		\includegraphics[trim=3.5cm 1.5cm 0.2cm 0.0cm,width=0.49\textwidth,clip]{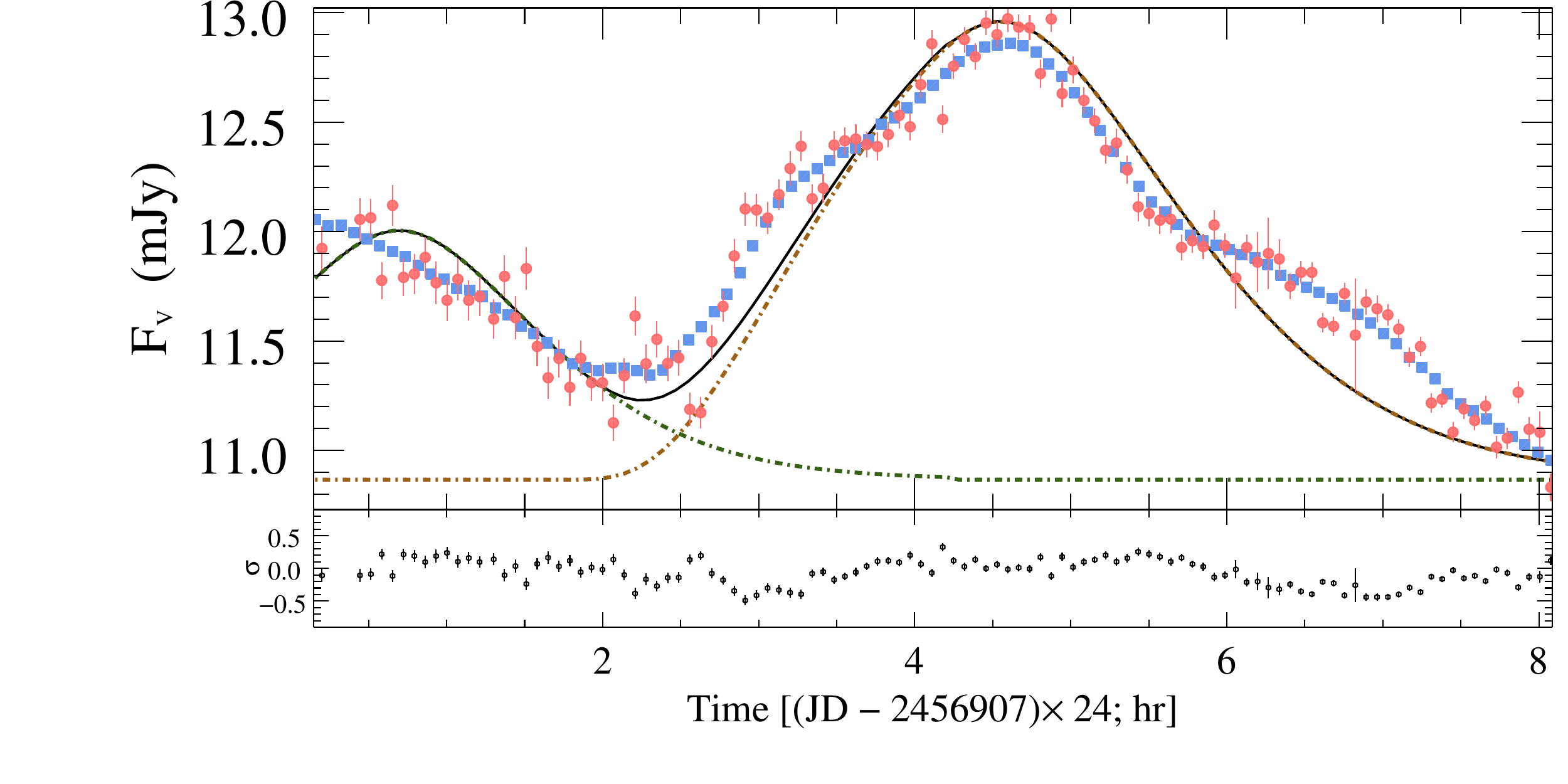}
		\includegraphics[trim=3.5cm 1.5cm 0.2cm 0.0cm,width=0.49\textwidth,clip]{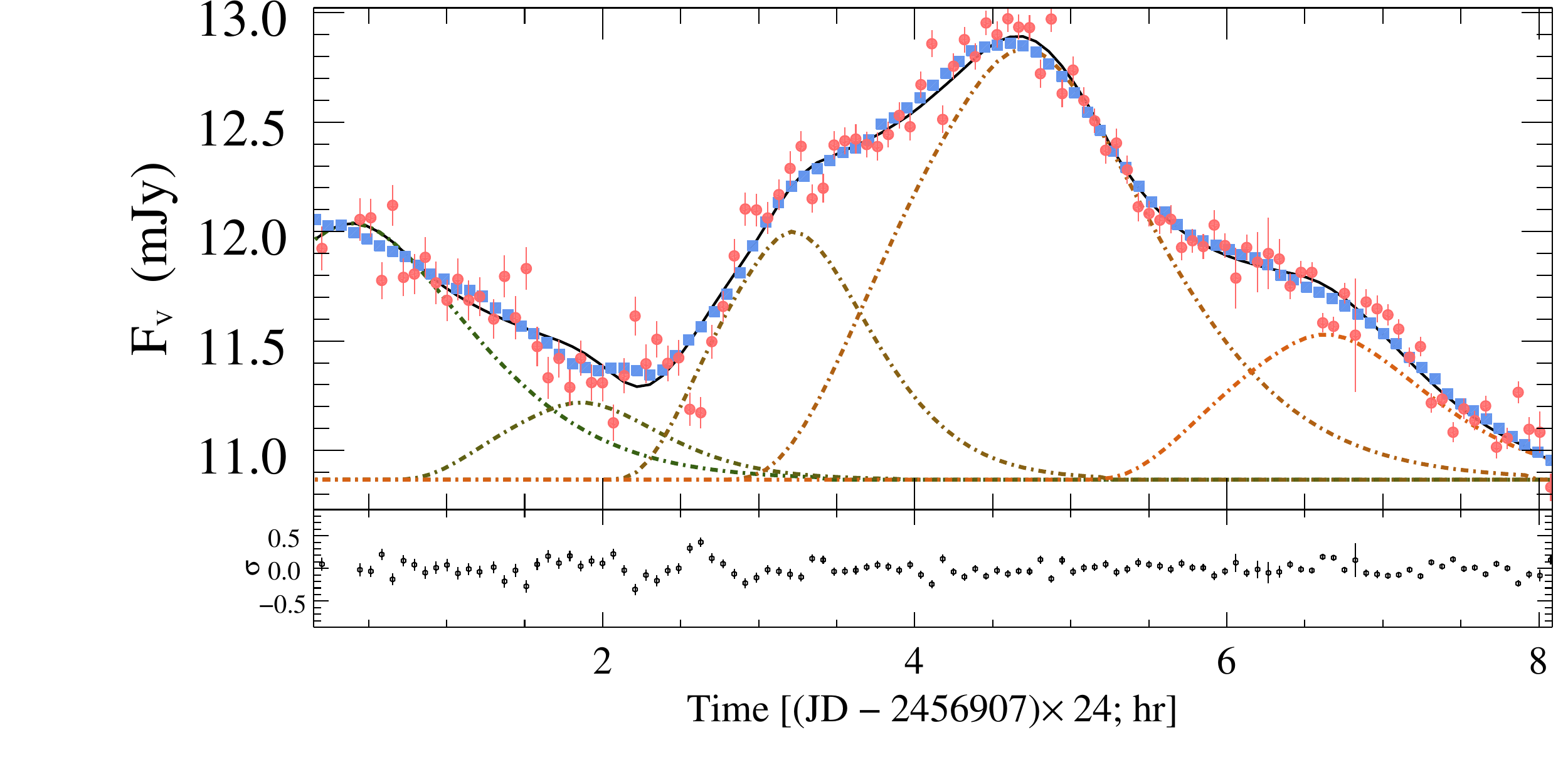}
		\includegraphics[trim=3.5cm 1.5cm 0.2cm 0.0cm,width=0.49\textwidth,clip]{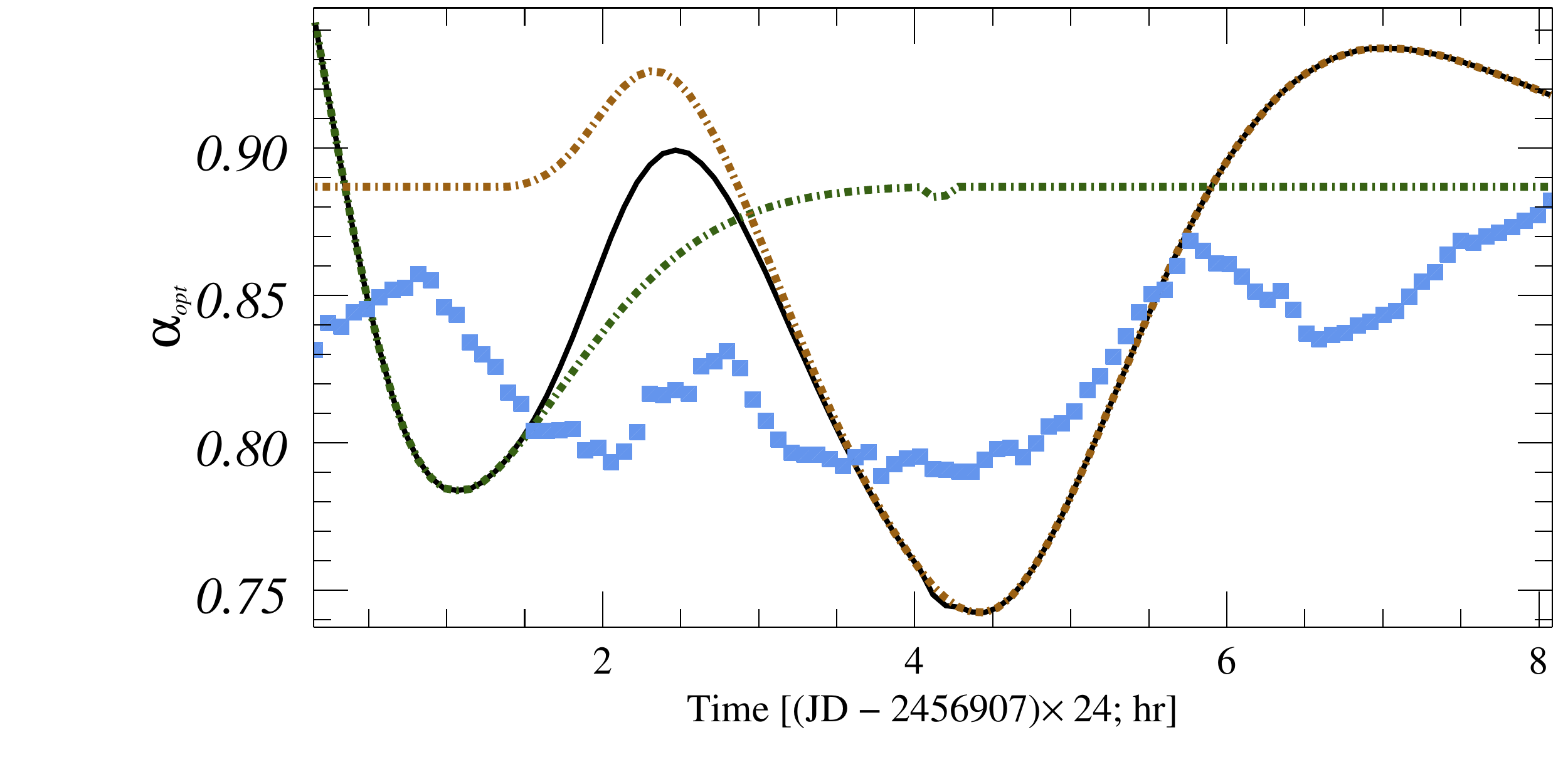}
		\includegraphics[trim=3.5cm 1.5cm 0.2cm 0.0cm,width=0.49\textwidth,clip]{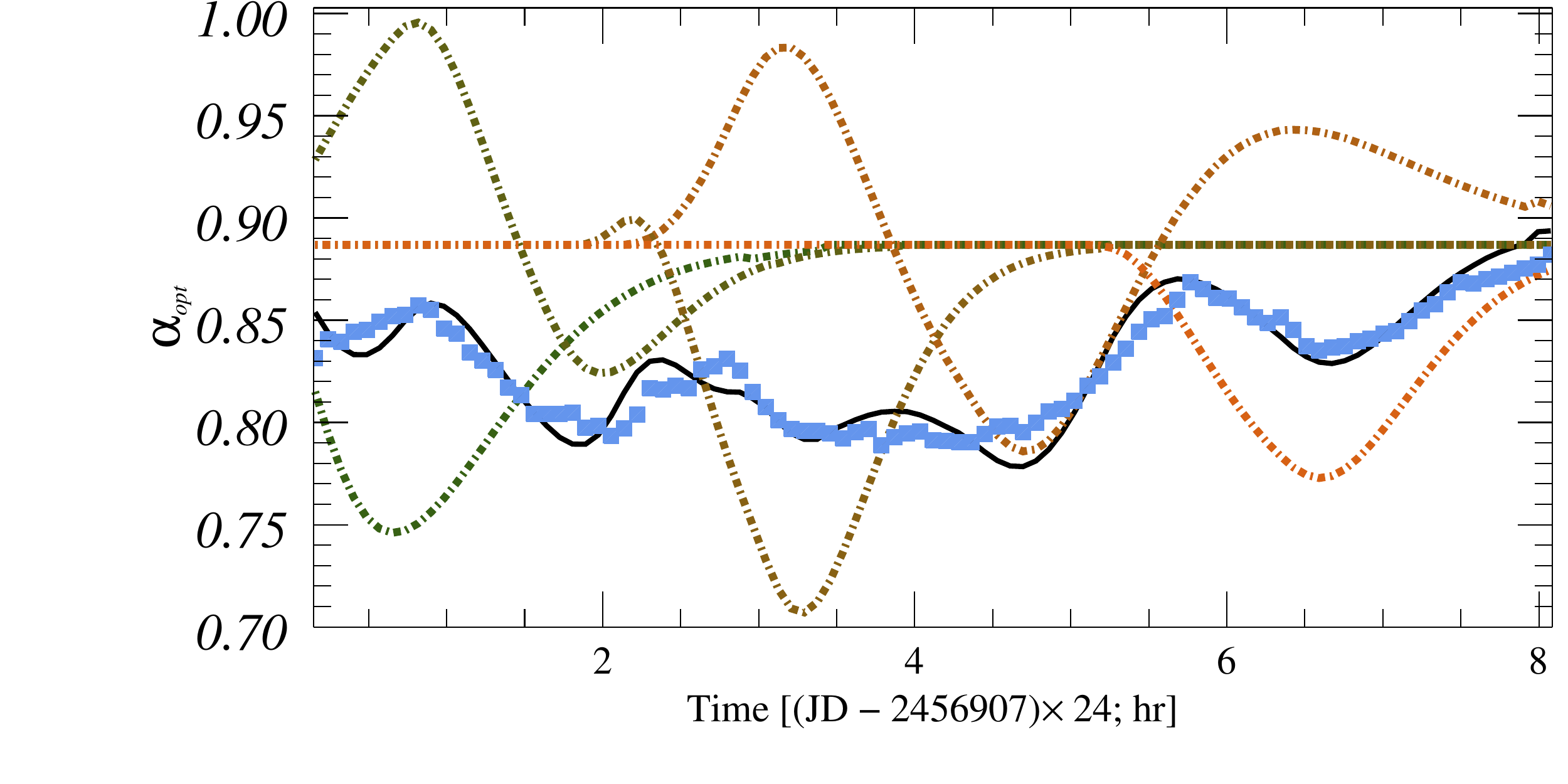}
		\includegraphics[trim=3.5cm 1.5cm 0.2cm 0.0cm,width=0.49\textwidth,clip]{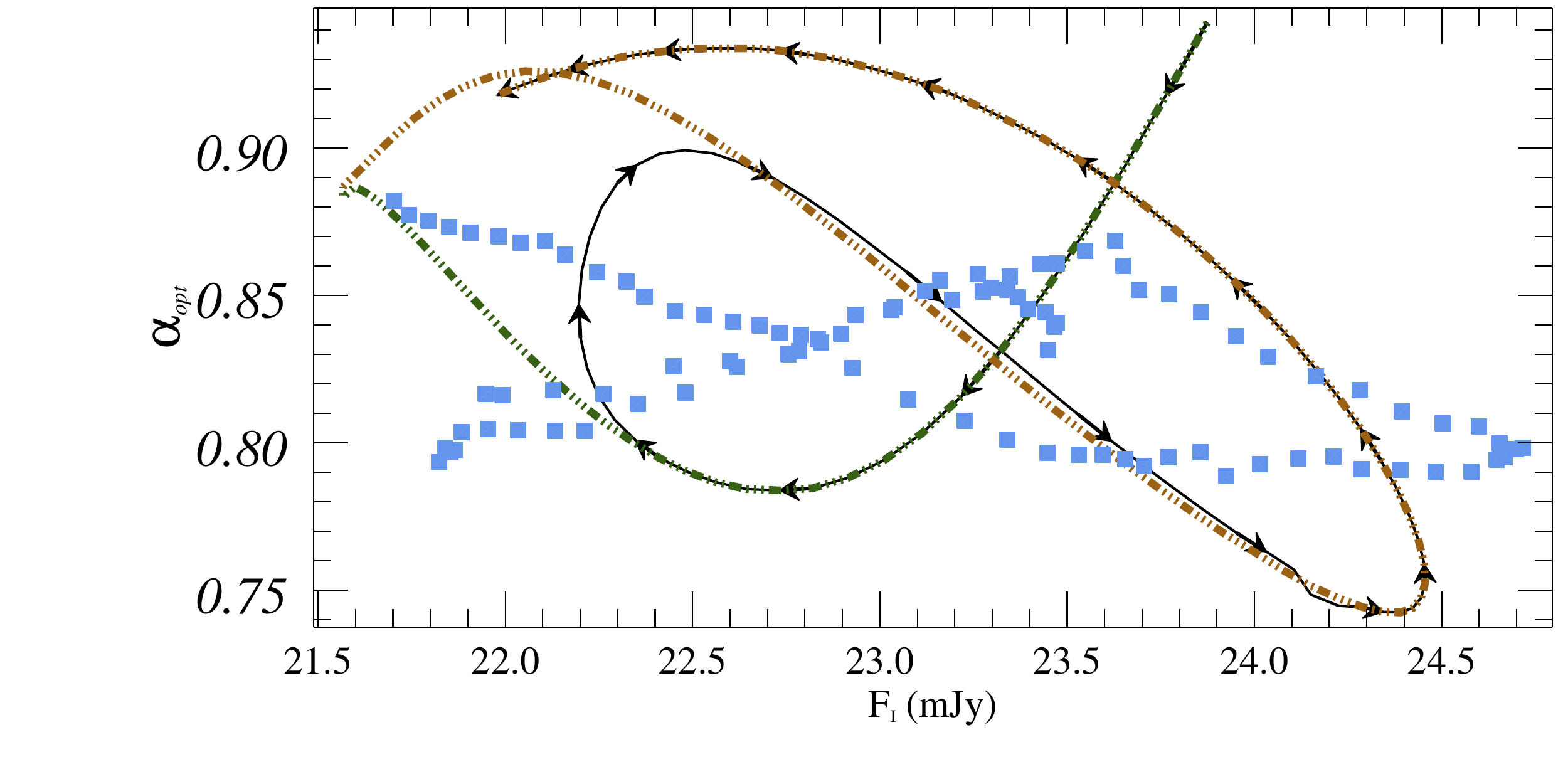}
		\includegraphics[trim=3.5cm 1.5cm 0.2cm 0.0cm,width=0.49\textwidth,clip]{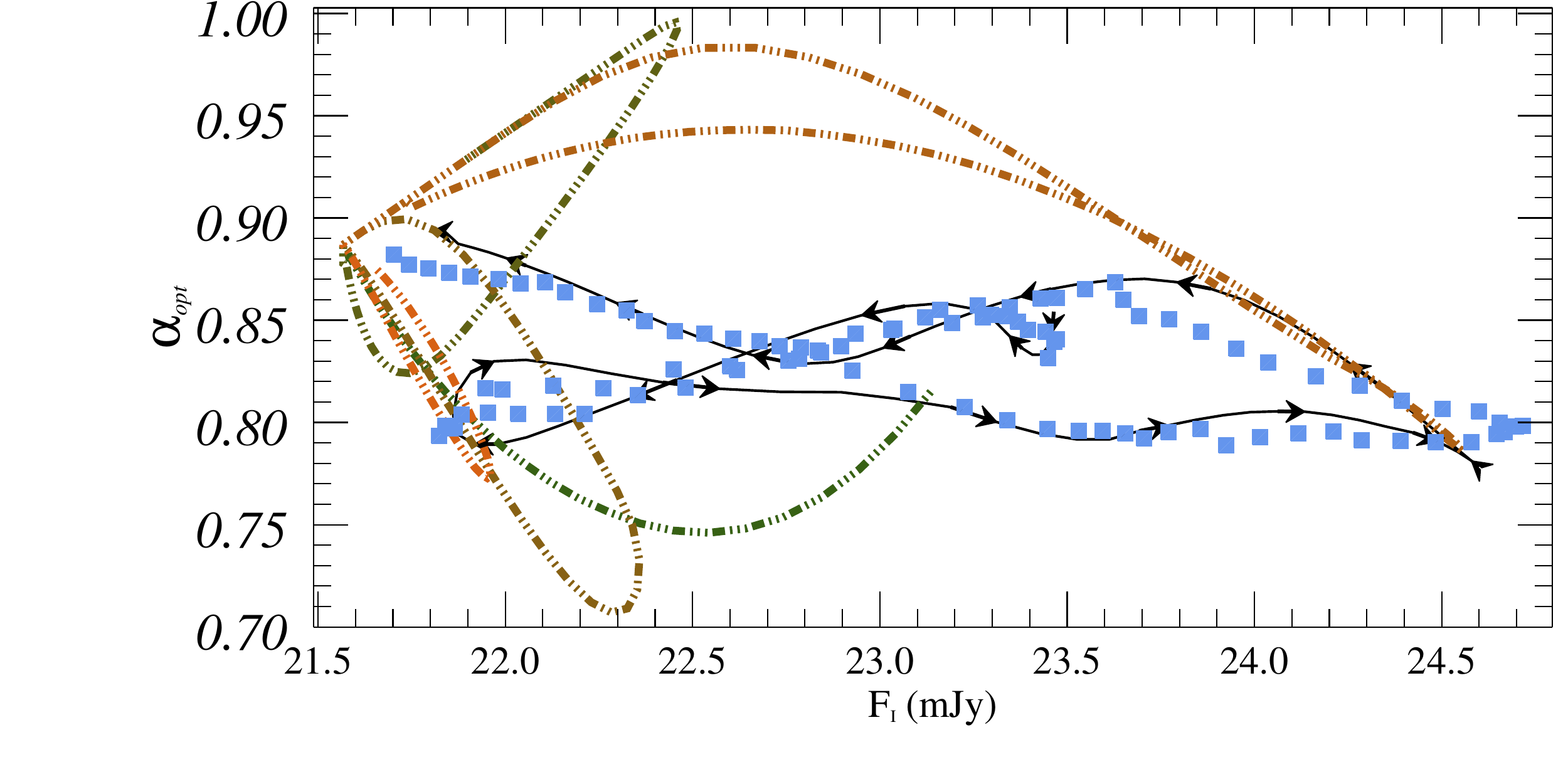}
	\end{minipage}\vspace{0.001cm}
	\caption{Examples of IDV fitting results. Left: Fitting results under single or double radiating regions as a comparison. Right: Best-fit results in the case of multiple radiating regions. From the top to the bottom, the fitting results of $I$, $R$, and $V$-band variability, the evolution of $\alpha_{opt}$, and the hardness-intensity diagrams are shown in order. In each panel, the red points are the original data, the blue points are smoothed data, and the black solid line is synthetic curves obtained by fitting the flare. The dashed lines in different colors show the independent flares obtained by the fit. The lower panel represents the residual distribution between the fitting data and the original data. All figures for all the data are available online.}
	\label{fig:fitexam}
\end{figure*}

Before fitting, to avoid the influence of introducing STV/LTV into the IDV fitting, we obtained the curve of basal components by fitting the LTV trend with polynomials. By fitting multiple pulses in numerous microvariability curves, we can obtain the size and the enhanced particle number density of the small-scale inhomogeneous emission region, and the distribution of acceleration parameters corresponding to these flares (assuming that other parameters such as shock velocity and magnetic field strength remain constant locally). These non-uniform regions exist in the form of superimposed background laminar flow, and the decomposition of background radiation components and radiation enhancing components will be further discussed in the following results analysis. The fitting process is described below.

Based on previous successful studies \citep[e.g.][]{Bha13, Xu19, Web21} that fitted independent pulse profile to microvariability curves, we have further refined the fitting method. By fitting the time-domain evolution of the optical multi-wavelength variability curves and spectral indices simultaneously, we obtain their flare parameters and thus calculate the physical parameters of the emission regions, unlike the previous paper where only the baseline model profile was used as a template for the fit. Here, simultaneous multi-band optical data after re-sampling of the original data are used as the sample, excluding the effect of small fluctuations on very short time scales. The lowest value of the intraday flux is taken as the baseline flux, and the average of the enhanced flux above it is taken as the average daily flare growth. Note that the parameters of emission zone obtained only from the observed optical variability features as a distinction are concurrent and thus their solutions are not unique, but for flares with different parameter sets still indicate that their origins may not be the same. Unlike the method described in the previous article, the number of flares by estimating the number of inflection points and extrema of the light curves at several different frequencies as well as the evolution of the $\alpha$, and the initial value of the flare simulation is thus obtained. 

Based on the above improvements, the automatic fitting method automatically fits the number of intraday flares, the amplitude and duration of individual flares by iterative random number and least squares methods to approximate the results that minimize the $\chi^{2}$ sum of the original data and the fitted curve. In this research, the light curves with at least 50 data points in each band on a single day were selected as the original data input, and the correlation coefficients between the light curves generated by the fitted flare superposition and the original data were all above 99\%. A full version of the fitted light curve plots is available online. The duration, amplitude, and central time of each pulse were directly obtained during the fitting of the flares. The size of the turbulence zone, the particle number density increment, and the relative position of the flares are also calculated.

 Some examples of the fitting for $I$, $R$, and $V$-band variability for BL Lacertae are presented in Figure \ref{fig:fitexam}. The top and bottom panels give the observed data and fit patterns, the evolution of the spectral indices, and the hardness-intensity diagrams, respectively. The spectral index $\alpha_{opt}$ is derived by fitting optical multi-band observations using a power-law function ($\nu F_{\nu} = A \nu^{-\alpha_{opt}}$). In each panel, the red dots are raw observations, and the blue square dots are the data points of multi-band simultaneity after smooth interpolation. The colored dashed lines indicate the different flares obtained by the fit, the solid black line is the total fit, and the horizontally dashed line indicates the constant background flux for a single day. The lower plot of each fitted pattern indicates the distribution of residual between the fitted results and the original data. The left panel of Fig. \ref{fig:fitexam} shows the fitting results for the single/dual emission region obtained by the general fitting method, while the right panel shows the fitting results for the multi-region obtained with our improved method. 

The correlation coefficients between observations and calculated curves for the $I$, $R$, and $V$-band obtained from the single/dual emission region are $99.24\%$, $99.21\%$, and $98.44\%$, respectively, and the correlation coefficient between the fitted result and the interpolated data of the hardness-intensity diagram is $74.22\%$. For comparison, the correlation coefficients of the light curves obtained by fitting the improved multi-region radiation model were $99.86\%$, $99.78\%$, and $99.72\%$, respectively, and the correlation coefficient of the hardness-intensity diagram was $95.58\%$. It is obvious that the flares in this IDV curve is more likely to originate from multiple non-uniform components than from similar emission regions, and the improved fitting method can explain the observed features more extensively.

\begin{table}
	\centering
	\caption{Flare Parameters Used to Fit the Data}
	\label{tab:example}
	\begin{tabular}{ccccccc}
		\hline \hline
		$Flare Index$ & $Band$ & $Center$ (hr) & $Amp$ (mJy) & $N \times$ $10^{-5}$ ($s^{-1} m^{-3}$) & $\tau_{flare}$ (hr) & $S_{cell}$ (AU) \\
		\hline
		    & $I$ & $-0.245$ & $1.82$  & $5.164$ & $1.92$  & $22.05$ \\
	  $1$   & $R$ & $0.318$  & $1.59$  & $5.441$ & $1.89$  & $21.71$ \\
		    & $V$ & $0.394$  & $1.17$  & $4.648$ & $1.86$  & $21.36$ \\
		    \hline
		    & $I$ & $0.868$  & $0.89$  & $2.530$ & $1.35$  & $15.44$ \\
	  $2$   & $R$ & $1.680$  & $0.19$  & $0.647$ & $1.32$  & $15.08$ \\
		    & $V$ & $1.862$  & $0.35$  & $1.400$ & $1.28$  & $14.73$ \\
		    \hline
		    & $I$ & $3.064$  & $0.79$  & $2.244$ & $1.35$  & $15.47$ \\
	  $3$   & $R$ & $3.052$  & $0.99$  & $3.391$ & $1.30$  & $14.87$ \\
		    & $V$ & $3.221$  & $1.13$  & $4.498$ & $1.25$  & $14.29$ \\
		    \hline
	    	& $I$ & $4.689$  & $2.99$  & $8.478$ & $2.81$  & $32.19$ \\
	  $4$   & $R$ & $4.648$  & $2.54$  & $8.653$ & $2.38$  & $27.32$ \\
		    & $V$ & $4.704$  & $1.97$  & $7.815$ & $2.02$  & $23.21$ \\
		    \hline
		    & $I$ & $6.714$  & $0.39$  & $1.115$ & $1.76$  & $20.15$ \\
	  $5$   & $R$ & $6.714$  & $0.62$  & $2.126$ & $1.69$  & $19.37$ \\
		    & $V$ & $6.631$  & $0.66$  & $2.633$ & $1.62$  & $18.62$ \\  
		\hline
	\end{tabular}
	\tablecomments{(This table is available in its entirety in machine-readable form.)}
\end{table}

The various flare durations, amplitudes, and center times obtained from the simulated IDV curves in Figure \ref{fig:fitexam} are shown in Table \ref{tab:example}. The full version of table \ref{tab:example} listing the flare parameters obtained from the BL Lacertae fitting is available online. Columns 1 and 2 of the table list the index of the flares and the observed bands, as well as columns 3 and 4 give the flare centers and amplitudes. The amplitude of the flare is then converted to the enhanced electron number density as listed in column 5. Column 6 gives the width of the flare in the observer's coordinate system ($\tau_{flare}^{obs}$). Column 7 indicates the size of the local turbulent cells, which correspond to the radiation timescale of the flare according to the relation
\begin{equation}
S_{cell} = \frac{\tau_{flare}^{obs}\beta_{s}c}{\Gamma(1-\beta\cos\theta)(1+z)}
\label{equ:Sizeofcell}
\end{equation}
where $\beta_{s}$c, $\Gamma$, $\beta$, and $\theta$ represent shock velocity, bulk Lorentz factor, normalized speed of the emission region, and angle with the line of sight of the jet, respectively.

\subsection{Statistical Analysis} \label{subsec:stasis}

\begin{figure*}
	\begin{minipage}{\textwidth}
		\centering
		\includegraphics[trim=1.0cm 1.1cm 0.5cm 0.9cm,width=0.99\textwidth,clip]{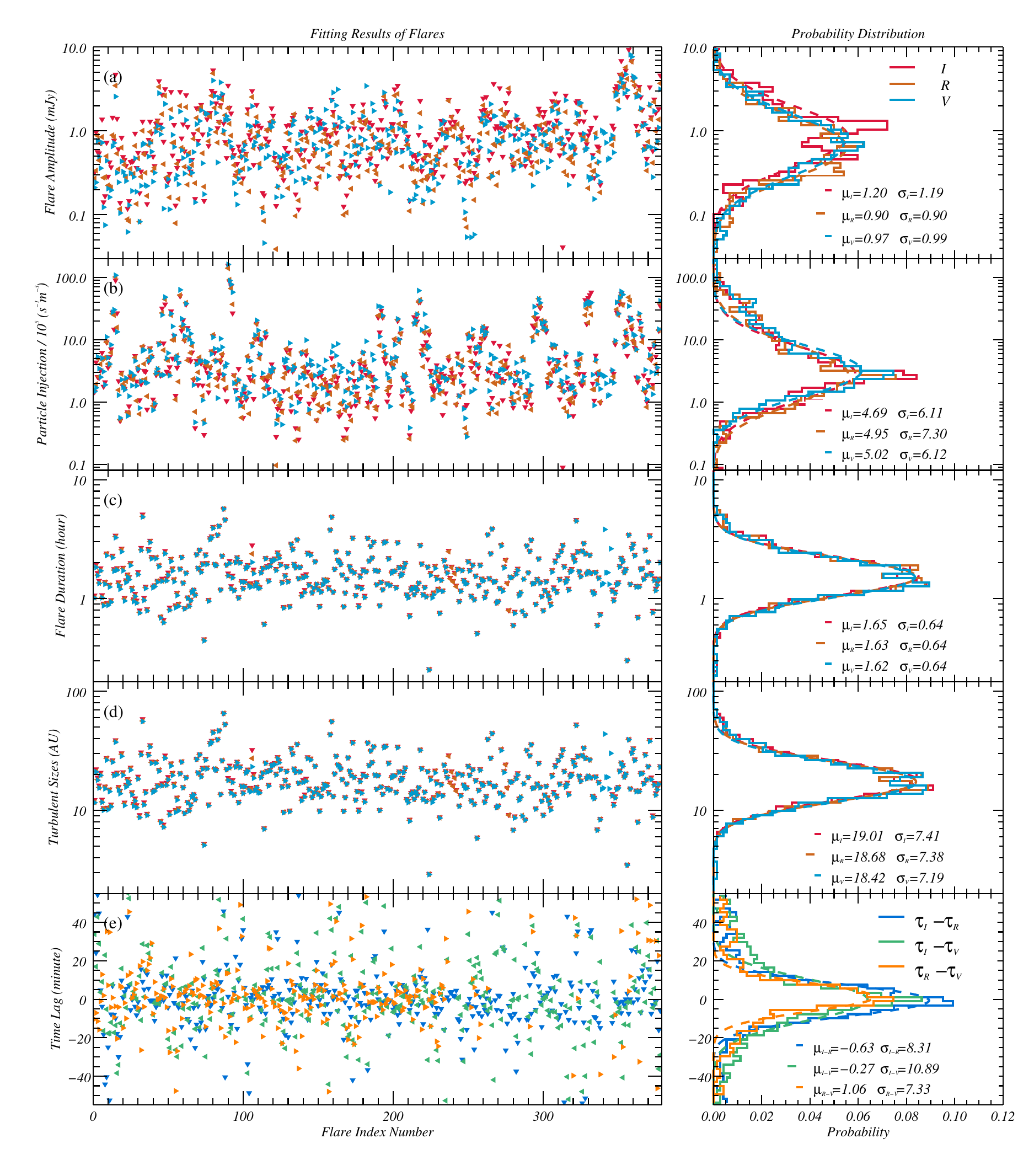}
	\end{minipage}\vspace{0.001cm}
	\caption{Left: The red, brown and blue triangular dots correspond to the flare parameters obtained by fitting in the $I$, $R$ and $V$-band, respectively. The blue, green and orange triangular dots correspond to the time delays between the $I$ and $R$-band, $I$ and $V$-band, and $R$ and $V$-band, respectively. Right: The histograms in different colors represent the probability distributions of the parameters obtained for the different bands. The dashed lines are the approximate distributions obtained from their probability distributions, as shown by the patterns, the particle injection rate, turbulence scale, and the amplitude and duration of the flares conform to a log-normal distribution, while the time delay conforms to a normal distribution.}
	\label{fig:fitresult}
\end{figure*}

\begin{figure*}
	\begin{minipage}{\textwidth}
		\centering
		\includegraphics[trim=2.2cm 1.5cm 0.2cm 0.2cm,width=0.49\textwidth,clip]{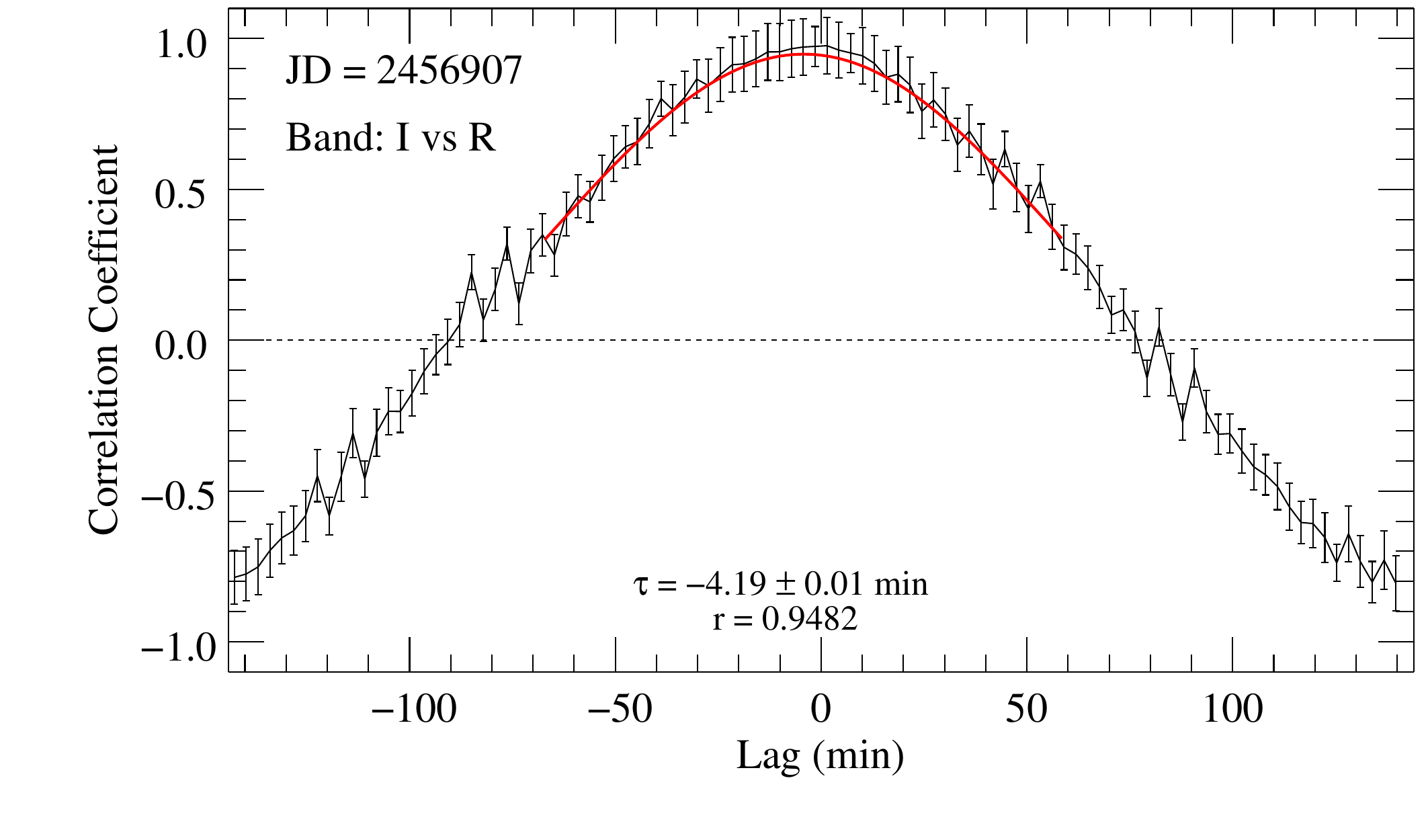}
		\includegraphics[trim=2.2cm 1.5cm 0.2cm 0.2cm,width=0.49\textwidth,clip]{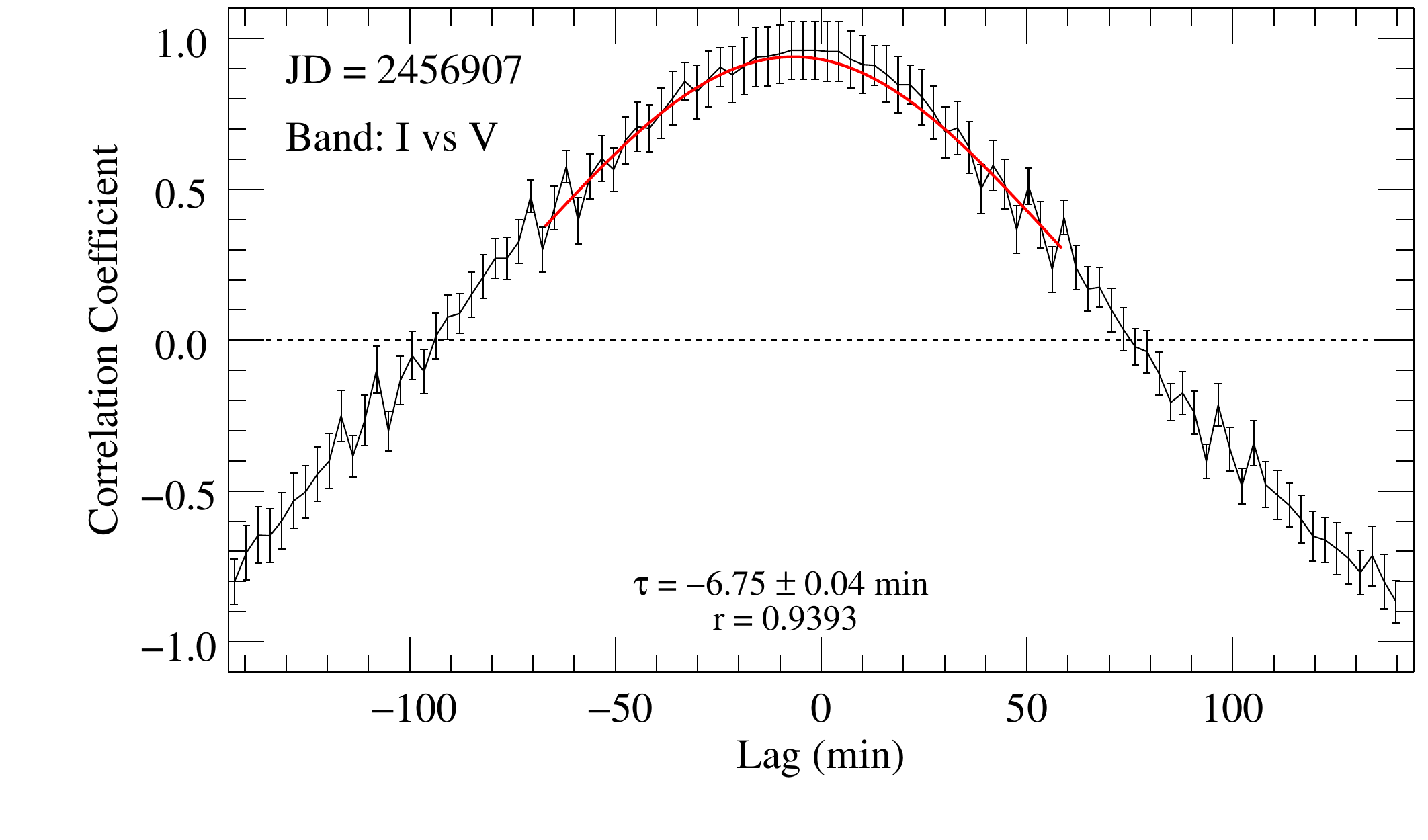}
	\end{minipage}\vspace{0.001cm}
	\caption{Examples of DCF correlation and fitting results between IDV curves in different bands. The red solid red lines show Gaussian fitting to the points. $\tau$ gives the lag result.}
	\label{fig:dcf}
\end{figure*}

\begin{figure*}
	\begin{minipage}{\textwidth}
		\centering
		\includegraphics[trim=1.0cm 1.1cm 0.9cm 0.9cm,width=0.99\textwidth,clip]{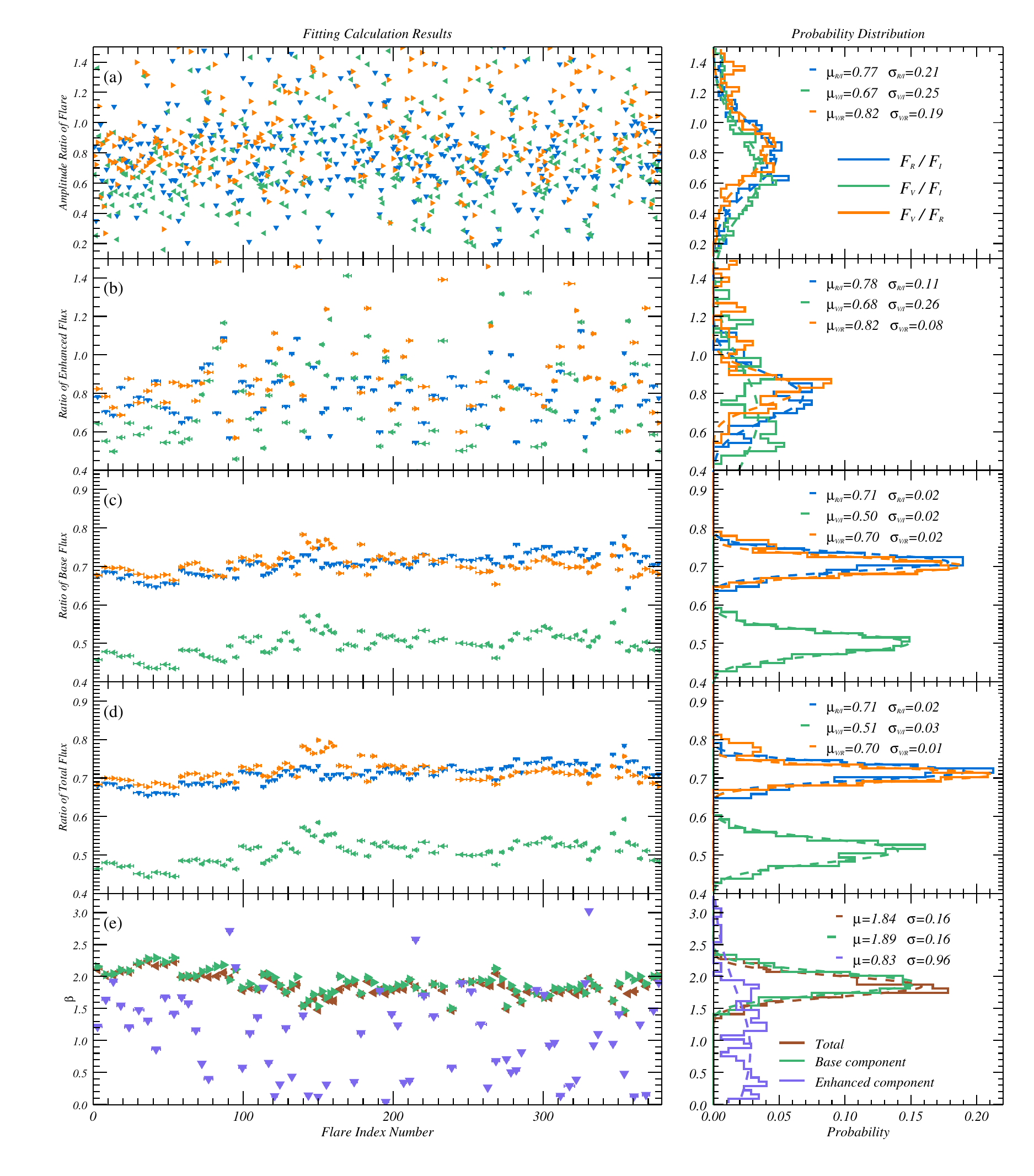}
	\end{minipage}\vspace{0.001cm}
	\caption{Left: The blue, green and orange triangular dots correspond to the flare (top), enhanced (upper-middle), basal (middle), and total (lower-middle) flux ratios between the $R$ and $I$-band, $V$ and $I$-band, and $V$ and $R$-band, respectively. The bottom panel shows the results of the spectral indices obtained with different compositions. Right: The histogram represents the probability distribution corresponding to the left panel. The dashed line is the probability distribution obtained by Gaussian approximation.}
	\label{fig:fitresult2}
\end{figure*}

After fitting and decomposition of the multi-band light curves for 89 nights selected for the BL Lacertae from $2009$ to $2021$, we obtained $378$ multi-wavelength flares, including $368$ in the $I$-band, $371$ in the $R$-band, and $369$ in the $V$-band. Using our fitting method and procedure, we directly obtained the pulse amplitude, duration, center time, and we calculated the enhanced particle number density, scale and relative position of the flare corresponding to the emission region. Since the multi-wavelength microvariability curves are obtained by simultaneous fitting, we can also infer the acceleration parameters based on the differences in the time delay and amplitude ratio of the flares at different wavelengths, and thus distinguish the composition of different emission regions.

Statistical distributions of the parameters that can be obtained directly from the fitting procedure are shown in Figure \ref{fig:fitresult}, where the left panel shows the distribution of the parameters for $378$ flares, with the colored triangular dots indicating the fitting results for different bands, and the right panel indicates the probability distributions obtained statistically for these fitted parameters. As can be seen from the statistical results, the distributions of the parameters of flare amplitude, duration, particle injection rate, and emission region scale conform a log-normal distribution, whose best fitting distribution is shown as a dashed line, with the expectation and variance indicated in the panel. Although the amplitudes and durations of the flares obtained by fitting curves in different bands are different, the particle injection rates and the scales of the radiation cells calculated from the parameters obtained in these different bands are very similar, which is consistent with the theoretical assumption that the radiation sources at different frequencies are in the same region. 

The size distributions of the radiation cells calculated from the fitting results for the $I$, $R$, and $V$-band are $18.30^{+7.58}_{-5.25}$ AU, $18.03^{+7.63}_{-5.25}$ AU, and $17.77^{+7.44}_{-5.14}$ AU, i.e., statistically $99.73\%$ of the turbulent cell sizes are from $2.40$ AU to $33.66$ AU. Note that the cell size obtained here is sensitive to the assumed parameters, including shock velocity, flow speed, and inclination. The energy spectrum fitting of turbulent cells obeys a power-law distribution $E(k) \propto k^{-p}$ obtained from the statistics of their turbulent cell distribution, where $p = 1.54^{+0.22}_{-0.21}$ and $k$ represents the wave number. This statistic is similar to the distribution obtained in relativistic turbulence simulations \citep{Zra12}, as well as with the $E(k) \propto k^{-5/3}$ distribution derived from Kolmogorov theory. Based on the theoretical parameters from previous studies  \citep[e.g.][]{Bar97, Sum12, Bar17} and the statistical analysis of turbulence scales, the Reynolds number is roughly estimated to be $R_{e} \sim 10^{3} - 10^{5}$. In addition, we find from the statistics that the distribution of particle injection rates may be a superposition of two Gaussian distributions, which predicts that the 378 flares we fitted may be located mainly in at least two large-scale plasma blobs at different locations during the observation time (about 10 years), a conjecture that will be analyzed and verified in further LTV studies.

The time delay of each flare at different frequencies are calculated based on the difference in the center time of the flares obtained by fitting the curves of different bands. The time delay distribution for 378 flares is presented at the bottom panel of Figure \ref{fig:fitresult}. It shows that the time delay distribution obtained by fitting approximately follows the normal distribution. The time delay distributions between $I$ and $R$-band, $I$ and $V$-band, and $R$ and $V$-band are $-0.84 \pm 8.40$ min, $-0.53 \pm 10.97$ min, and $1.22 \pm 6.95$ min, respectively. The obtained minute-scale time delays from the fitting align with the theoretical predictions in Section \ref{subsec:ParameterA}. This suggests that the minute-scale time delays between IDV flares may originate from accelerating or radiating processes. Furthermore, the parameter-dependent time delays can cause fluctuations around the zero-delay, where co-spatiality plays a role. For comparison, we also calculated the discrete correlation function \citep[DCF;][]{Edl88}. The Figure \ref{fig:dcf} gives an example of the DCF obtained from the single-night IDV curve, which shows a clear positive peak. In order to estimate the most likely value of time delay, we fitted a Gaussian function over a limited range of time delays (red dashed line). Statistical time delay distributions were obtained for all 89 nights of independent DCF results between the $I$ and $R$-band, and between the $I$ and $V$-band, producing $\bar{\tau} = -0.33 \pm 5.81$ min and $\bar{\tau} = -1.19 \pm 5.88$ min, respectively. The results from the DCF calculations indicate that the time delays converge to the order of minutes, consistent with the time delays obtained by fitting flares.

From the parametric analysis, it can be seen that the ratio of the amplitudes of the flares at different frequencies depends on the physical parameters of the acceleration zone, and considering that the variation of the spectral index and its evolutionary trend originates from the amplitude ratios of the flares at different frequencies, we speculate that the difference between the spectral index and the amplitude ratios can reflect the similarities and differences of the acceleration zone in which the flares are located. According to the model described in Section \ref{sec:model}, we decompose the variability into long-term components and short-term flares. Based on the fitting results, we have performed statistics on the observed flux ratios and the flux ratios of the different components obtained from the decomposition, as shown in Figure \ref{fig:fitresult2}. The statistical results show that the distribution of the amplitude ratios of each component can be approximated as Gaussian distribution. The distribution of the flare flux ratio, the flux ratio of the intraday enhancement, the flux ratio of the basal component, and the observed flux ratio between the $R$ and $I$-band are $0.77\pm0.21$, $0.78\pm0.11$, $0.71\pm0.02$, and $0.71\pm0.02$, respectively. From the statistical results, the distribution of the intraday enhanced component is similar to that of the decomposed flare component, while the distribution of the flux ratio of the basal component is similar to that of the total flux ratio, which is consistent with the hypothesis of the physical model about the multi-component decomposition. 

Based on the flux ratios of various components, we further calculate the distribution of the spectral indices, as shown in the bottom panel of Figure \ref{fig:fitresult2}. The distribution of the spectral indices demonstrates that the basal component remains consistent with the overall distribution of about $1.84\pm0.16$, while the distribution of the spectral indices obtained from the intraday enhanced fluxes is in a broad range of $0.83\pm0.96$. These spectral index distributions illustrate that the dominant trend of the spectral index variation can be determined by the LTV, and that the observed evolutionary trends in the various diversities of intraday spectral indices may be due to the superposition of several different regions with their different acceleration environments. Furthermore, we find that the spectral index distribution obtained from the intraday increments can be approximately decomposed into two Gaussian distributions, which may provide support for a conjecture that the dominant optical emission region may be located in at least two large-scale blobs during the period 2009 to 2021.

\subsection{Correlation Analysis} \label{subsec:corstudy}

\begin{figure*}
	\begin{minipage}{\textwidth}
		\centering
		\includegraphics[trim=2.2cm 1.5cm 0.3cm 0.0cm,width=0.49\textwidth,clip]{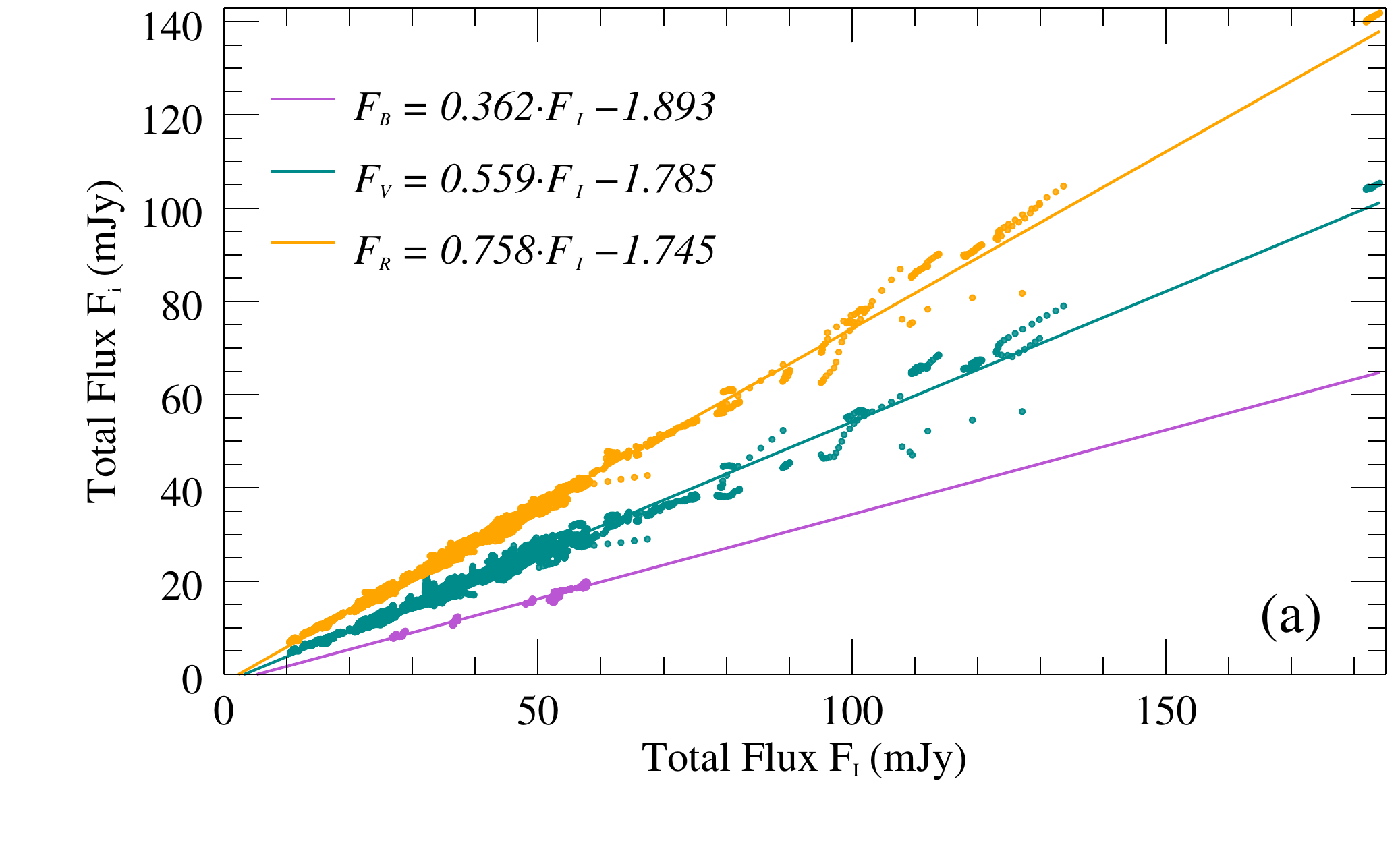}
		\includegraphics[trim=2.2cm 1.5cm 0.3cm 0.0cm,width=0.49\textwidth,clip]{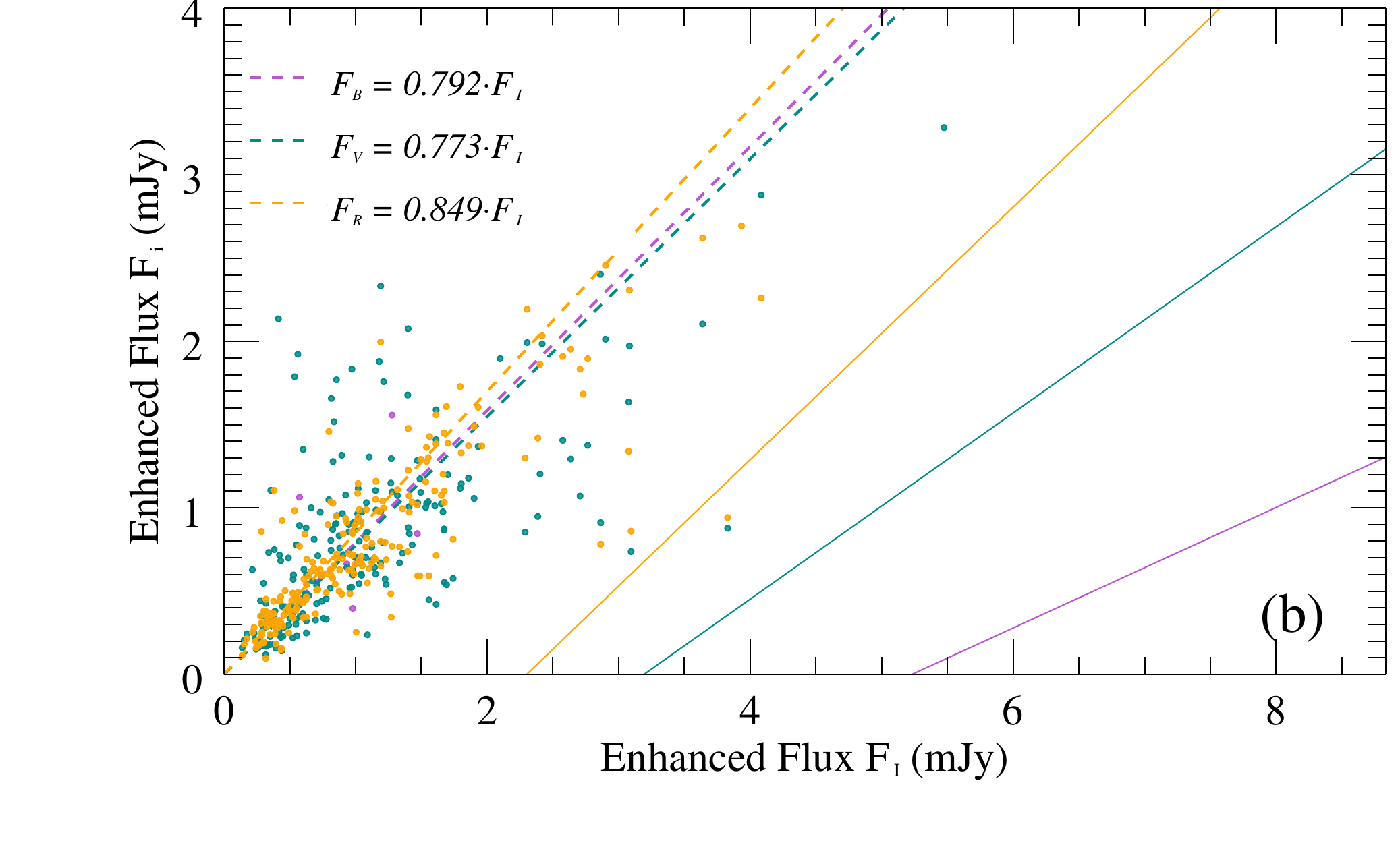}
		\includegraphics[trim=2.2cm 1.5cm 0.3cm 0.0cm,width=0.49\textwidth,clip]{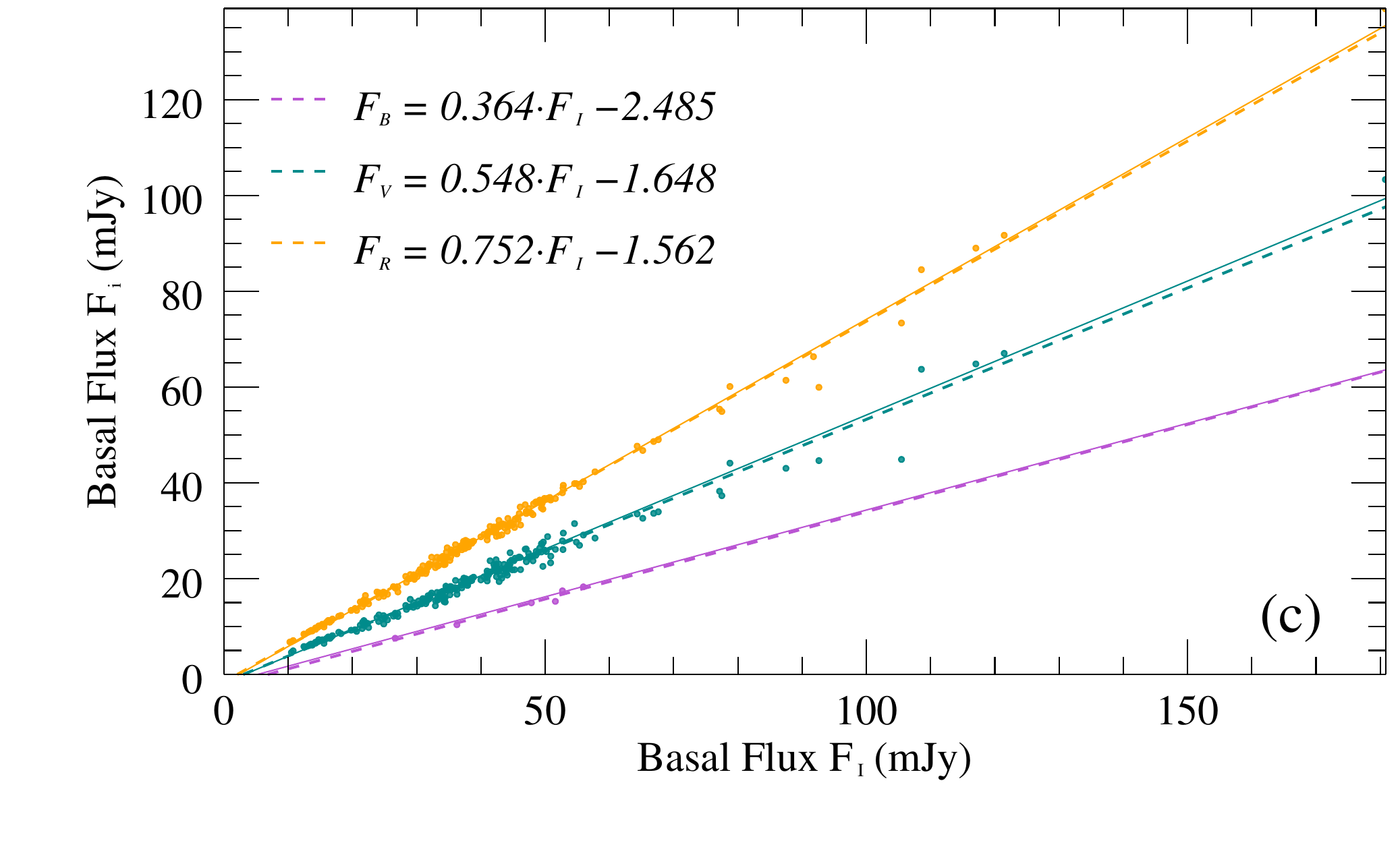}
		\includegraphics[trim=2.2cm 1.5cm 0.3cm 0.0cm,width=0.49\textwidth,clip]{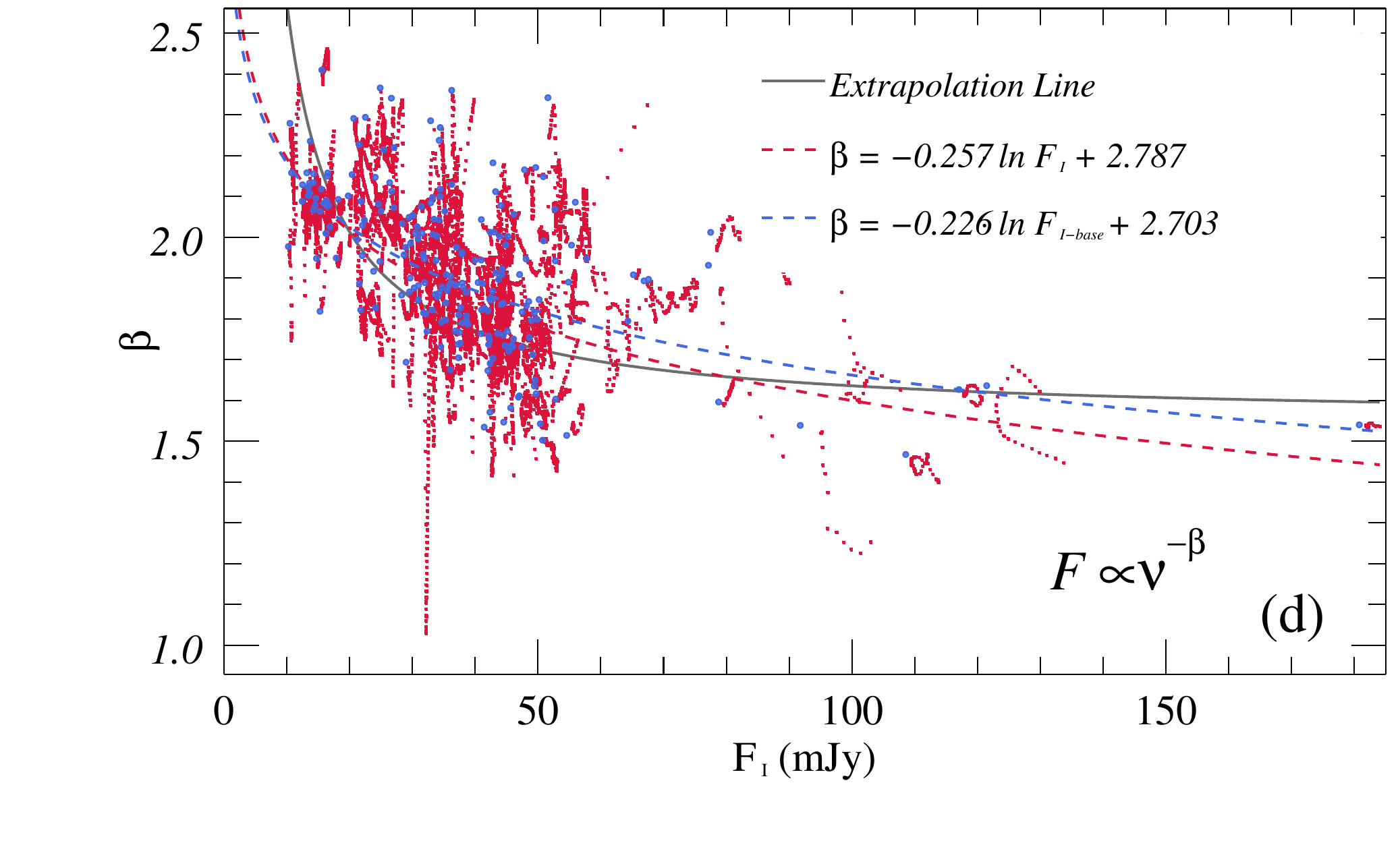}
	\end{minipage}\vspace{0.001cm}
	\caption{Flux–flux diagrams based on observed data (top left panel), basal components assumed to be constant over a single day (bottom left panel), and enhanced components (top right panel), respectively. Colored markers indicate the $F_{B}$, $F_{V}$, $F_{R}$ relative to $F_{I}$. Solid lines and dashed lines show linear fitting of flux-flux relationships obtained from observed data points and different components, respectively. The bottom right panel shows spectral index depending on $F_{I}$, based on observed data (red dots) and basal components (blue dots), respectively. The dashed lines show the approximate trends obtained by fitting, while the solid line shows the inferred line obtained from the flux–flux diagram. Both observations and extrapolations show that there is a dominant BWB trend under LTV.}
	\label{fig:ffdiagram}
\end{figure*}

According to theoretical analysis from Section \ref{subsec:ParameterA} and the parametric simulations from Section \ref{subsec:SVP}, it is shown that different spectral index evolution patterns are obtained for different acceleration parameters, which means different variable components may have different color behaviors. In order to study the evolution of the spectral indices of various components (e.g., constant and variable components) and their correlation with flux variations, we used the methods of \cite{Hag97} and \cite{Lar08} to plot the flux-flux and hardness-intensity diagrams (HID) of various components, as shown in Figure \ref{fig:ffdiagram}. The flux-flux diagram demonstrates that the basal component (i.e., the constant component) shows almost identical linear dependence to the slope of the observed total flux, while the slope of the flux-flux dependence of the variable component is slightly different compared to them. Meanwhile, according to the HID, it demonstrates an overall BWB trend, while the local spectral index evolution is not a single linear dependence, and the presence of a loop pattern similar to that in the simulation of Figure \ref{fig:svp} can be clearly found in numerous HIDs obtained from IDV curves. 

Therefore, based on these observed features combined with the theoretical simulation results, we interpret this correlated evolutionary feature of the flux and spectral index as the result of the superposition of multiple components. As the LTV component that accounts for most of the flux, it dominates the overall dependence of the HID and observed flux-flux diagram, while these STV produced by small-scale emission regions in the jet are superimposed on these long-term components and determine the short-term spectral index evolution of the observed objects according to their relative contribution to the total flux. The evolution of the spectral indices of the large scale blobs in the jet will produce similar evolution due to the change of the Doppler factor or other parameters. In contrast, the HID evolution of the small-scale radiation cells will produce various specific patterns depending on the acceleration parameters, which can explain the evolution characteristics of most observed flux-flux diagrams and HID under this model conjecture.

\begin{figure*}
	\begin{minipage}{\textwidth}
		\centering
		\includegraphics[trim=2.0cm 0.4cm 0.3cm 0.1cm,width=0.99\textwidth,clip]{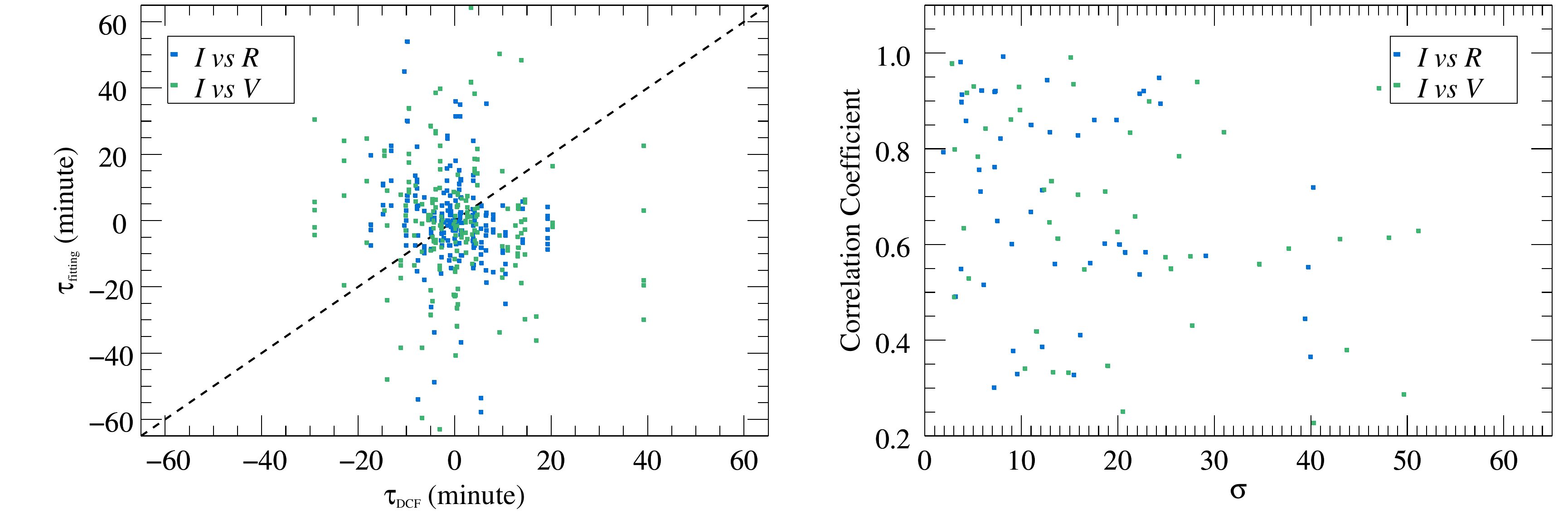}
	\end{minipage}\vspace{0.001cm}
	\caption{Left: Comparison between the time delays obtained by DCF and those obtained by fitting the flare. Right: Mean squared deviation calculated from the time delays obtained by the different methods vs. the correlation coefficients between their corresponding variability curves at different wavelengths.}
	\label{fig:tlagcompare}
\end{figure*}

The overall time delay between different bands is obtained by calculating the DCF for the IDV curves, and the definition of the time delay obtained by this method is different from that of the time delay we obtained by fitting the flares. The set of flares obtained during the decomposition of the variability curve is not unique and only one possible solution is provided here, thus there is a selection effect in the method of obtaining time delays by fitting. In order to compare the similarities and differences of the time delays obtained by these two methods, a statistical analysis of the DCF with the time delays obtained by fitting the flares is performed, as shown in Figure \ref{fig:tlagcompare}. As can be seen from the left panel, the time delay of each individual flare is closer to its overall time delay and fluctuates around its value. The right panel shows that, the larger the difference between the overall time delay and the time delay of each flare obtained from model decomposition, the worse the correlation of its IDV tends to be between different bands. 

The similarities and differences between the time delay of the flare and the IDV curves can be illustrated by this model, where the overall IDV fluctuations are generated by the superposition of radiation from many different small-scale emission regions. If these areas have similar physical environments, the time delays between the different bands will be similar, and thus the IDVs of different bands produced by the superposition will have a high correlation coefficient. Conversely, because the physical environment differs greatly, the different flares at different frequencies will have different time delays and amplitude ratios, and thus the correlation coefficients will tend to become worse. However, the correlation between IDVs of different bands is also affected by other reasons and the above inference is only taken as a possibility.

\begin{figure*}
	\begin{minipage}{\textwidth}
		\centering
		\includegraphics[trim=2.2cm 0.6cm 0.5cm 0.4cm,width=0.99\textwidth,clip]{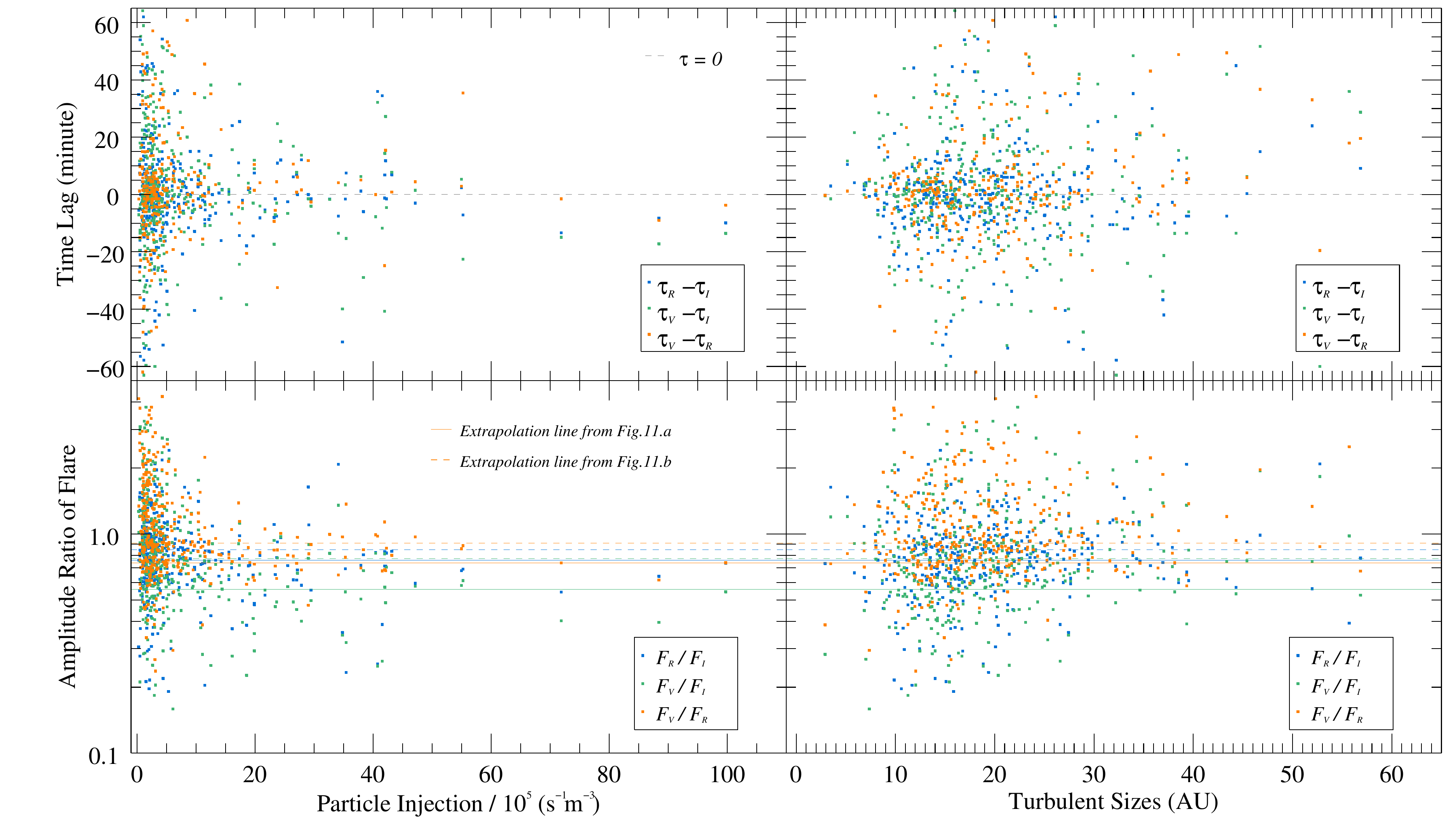}
		\includegraphics[trim=2.2cm 0.6cm 0.5cm 0.2cm,width=0.99\textwidth,clip]{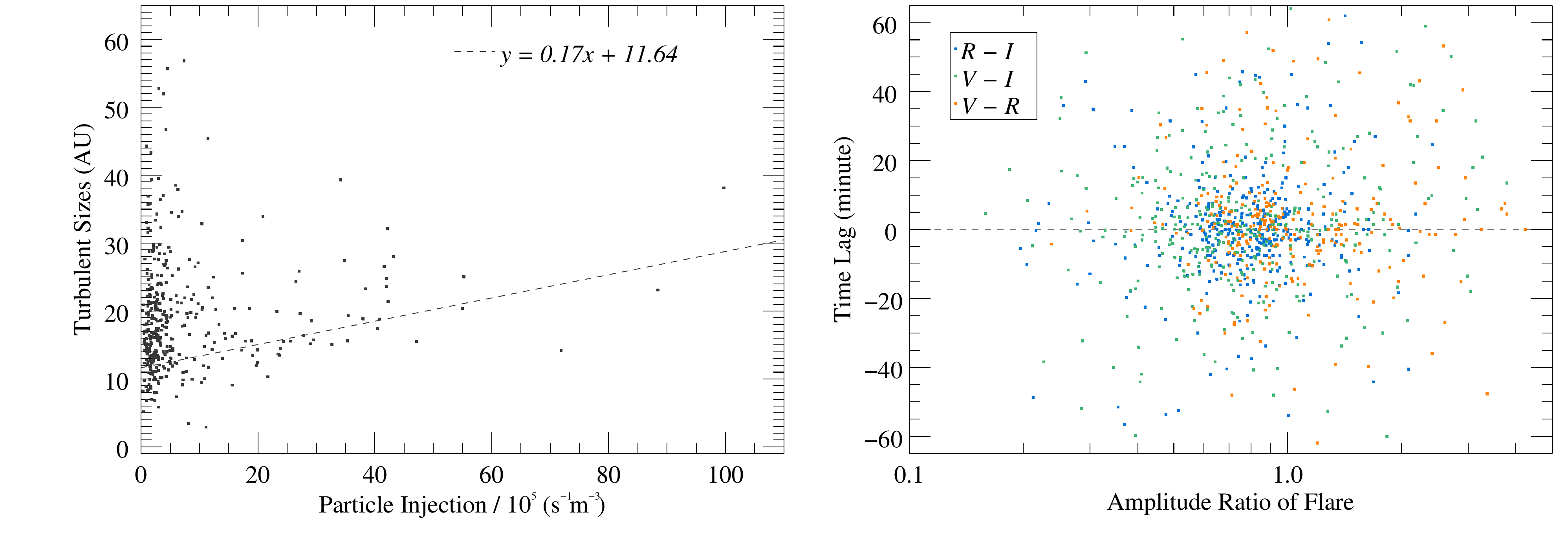}
	\end{minipage}\vspace{0.001cm}
	\caption{Statistical distribution of the correlation between particle injection rate, turbulence size, time delay, and amplitude ratio with each other. Here the particle injection rate and turbulence size are the mean values obtained from multi-band fitting. The blue, green and orange marked points indicate the statistical points of the fitting results for the time delay and amplitude ratio between different bands. The colored dashed and solid lines correspond to the amplitude ratios of different bands determined from the observed and enhanced fluxes in Fig.\ref{fig:ffdiagram}, respectively. The dotted lines indicate the possible approximations obtained by fitting.}
	\label{fig:relation}
\end{figure*}

In order to investigate whether there is a correlation between the parameters obtained from the fitting results, we have statistically analyzed the parameters such as the turbulence size of the flare, the particle injection rate, the amplitude ratio of different bands and the time delay, whose distributions are shown in Figure \ref{fig:relation}. From the statistical results, it can be seen that the distribution of the time delay is independent of the scale of the emission regions and the number of enhanced particles. However, the distribution of flare amplitude ratios in different bands differs in regions with different particle number density enhancements, and the amplitude ratio tends to a fixed value for larger particle number enhancement. We find that this value is close to the inferred value obtained from the flux-flux diagram, while the amplitude ratios are spread over a wide range when the injection rate is small. This suggests that the regions with stronger particle number enhancement are closer to or in large scale blobs in the jet, while the small turbulent regions may be widely distributed at different locations in the jet. 

In addition, we compared the enhancement of the particle number density and size of the cells obtained by the fitting. Although there is no strict dependence between them, statistically the particle injection rate tends to be larger as the cell scale increases. Our statistical results are in agreement with the study of \cite{Zra12}, where it was shown that the magnetic cells become larger as the turbulence grows. The size, location distribution, and degree of turbulence of these non-uniform regions have been a popular issue in microvariability studies in recent years, and the precise parameters of these regions need to be further investigated and verified.

\section{Summary and Discussion} \label{sec:sum}

We use the model of synchrotron cells in jets proposed and developed by \cite{Kir98}, \cite{Bha13}, and \cite{Xu19}. We apply the model to STV and develop a multi-region radiation model with multiple small-scale nonuniform regions in large scale plasma blobs. We assume that the LTV is dominated by the emission of the jet as a whole and the large-scale blobs within it, while the STV originates from the superposition of emission from several small-scale inhomogeneous regions in the jet. We interpret turbulent cells with various acceleration parameters as originating at different locations in the jet or blobs, while turbulent cells with similar parameters originate at the same or similar spatial locations. We provide constraints on these turbulent cells according to the evolution of the variability curves and the spectral indices.

We study the flare properties obtained from the KRM equation at various frequencies using different parameters (e.g., $\nu$, $B$, $\eta_{f}$, $\gamma_{0}$, $\gamma_{max}$, $t_{f}$, $t_{b}$ and $t_{esc}/t_{acc}$). The calculation and simulation results show that the flares have different profiles for different acceleration parameters, and different amplitude ratios, durations and peak times for different frequencies. The differences in multi-frequency amplitude ratios depending on the acceleration parameters indicate the diversity of the evolution of the spectral indices. The flare durations obtained at various frequencies differ slightly for the same spatially scaled acceleration and emission regions at optical wavelengths. The different peak times of flares obtained from the same excitation source at different frequencies imply that there may be inherent time delays between flares at different frequencies. The time delays between the I-band and the R, V, B, and U-bands were obtained as $0.41_{-0.41}^{+0.21}$ min, $0.91_{-0.20}^{+0.10}$ min, $1.56_{-0.40}^{+0.17}$ min, and $3.57_{-1.84}^{+1.16}$ min, respectively, from a 1-hour flare simulation using the baseline model parameter set. The time delays provided here are presented only as examples. The actual time delays depend strongly on the choice of parameter set used. However, in general, the time delays between optical bands for hourly scale flares are typically within a few minutes.

It was found that $t_{esc}/t_{acc}$, $\gamma_{0}$ and $\gamma_{max}$ are the main parameters affecting the amplitude ratios of flares. These parameters are related to the acceleration and escape rates, i.e. they correspond to the main parameters affecting the trend of the spectral index evolution. While other factors such as the Doppler factor may also lead to peak frequency shifts and influence the spectral index evolution, their influence on the evolutionary trend of the spectral index in a more monotonous way and are not the essential origin of the evolution of the spectral index. We suspect that variations in parameters including initial, maximum and broken energies of the accelerated particle population, as well as the acceleration and escape rates, comprise the primary factors responsible for the differing patterns in spectral index evolution. The superposition of the specificity of multiple small-scale emission regions leads to the diversity of the evolution of the spectral index at short timescales, while the overall evolution of the spectral index at LTV may be limited by the nature of the jet as a whole or the dominant emission region within it, for example, it may result from the change of the Doppler factor due to the change of the overall motion of the large-scale blobs in the jet.

We processed the observations of the BL Lacertae at Weihai Observatory of Shandong University between 2009 and 2021, and selected the well-sampled IDV curves of 89 nights for model fitting. Based on the previous study \citep{Xu19}, we improved the method for automatically fitting the IDV curves with considering the spectral index evolution pattern while fitting the multi-band variability curves simultaneously, that allows to constrain the number and location of the emission regions and the physical parameters. The number, center time, duration and amplitude of the flares in the IDV are obtained by model fitting. Based on this, we can calculate the amplitude ratio and time delay of the flares at different frequencies, thus limiting the acceleration parameters, the size of the turbulent cells, and the enhanced particle density, so that we can determine whether the flares originate from similar regions. Based on the model fitting results, we can well track the time-domain evolution of each flare as well as the trajectory of the spectral index evolution, which can well explain the various observed phenomena of BWB and/or RWB.

We have performed a statistical analysis of the fitted results, and the results show that the flare amplitudes and duration obtained from the decomposition of the IDV curves are consistent with a log-normal distribution. Among them, two peaks appear in the particle injection rate as well as in the distribution of the flare spectral index, and we speculate that at least two large-scale blobs exist as the main emission regions of the LTV during this ten-year observation period, while these observed short-term flares correspond to the small-scale non-uniform regions. The time delays between each optical band are about several minutes and they conform to the normal distribution. This is similar to the intrinsic time delays obtained from the theoretical model analysis. In addition, we obtained that the time delays of the IDV curves between different frequencies calculated by both methods, DCF and fitting procedure, are similar. We speculate that the variance of the time delays obtained by these two methods may be related to the correlation coefficients of the IDV curves between different frequencies, which is self-consistent with the assumptions of the multi-region radiation model. The larger the difference in the physical environment of the multiple emission regions included in the light-variance profile, the poorer the correlation of their superimposed resulting radiation fluxes. Considering that the fitting results are not unique, the above conjecture we present as a possibility only, and we will further justify it in future studies of LTV.

\begin{acknowledgments}
We thank the anonymous referee, Defu Bu and Suoqing Ji which help to polish of the manuscript. This work is supported by the Natural Science Foundation of China under grant No. 11873035, the Natural Science Foundation of Shandong province (No. JQ201702), and the Young Scholars Program of Shandong University (No. 20820162003).
\end{acknowledgments}

%%\bibliography{sample631}{}

\begin{thebibliography}{}
	\bibitem[Abeysekara et al.(2018)]{Abe18} Abeysekara, A.~U., Benbow, W., Bird, R., et al.\ 2018, \apj, 856, 95. doi:10.3847/1538-4357/aab35c
	\bibitem[Ackermann et al.(2011)]{Ack11} Ackermann, M., Ajello, M., Allafort, A., et al.\ 2011, \apj, 743, 171. doi:10.1088/0004-637X/743/2/171
	\bibitem[Agarwal \& Gupta(2015)]{Aga15} Agarwal, A. \& Gupta, A.~C.\ 2015, \mnras, 450, 541. doi:10.1093/mnras/stv625
	\bibitem[Ball \& Kirk(1992)]{Bal92} Ball, L. \& Kirk, J.~G.\ 1992, \apjl, 396, L39. doi:10.1086/186512
	\bibitem[Baring et al.(1997)]{Bar97} Baring, M.~G., W. Ogilvie, C. Ellison, et al.\ 1997, \apj, 476, 889. doi:10.1086/303645
	\bibitem[Baring et al.(2017)]{Bar17} Baring, M.~G., B{\"o}ttcher, M., \& Summerlin, E.~J.\ 2017, \mnras, 464, 4875. doi:10.1093/mnras/stw2344
	\bibitem[Bertaud et al.(1969)]{Ber69} Bertaud, C., Dumortier, B., V{\'e}ron, P., et al.\ 1969, \aap, 3, 436
	\bibitem[Bessell(1979)]{Bes79} Bessell, M.~S.\ 1979, \pasp, 91, 589. doi:10.1086/130542
	\bibitem[Bhatta et al.(2013)]{Bha13} Bhatta, G., Webb, J.~R., Hollingsworth, H., et al.\ 2013, \aap, 558, A92. doi:10.1051/0004-6361/201220236
	\bibitem[Bhatta et al.(2018)]{Bha18a} Bhatta, G., Mohorian, M., \& Bilinsky, I.\ 2018, \aap, 619, A93. doi:10.1051/0004-6361/201833628
	\bibitem[Bhatta \& Webb(2018)]{Bha18b} Bhatta, G. \& Webb, J.\ 2018, Galaxies, 6, 2. doi:10.3390/galaxies6010002
	\bibitem[Bhatta(2021)]{Bha21} Bhatta, G.\ 2021, \apj, 923, 7. doi:10.3847/1538-4357/ac2819
	\bibitem[Bloom et al.(1997)]{Blo97} Bloom, S.~D., Bertsch, D.~L., Hartman, R.~C., et al.\ 1997, \apjl, 490, L145. doi:10.1086/311035
	\bibitem[B{\"o}ttcher et al.(2013)]{Bot13} B{\"o}ttcher, M., Reimer, A., Sweeney, K., et al.\ 2013, \apj, 768, 54. doi:10.1088/0004-637X/768/1/54
	\bibitem[Camenzind \& Krockenberger(1992)]{Cam92} Camenzind, M. \& Krockenberger, M.\ 1992, \aap, 255, 59
	\bibitem[Cardelli et al.(1989)]{Car89} Cardelli, J.~A., Clayton, G.~C., \& Mathis, J.~S.\ 1989, \apj, 345, 245. doi:10.1086/167900
	\bibitem[Celotti \& Ghisellini(2008)]{Cel08} Celotti, A. \& Ghisellini, G.\ 2008, \mnras, 385, 283. doi:10.1111/j.1365-2966.2007.12758.x
	\bibitem[Chatterjee et al.(2012)]{Cha12} Chatterjee, R., Bailyn, C.~D., Bonning, E.~W., et al.\ 2012, \apj, 749, 191. doi:10.1088/0004-637X/749/2/191
	\bibitem[Chen et al.(2014)]{Che14} Chen, X., Hu, S.~M., Guo, D.~F., et al.\ 2014, \apss, 349, 909. doi:10.1007/s10509-013-1693-x
	\bibitem[D'Ammando et al.(2019)]{Dam19} D'Ammando, F., Raiteri, C.~M., Villata, M., et al.\ 2019, \mnras, 490, 5300. doi:10.1093/mnras/stz2792
	\bibitem[Edelson \& Krolik(1988)]{Edl88} Edelson, R.~A. \& Krolik, J.~H.\ 1988, \apj, 333, 646. doi:10.1086/166773
	\bibitem[Fiorucci \& Tosti(1996)]{Fio96} Fiorucci, M. \& Tosti, G.\ 1996, \aaps, 116, 403
	\bibitem[Gaur et al.(2015)]{Gau15} Gaur, H., Gupta, A.~C., Bachev, R., et al.\ 2015, \mnras, 452, 4263. doi:10.1093/mnras/stv1556
	\bibitem[Giannios et al.(2009)]{Gia09} Giannios, D., Uzdensky, D.~A., \& Begelman, M.~C.\ 2009, \mnras, 395, L29. doi:10.1111/j.1745-3933.2009.00635.x
	\bibitem[Giebels \& Degrange(2009)]{Gie09} Giebels, B. \& Degrange, B.\ 2009, \aap, 503, 797. doi:10.1051/0004-6361/200912303
	\bibitem[Hagen-Thorn(1997)]{Hag97} Hagen-Thorn, V.~A.\ 1997, Astronomy Letters, 23, 19
	\bibitem[Hagen-Thorn et al.(2002)]{Hag02} Hagen-Thorn, V.~A., Larionova, E.~G., Jorstad, S.~G., et al.\ 2002, \aap, 385, 55. doi:10.1051/0004-6361:20020145
	\bibitem[Hu et al.(2014)]{Hu14} Hu, S.-M., Han, S.-H., Guo, D.-F., et al.\ 2014, Research in Astronomy and Astrophysics, 14, 719-732. doi:10.1088/1674-4527/14/6/010
	\bibitem[Jorstad et al.(2013)]{Jor13} Jorstad, S.~G., Marscher, A.~P., Smith, P.~S., et al.\ 2013, \apj, 773, 147. doi:10.1088/0004-637X/773/2/147
	\bibitem[Kirk et al.(1998)]{Kir98} Kirk, J.~G., Rieger, F.~M., \& Mastichiadis, A.\ 1998, \aap, 333, 452
	\bibitem[Larionov et al.(2008)]{Lar08} Larionov, V.~M., Jorstad, S.~G., Marscher, A.~P., et al.\ 2008, \aap, 492, 389. doi:10.1051/0004-6361:200810937
	\bibitem[Madejski et al.(1999)]{Mad99} Madejski, G.~M., Sikora, M., Jaffe, T., et al.\ 1999, \apj, 521, 145. doi:10.1086/307524
	\bibitem[Marscher \& Gear(1985)]{Mar85} Marscher, A.~P. \& Gear, W.~K.\ 1985, \apj, 298, 114. doi:10.1086/163592
	\bibitem[Marscher et al.(1992)]{Mar92} Marscher, A.~P., Gear, W.~K., \& Travis, J.~P.\ 1992, Variability of Blazars, 85
	\bibitem[Marscher et al.(2008)]{Mar08} Marscher, A.~P., Jorstad, S.~G., D'Arcangelo, F.~D., et al.\ 2008, \nat, 452, 966. doi:10.1038/nature06895
	\bibitem[Miller et al.(1978)]{Mil78} Miller, J.~S., French, H.~B., \& Hawley, S.~A.\ 1978, \apjl, 219, L85. doi:10.1086/182612
	\bibitem[Nilsson et al.(2018)]{Nil18} Nilsson, K., Lindfors, E., Takalo, L.~O., et al.\ 2018, \aap, 620, A185. doi:10.1051/0004-6361/201833621
	\bibitem[Potter \& Cotter(2012)]{Pot12} Potter, W.~J. \& Cotter, G.\ 2012, \mnras, 423, 756. doi:10.1111/j.1365-2966.2012.20918.x
	\bibitem[Potter(2018)]{Pot18} Potter, W.~J.\ 2018, \mnras, 473, 4107. doi:10.1093/mnrasx2371
	\bibitem[Quirrenbach et al.(1991)]{Qui91} Quirrenbach, A., Witzel, A., Wagner, S., et al.\ 1991, \apjl, 372, L71. doi:10.1086/186026
	\bibitem[Raiteri et al.(2013)]{Rai13} Raiteri, C.~M., Villata, M., D'Ammando, F., et al.\ 2013, \mnras, 436, 1530. doi:10.1093/mnras/stt1672
	\bibitem[Sakimoto et al.(2013)]{Sak13} Sakimoto, K., Uemura, M., Sasada, M., et al.\ 2013, \pasj, 65, 35. doi:10.1093/pasj/65.2.35
	\bibitem[Sambruna et al.(1999)]{Sam99} Sambruna, R.~M., Ghisellini, G., Hooper, E., et al.\ 1999, \apj, 515, 140. doi:10.1086/307005
	\bibitem[Schlafly \& Finkbeiner(2011)]{Sch11} Schlafly, E.~F. \& Finkbeiner, D.~P.\ 2011, \apj, 737, 103. doi:10.1088/0004-637X/737/2/103
	\bibitem[Summerlin \& Baring(2012)]{Sum12} Summerlin, E.~J. \& Baring, M.~G.\ 2012, \apj, 745, 63. doi:10.1088/0004-637X/745/1/63
	\bibitem[Uttley et al.(2002)]{Utt02} Uttley, P., McHardy, I.~M., \& Papadakis, I.~E.\ 2002, \mnras, 332, 231. doi:10.1046/j.1365-8711.2002.05298.x
	\bibitem[Villata et al.(2002)]{Vil02} Villata, M., Raiteri, C.~M., Kurtanidze, O.~M., et al.\ 2002, \aap, 390, 407. doi:10.1051/0004-6361:20020662
	\bibitem[Villata et al.(2009)]{Vil09} Villata, M., Raiteri, C.~M., Larionov, V.~M., et al.\ 2009, \aap, 501, 455. doi:10.1051/0004-6361/200912065
	\bibitem[Weaver et al.(2020)]{Wea20} Weaver, Z.~R., Williamson, K.~E., Jorstad, S.~G., et al.\ 2020, \apj, 900, 137. doi:10.3847/1538-4357/aba693
	\bibitem[Webb et al.(2010)]{Web10} Webb, J.~R., Bhatta, G., \& Hollingsworth, H.\ 2010, \aas
	\bibitem[Webb et al.(2021)]{Web21} Webb, J.~R., Arroyave, V., Laurence, D., et al.\ 2021, Galaxies, 9, 114. doi:10.3390/galaxies9040114
	\bibitem[Xu et al.(2019)]{Xu19} Xu, J., Hu, S., Webb, J.~R., et al.\ 2019, \apj, 884, 92. doi:10.3847/1538-4357/ab3e50
	\bibitem[Zrake \& MacFadyen(2012)]{Zra12} Zrake, J. \& MacFadyen, A.~I.\ 2012, \apj, 744, 32. doi:10.1088/0004-637X/744/1/32
	
\end{thebibliography}
%%\bibliographystyle{aasjournal}

\end{document}